\documentclass[12pt, lot, lof]{puthesis}
\newcommand{\proquestmode}{}

\title{Chasing Hamiltonian structure in gyrokinetic theory}
\author{J. W. Burby}
\adviser{H. Qin}
\submitted{September 2015}
\copyrightyear{2015}
\department{Astrophysical Sciences\\Program in Plasma Physics}

\usepackage{natbib}
\usepackage[english]{babel}
\usepackage[svgnames]{xcolor}
\usepackage{bm,graphicx, fancyhdr, setspace, times, multicol, multirow, hhline, latexsym, amssymb, amsthm, amsmath,bm}
\usepackage{dcolumn, float, rotating, chapterbib, comment, longtable, verbatim, booktabs, lgreek, epsfig, tabulary, threeparttable}
\usepackage{calligra}
\usepackage[colorlinks=true, linkcolor=blue]{hyperref}
\usepackage{xfrac}
\usepackage{amsthm} 

\ifdefined\printmode

	\usepackage{url}

\else

	\ifdefined\proquestmode 

		\usepackage{hyperref}
		\hypersetup{bookmarksnumbered}

		\makeatletter
		\hypersetup{pdftitle=\@title,pdfauthor=\@author}
		\makeatother

		\usepackage[hyphenbreaks]{breakurl}

	\else	

		\usepackage{hyperref}
		\hypersetup{colorlinks, bookmarksnumbered}

		\makeatletter
		\hypersetup{pdftitle=\@title,pdfauthor=\@author}
		\makeatother

		\usepackage[hyphenbreaks]{breakurl}

	\fi 

\fi 

\usepackage{algorithmic,eqparbox}
\usepackage[chapter]{algorithm}

\DeclareMathAlphabet{\mathcalligra}{T1}{calligra}{m}{n}
\DeclareFontShape{T1}{calligra}{m}{n}{<->s*[2.1]callig15}{}

\newcommand{\calr}{\mathcalligra{r}}

\newcommand{\ddt}{\frac{\mathrm{d}}{\mathrm{d}t}}
\newcommand{\B}[1]{\mathbf{#1}}
\newcommand{\T}[1]{\tilde{#1}}

\newtheorem{theorem}{Theorem}

\abstract{

Hamiltonian structure is pursued and uncovered in collisional and collisionless gyrokinetic theory. A new Hamiltonian formulation of collisionless electromagnetic theory is presented that is ideally suited to implementation on modern supercomputers. The method used to uncover this structure is described in detail and applied to a number of examples, where several well-known plasma models are endowed with a Hamiltonian structure for the first time. The first energy- and momentum-conserving formulation of full-F collisional gyrokinetics is presented. In an effort to understand the theoretical underpinnings of this result at a deeper level, a \emph{stochastic} Hamiltonian modeling approach is presented and applied to pitch angle scattering. Interestingly, the collision operator produced by the Hamiltonian approach is equal to the Lorentz operator plus higher-order terms, but does not exactly conserve energy. Conversely, the classical Lorentz collision operator is provably not Hamiltonian in the stochastic sense.

}
\acknowledgements{

It's been much harder to learn how to \emph{disengage} from my thesis research than it was to become engrossed in it. In part, this is due to the complete autonomy conferred on me by my advisor, Hong; it's a symptom of working on very little, other than topics dear to my heart. In equal measure, my addictive personality is to blame. For all of the days I spent happily fiddling with Poisson brackets and variational principles, there were nights when I wished I could forget about such things, and just sleep.

For the pleasant times I've had as a paid free thinker at Princeton, I need to express my extreme gratitude to the PPPL staff who continually encouraged and cultivated my interests in hard-to-market, mathematically-oriented basic theory problems. Hong Qin, John Krommes, Bill Tang, Roscoe White, Ilya Dodin, Nat Fisch, Cynthia Phillips, Doug Darrow, Amitava Bhattacharjee, and Gerrit Kramer: thank you, my interactions with you have been incalculably helpful, and genuinely enjoyable. Barbara, I can't thank you enough for your detailed and patient descriptions of how to do normal things at the lab without suffering the wrath of Uncle Sam. Beth, you've made the process of  attempting to graduate a real pleasure.

For pulling me away from research when I really needed to be, I need to thank my close friends. Jono and Seth, I can't imagine better housemates than you guys; you've made living in Princeton a lot less stressful and lonely than it might have been in your absence. Alex, while I only had the pleasure of being your housemate for a year, it was a year well-spent; I wish you, Jack, Jono, Seth and I lived together sooner. Lei and Yao, all of the time you spent patiently introducing me to Chinese food in the US and in China hasn't been wasted; our numerous visits to the Chinese restaurants around Princeton kept me sane and well-fed. Lee, you've been my brother at Princeton. I'm sad our career paths will no longer be coincident, but I'm excited to see where you end up. Mike, you've helped me appreciate and laugh about many of the absurd aspects of academia and life in general. Lan, our runs have been a necessary distraction for me (and I hope your foot gets better.)

Looking ahead, I'm glad that I'll be spending the next few years fairly close to my parents, brothers, and dogs in Saratoga Springs. I relish the thought of being near enough to see my parents retire, my brothers settle into jobs they enjoy, and my dogs grow out of puppyhood.

}
\dedication{If it's in a word, or it's in a look, you can't get rid of the Babadook.}

\begin{document}

\pagestyle{fancy}
\headheight 15pt
\fancyhf{}
\fancyhead[RE,RO]{\thepage}
\fancyhead[LE,LO]{\scshape\nouppercase \leftmark}
\makefrontmatter

\pagestyle{fancy}
\headheight 15pt
\fancyhf{}
\fancyhead[RE,RO]{\thepage}
\fancyhead[LE,LO]{\scshape\nouppercase \leftmark}



\chapter[Introduction]{Introduction}
\label{ch:intro}

On microscopic scales, the physics of plasmas is Hamiltonian in nature. Neglecting quantum, relativistic, and radiative effects for simplicity's sake, the microscopic description of a plasma consists of a separate instance of the Lorentz force law for each plasma particle along with Maxwell's equations to couple everything together. This system of equations can be derived from a variational principle, which in turn can be used to derive a microscopic Hamiltonian functional and Poisson bracket.  These equations governing microscopic plasma physics have been called the Klimontovich-Maxwell system, and their Lagrangian formulation is described for instance in \cite{Qin_PRE_2014}.

With Hamiltonian structure ingrained so deeply in the foundation of the subject, plasma theory ought to be some grand exercise in the broader theory of Hamiltonian systems. And when viewed from a great distance, it is! However, for those in the trenches, studying plasma theory on a day-to-day basis, things seem different. The equations governing microscopic plasma theory are so hopelessly complicated that \emph{reduced} plasma models are typically preferable to the Klimontovich-Maxwell model. These reduced models are obtained by carefully and cleverly applying Occam's razor in order to tame the mathematical morass presented by the microscopic equations of motion. Sometimes, for instance in the case of ideal magnetohydrodynamics, the reduced model is provably Hamiltonian in nature\, \cite{Morrison_MHD_1980} \footnote{In the case of ideal MHD, as well as many other Hamiltonian reduced models, it is still unclear how the reduced model's Hamiltonian structure is related to the microscopic Hamiltonian structure.}. However, often times the connection between the reduced model and Hamiltonian mechanics is hazy at best. The most striking examples of this divorce from Hamiltonian mechanics arise when collisions must be accounted for within the confines of a continuum model. For instance the Landau collision operator probably cannot be derived from a conventional variational principle. The same is true of the more-fundamental Balescu-Lenard collision operator. More generally, there is a common feeling amongst plasma physicists that ``dissipative dynamics are not Hamiltonian." Even when collisions are neglected, there are reduced models that either fail to be Hamiltonian in nature, or so far have resisted attempts to be cast in Hamiltonian form.

This thesis is devoted to revealing some new connections between Hamiltonian mechanics and the particularly interesting reduced plasma model known as gyrokinetics. I believe that it illustrates two general points in connection with the ``Hamiltonian dichotomy" between reduced plasma models and the microscopic plasma model just described. First, Hamiltonian mechanics can be surprisingly useful in the study of reduced plasma models; the benefits of exploiting the Hamiltonian formalism to formulate and study reduced models can be unpredictable. Second, the connection between collisional plasma models and Hamiltonian mechanics is surprisingly deep; while collision operators do not fit within the traditional Hamiltonian framework, they may very well fit within a \emph{stochastic} Hamiltonian framework\, \cite{Lazaro-Cami_2008}. The first point is covered roughly by Chapters 2 through 4, while the second point is discussed in Chapter 5.

In Chapter 2, I present the results of an attempt to cast collisionless electromagnetic gyrokinetics in Hamiltonian form, a theory that already enjoys several \emph{Lagrangian} formulations. The earliest of the Lagrangian formulations are given in \cite{Sugama_2000,Brizard_PRL_2000,Brizard_POP_2000}, while more recent additions can be found in \cite{Pfirsch_2004,Squire_GK_2013}. Given the typical intimate relationship between the Hamiltonian and Lagrangian 
formalisms, it is tempting to believe that the Hamiltonian structure of electromagnetic gyrokinetics should be straightforward to obtain. This was not the case. The usual Legendre transform technique fails when applied to gyrokinetic Lagrangians (one problem is these Lagrangians are linear in the phase space velocity of plasma species). Nevertheless, the existence of Lagrangian formulations of electromagnetic gyrokinetics suggested that a Hamiltonian formulation \emph{should} exist; the riddle was how to find it. In the process of searching for this Hamiltonian structure, and with guidance from Professor Morrison, I found it technically convenient to slightly reformulate electromagnetic gyrokinetics in order to work with a manifestly gauge-invariant\footnote{Previous work on Lagrangian electromagnetic gyrokinetics was usually done in the Coulomb gauge.} theory along the lines of \cite{Morrison_lifting_2013}. Surprisingly, this reformulation turned out to have several features that make it especially well-suited to simulation on modern supercomputers. This is the first example in the thesis of a surprising consequence of pursuing Hamiltonian structure in reduced plasma models.

In chapter 3 I give an extended account of the theoretical machine used to derive the gyrokinetic Poisson bracket from Chapter 2. This machine, which eats (possibly degenerate) Lagrangians and spits out Poisson brackets, is very closely related to the Peierls bracket formalism  \cite{Peierls_1952}, as well as the Dirac constraint formalism used in \cite{Squire_GK_2013}. Nevertheless, several of the examples worked out in this chapter (besides electromagnetic gyrokinetics) are new. The first example that contains a new result concerns the Vlasov-Darwin system. This system had previously been cast in Hamiltonian form in \cite{Krauss_2007} using position-canonical momentum coordinates on the single-particle phase space. The novelty of the example in this Chapter is that the derivation of the bracket is done using position-velocity coordinates on the single-particle phase space (which leads to a different expression for the bracket.) The second novel example is concerned with deriving a bracket for the (quasi) neutral Vlasov system introduced by C. Tronci and E. Camporeale in \cite{Tronci_2015}. Tronci and Camporeale provide a Lagrangian formulation of this reduced model, but stop short of passing to the Hamiltonian side. Thus, this example presents a Poisson bracket and Hamiltonian functional for the neutral Vlasov model for the first time. The final novel example derives a Poisson bracket for force-free electrodynamics\, \cite{Gralla_2014}. Previously, a canonical bracket for this system was given in terms of an Euler potential representation of the electromagnetic field. The non-canonical bracket presented here does not use the Euler potential representation, and so is more general and potentially more useful for future stability analyses.

Chapter 4 presents a result obtained in conjunction with Professor Brizard on an energy and momentum conserving nonlinear collision operator for full-f gyrokinetics. This work serves as the second example of a surprising consequence of applying the Hamiltonian formalism. The key insight that lead to this collision operator was a peculiar way of expressing the particle-space Landau operator in terms of single-particle Poisson brackets. A deep reason as to why the Poisson bracket representation is as useful as it appears to be is still missing, and this is why the result is somewhat surprising. 

Finally, Chapter 5 contains the beginnings of a search for Hamiltonian structure underlying reduced collisional models at a deeper level than discussed in Chapter 4. It begins by describing generally how stochastic acceleration problems can be formulated in terms of \emph{stochastic Hamiltonian mechanics}\, \cite{Lazaro-Cami_2008}. Where ordinary Hamiltonian mechanics is concerned with one-parameter subgroups of the phase space symplectomorphism group, stochastic Hamiltonian mechanics is concerned with Brownian motion on the symplectomorphism group. The remainder of the chapter is then devoted to applying these ideas to the pitch angle scattering problem, which can be formulated as an example of stochastic acceleration. A surprise here is that there is a tension between energy conservation and the stochastic Hamiltonian formalism. The collision operator produced using the Hamiltonian approach is equal to the Lorentz collision operator plus higher-order terms, but does not exactly conserve kinetic energy. Conversely, the classical Lorentz operator is provably outside the realm of stochastic Hamiltonian mechanics. As I discuss at the end of the chapter, it seems likely that a way to overcome this problem is to slighly relax the Markov approximation. However, this realization came too late in my stay at Princeton, and so I was not able to explore it more fully in this thesis.


\chapter[Hamiltonian formulation of the gyrokinetic Vlasov-Maxwell equations]{Hamiltonian formulation of the gyrokinetic Vlasov-Maxwell equations}
\label{ch:GKVM}

\section{Introduction}
Electromagnetic gyrokinetic theory (EMGT) is a model used to describe the turbulent transport of particles and heat induced by fluctuating electric and magnetic fields in strongly magnetized plasmas. EMGT is, in many ways, a more utilitarian  tool than the more-fundamental Vlasov-Maxwell kinetic theory (VMKT). However, VMKT enjoys two important advantages over existing formulations of EMGT. (I) When simulated on a computer, the VMKT field solve is local; advancing the electromagnetic field in time at a given grid point only requires communication with nearby grid points \cite{Bowers_2009}. (II) There is an energy principle for assessing the stability of Vlasov-Maxwell equilibria \cite{Morrison_Pfirsch_1989} (also see \cite{Kruskal_Oberman_1958,Holm_stability_1985,Morrison_fluid_1998,Andreussi_2012,Andreussi_2013} for similar energy principles in other contexts). In contrast, modern EMGT simulations require global Poisson-like field solves at each time step. This prevents EMGT simulations from scaling as favorably \cite{Madduri_2011} as VMKT simulations when the number of processing cores is increased at fixed problem size. Likewise, the free energy of perturbations to EMGT equilibria is unknown. Thus, the basic tool for studying the stability of EMG equilibria by way of an energy principle is unavailable. The purpose of this Chapter is to describe a new formulation of electromagnetic gyrokinetics that enjoys properties (I) and (II). The new formulation, which we will refer to as the gyrokinetic Vlasov-Maxwell (GVM) system,  enjoys a local field solve and has an energy principle, while retaining  the traditional advantages of gyrokinetic theory.

\section{The new formulation }The gyrokinetic Vlasov-Maxwell equations are given by
\begin{subequations}
\label{gvm}
\begin{align}
\frac{\partial f_s}{\partial t}&=-L_{V_s^{\text{gy}}}f_s\label{gk_vlasov}\\
\frac{1}{c}\frac{\partial\bm{D}}{\partial t}&=\nabla\times\bm{H}-\frac{4\pi}{c}\bm{J}_{\text{gy}}\label{gk_ampere}\\
\frac{1}{c}\frac{\partial\bm{B}}{\partial t}&=-\nabla\times\bm{E}\label{gk_faraday}\\
\nabla\cdot\bm{D}&=4\pi\rho_{\text{gy}}\label{gk_gauss}\\
\nabla\cdot\bm{B}&=0.\label{gk_divb}
\end{align}
\end{subequations}
$f_s$ is the gyrocenter volume form of species $s$, $V_s^{\text{gy}}$ is the gyrocenter phase space velocity, $L_{V_s^\text{gy}}$ denotes the Lie derivative along the gyrocenter phase space velocity, $\bm{J}_{\text{gy}}$ is the gyrocenter current density, $\rho_{\text{gy}}$ is the gyrocenter charge density, $\bm{E},\bm{B}$ are the fluctuating electric and magnetic fields, and $\bm{D},\bm{H}$ are the auxiliary electric and magnetic fields. The volume form $f_s$ is defined by requiring that the number of particles of species $s$ in a region of phase space $U$ be given by $\int_Uf_s$. The gyrocenter phase space velocity is specified by the time-dependent tensor form of Hamilton's equations,
\begin{align}\label{velocity}
\text{i}_{V_s^{\text{gy}}}\omega_s^{\text{gy}}=\mathbf{d}K_s-e_s\bm{E}\cdot d\bm{X},
\end{align}
where $\omega_s^{\text{gy}}$ is the gyrocenter symplectic form, $K_s$ is the gyrocenter kinetic energy, and $d\bm{X}$ denotes the vector line element in the space of gyrocenter positions. The gyrocenter symplectic form  is the sum of the guiding center symplectic form \cite{Cary_2009,Burby_gc_2013} and the fluctuating magnetic flux,
\begin{align}
\omega_s^{\text{gy}}=\omega_s^{\text{gc}}-\frac{e_s}{c}\bm{B}\cdot dS,
\end{align}
where $dS$ is the surface element in the space of gyrocenter positions. The gyrocenter kinetic energy is a functional of the fluctuating electric and magnetic fields, and is related to the gyrocenter Hamiltonian by $H_s^{\text{gy}}=K_s+e_s\phi$ (an explicit expression for $K_s$ will be given near the end of this Chapter). 
The auxiliary fields $\bm{D},\bm{H}$ are related to $\bm{E},\bm{B}$ by using  relations that emerge from the Hamiltonian theory developed in \cite{Morrison_lifting_2013}, i.e.,  the constitutive relations are given by
\begin{align}
\bm{D}&=\bm{E}-4\pi\frac{\delta\mathcal{K}}{\delta \bm{E}}\\
\bm{H}&=\bm{B}+4\pi\frac{\delta\mathcal{K}}{\delta \bm{B}},
\end{align}
where $\mathcal{K}(f,\bm{E},\bm{B})=\sum_s\int f_sK_s(\bm{E},\bm{B})$. 

Following \cite{Morrison_lifting_2013}  the  system  above constitutes  an infinite-dimensional Hamiltonian system with  dynamical variables $f$, $\bm{D}$, and $\bm{B}$, and   Hamiltonian functional  given by
\begin{align}\label{gvm_hamiltonian}
\mathcal{H}(f,\bm{D},\bm{B})&=\mathcal{K}(f,\hat{\bm{E}},\bm{B})+\int\hat{\bm{P}}\cdot\hat{\bm{E}}\,d^3\bm{X}\nonumber\\
&+\frac{1}{8\pi}\int\bigg(\hat{\bm{E}}\cdot\hat{\bm{E}}+\bm{B}\cdot\bm{B}\bigg)\,d^3\bm{X},
\end{align}
where $\hat{\bm{E}}=\hat{\bm{E}}(f,\bm{D},\bm{B})$ is the electric field operator defined implicitly by the equation
\begin{align}
\bm{D}&=\hat{\bm{E}}(f,\bm{D},\bm{B})-4\pi\frac{\delta\mathcal{K}}{\delta \bm{E}}(f,\hat{\bm{E}}(f,\bm{D},\bm{B}),\bm{B}),
\end{align}
and $\hat{\bm{P}}=\hat{\bm{P}}(f,\bm{D},\bm{B})$ is the gyrocenter polarization operator given by
\begin{align}
\hat{\bm{P}}(f,\bm{D},\bm{B})=\frac{1}{4\pi}(\bm{D}-\hat{\bm{E}}(f,\bm{D},\bm{B})).
\end{align}
The noncanonical Poisson bracket is given by
\begin{align}\label{the_bracket}
&[\mathcal{F},\mathcal{G}]=\nonumber\\
&\sum_{s=1}^{N_s}\int\mathcal{B}_{s}^{\text{gy}}\left(\mathbf{d}\frac{\delta\mathcal{F}}{\delta f_s}-4\pi e_s\frac{\delta\mathcal{F}}{\delta \bm{D}}\cdot d\bm{X},\mathbf{d}\frac{\delta\mathcal{G}}{\delta f_s}-4\pi e_s\frac{\delta\mathcal{G}}{\delta\bm{D}}\cdot d\bm{X}\right)\,f_s\nonumber\\
&+4\pi c\int\bigg(\frac{\delta\mathcal{F}}{\delta \bm{D}}\cdot\nabla\times\frac{\delta\mathcal{G}}{\delta\bm{B}}-\frac{\delta\mathcal{G}}{\delta \bm{D}}\cdot\nabla\times\frac{\delta\mathcal{F}}{\delta \bm{B}}\bigg)\,d^3\bm{X}.
\end{align} 
Here $\mathcal{B}_s^{\text{gy}}$ is the gyrocenter Poisson tensor, which is defined as follows. If $z^a$ is a coordinate system on the gyrocenter phase space and $\alpha,\beta$ are $1$-forms on the same space, $\mathcal{B}_s^{\text{gy}}(\alpha,\beta)=\alpha_a\beta_b\{z^a,z^b\}_s^{\text{gy}}$, where $\{\cdot,\cdot\}_s^{\text{gy}}$ is the gyrocenter Poisson bracket. Note that a Poisson bracket for electrostatic gyrokinetics was given in \cite{Squire_GK_2013}. The complexity of that bracket should be contrasted with the relative simplicity of the bracket given here for electromagnetic gyrokinetics. This bracket, which  has  a form akin to that of  \cite{Morrison_lifting_2013}, is to our knowledge the first demonstration of Hamiltonian structure  for {\em any} electromagnetic gyrokinetic theory. 

\section{Origins and comparisons }We arrived at this electromagnetic gyrokinetic system by modifying the standard variational derivation of electromagnetic gyrokinetics \cite{Sugama_2000,Brizard_PRL_2000,Brizard_POP_2000,Brizard_2007,Squire_GK_2013}. In the standard approach, a gyrokinetic system Lagrangian is constructed by adding a gauge-dependent \cite{Xu_Wang_2013} net gyrocenter Lagrangian to a non-relativistic limit (known as the Darwin limit) of the free Maxwell field Lagrangian. Applying appropriate variations to the system Lagrangian then produces the standard equations of EMGT. Roughly speaking, adopting a gyrocenter Lagrangian instead of a particle Lagrangian amounts to dropping terms from the particle equations of motion. Likewise, adopting the Darwin approximation amounts to dropping terms from Maxwell's equations.  We modified this approach by adding a manifestly gauge invariant net gyrocenter Lagrangian \cite{Pfirsch_2004} to the full free Maxwell field Lagrangian to produce the system Lagrangian. Thus, in the modified approach, fewer terms are dropped from Maxwell's equations. While dropping these terms as in the standard approach would be justified (using the assumption of non-relativistic particles), doing so is not necessary. We therefore conclude that the GVM equations are no less accurate than standard EMGT.

\section{Computational benefits }The usual argument for invoking the Darwin approximation in EMGT is that doing so eliminates light waves. This may seem to be an especially compelling argument from a computational point of view. After all, the presence of traveling waves with phase velocity $c$ leads to a very restrictive CFL condition for explicit integration schemes. Therefore, avoiding the Darwin approximation as we have done may appear objectionable in a practical sense.

On the other hand, this numerical argument supporting the Darwin approximation is not as strong as it appears. As is evident from the form of the GVM equations given above, avoiding the Darwin approximation does not lead to Maxwell's equations, but Maxwell's equations in a polarized and magnetized medium. Therefore, the light waves supported by these equations do not travel at the speed of light in vacuum.

It is well known \cite{Krommes_1993} that the dielectric constant resulting from gyrocenter polarization is large, which implies that the speed of light is much smaller than $c$ in a gyrokinetic plasma (this is consistent with the notion of a so-called ``gyrokinetic vacuum"). Using the long-wavelength limit of the gyrokinetic dielectric function, $\omega_{pi}^2/\omega_{ci}^2$, as a rough approximation, we find that light waves in the GVM equations propagate at the Alfv\'en speed. Thus, the CFL constraint imposed by light waves in the GVM equations is not nearly as strict as the usual argument might suggest \footnote{Strictly speaking, it is only light waves that travel perpendicular to the magnetic field that experience a reduced propgation speed. Those that travel along the magnetic field lines may still travel near the speed of light in vacuum. However, the numerical grids appropriate for gyrokinetic simulations are significantly elongated along the field lines, which substantially reduces the parallel CFL condition. }. 

An even stronger case can be made for the computational viability of this new formulation of electromagnetic gyrokinetics. We first make the following simple observation. A familiar calculation shows that if $\nabla\cdot\bm{D}=4\pi\rho_\text{gy}$ and $\nabla\cdot\bm{B}=0$ at $t=0$, then these equations will also be satisfied for all subsequent times. This means that the evolution of the magnetic field and the auxiliary electric field is completely determined by the Amp\`ere equation and the Faraday equation. Interestingly, it can be shown that this property arises as a direct consequence of employing a gauge-invariant gyrocenter Lagrangian; the quantity $\nabla\cdot\bm{D}-4\pi\rho_{\text{gy}}$ is the conserved quantity associated with gauge symmetry by Noether's theorem.

Now suppose the Amp\`ere and Faraday equations were used to advance $\bm{D}$ and $\bm{B}$ in time on a computer. Employing a simple explicit scheme, the following steps would have to be taken at each time step. (1) Using the constitutive relations, compute $\bm{E}$ and $\bm{H}$ from the known values of $\bm{D}$ and $\bm{B}$. (2) Compute $\nabla\times\bm{H}$ and $\nabla\times\bm{E}$. (3) Using a finite difference approximation for the partial time derivative, solve for the new $\bm{D}$ and $\bm{B}$. 

Steps (2) and (3) clearly require only local operations, and so represent nearly embarrassingly parallel computations. Again invoking the long wavelength limit, step (1) can also be seen to be local. In this limit, there is a simple algebraic relationship between $\bm{D}$ and $\bm{E}$ (see \cite{Brizard_2013}, for example) that can be inverted analytically. Thus, the entire field solve step in an explicit time marching scheme for the GVM equations is nearly embarrassingly parallel. Such a field solve is preferable to the nonlocal Poisson-like solves necessary in conventional EMGT, especially when performing parallel simulations with very few particles per processing core.

\section{Theoretical benefits }We will now turn from numerical benefits offered by the GVM equations in order to discuss their analytical benefits. First, we mention the system's conservative properties. An immediate consequence of the GVM Poisson bracket structure is conservation of the Hamiltonian functional (this follows from antisymmetry of the bracket). It is also not difficult to show that there is a conserved momentum functional for each rotation or translation symmetry of the background magnetic field. Finally, there is a large family of conserved functionals given by the Poisson bracket's Casimirs. These are functionals $C$ that Poisson commute with every other functional, i.e. $\forall \mathcal{F}$, $[C,\mathcal{F}]=0$. Systems of gyrokinetic equations (electromagnetic or electrostatic) with exact energy and momentum conservation laws can also be derived using the standard variational approach \cite{Scott_2010,Sugama_2000,Brizard_PRL_2000,Brizard_POP_2000,Pfirsch_2004,Squire_GK_2013}. Indeed, this was the main motivation for developing the standard variational formulations of gyrokinetics. However, variational approaches do not readily produce the Casimir invariants (nor has it been shown that the usual variational formulations of EMGT possess Poisson brackets and Casimir invariants at all).

Many of the GVM bracket's Casimirs are given as follows.  Let
\begin{align}
\Omega_s=-\frac{1}{3!}\omega_s^{\text{gy}}\wedge\omega_s^{\text{gy}}\wedge\omega_s^{\text{gy}}
\end{align}
be the Liouville volume form defined by the gyrocenter symplectic form and introduce the gyrocenter distribution \emph{function}, $F_s$, where
\begin{align}
f_s=F_s\Omega_s,
\end{align} 
then
\begin{align}
C_h=\sum_{s=1}^{N_s}\int_{TQ}h_s(F_s)\,\Omega_s
\end{align}
is a Casimir for each function of a single real variable $h_s$. Moreover, any functional of $\nabla\cdot\bm{D}-4\pi\rho_{\text{gy}}$ is a Casimir, which is one way of seeing that Eq.\,(\ref{gk_gauss}) is satisfied in the Hamiltonian formulation of the GVM equations.

Another advantage the Poisson bracket formulation of the GVM equations provides, which  a variational formulation does not,  is immediate access to the theory of dynamically accessible variations \cite{Morrison_Pfirsch_1989} (see also \cite{Morrison_fluid_1998,Andreussi_2013}). Suppose we perturb a GVM equilibrium by switching on a small time-dependent term in the Hamiltonain, i.e. $\mathcal{H}\rightarrow \mathcal{H}+\delta\mathcal{H}_t$, where $\delta\mathcal{H}_t$ is a time-dependent functional that is non-zero only in a brief interval of time after $t=0$. Using the Poisson bracket, we can give an energy principle for assessing the stability of this perturbation in the limit where the kick caused by switching on $\delta\mathcal{H}_t$ is infinitesimal. 

In this limit, and accounting for the fact  that the perturbation is generated by altering the Hamiltonian, we find that the perturbed distribution function, auxiliary electric field, and magnetic field must have the form
\begin{align}
\delta f_s&=-L_{\xi_s}f_s\\
\delta \bm{D}&=-4\pi \bm{J}(\xi,f)+4\pi c\,\nabla\times\bm{\beta}\\
\delta\bm{B}&=-4\pi c\,\nabla\times\bm{\alpha},
\end{align} 
where $\bm{\alpha},\bm{\beta}$ are arbitrary vector fields on configuration space, the phase space fluid displacement vector $\xi_s$ is determined by Hamilton's equations,
\begin{align}
\text{i}_{\xi_s}\omega_s^{\text{gy}}=\mathbf{d}\chi_s+4\pi e_s\bm{\alpha}\cdot d\bm{X},
\end{align} 
with $\chi_s$ an arbitrary function on gyrocenter phase space, and $\bm{J}(\xi,f)$ is the gyrocenter current density generated by fiducial gyrocenters with phase space velocity $\xi_s$ and distribution $f_s$. Appealing to the general theory of dynamically accessible variations (see e.g.\ \cite{Morrison_fluid_1998}), our perturbation will be stable if the \emph{free energy functional} $\delta^2 F(\bm{\alpha},\bm{\beta},\chi)$ is positive whenever $\delta f_s,\delta\bm{D}$, and $\delta\bm{B}$ are not each zero. The free energy functional is defined by
\begin{align}
\delta^2 F(\bm{\alpha},\bm{\beta},\chi)=\frac{1}{2}[[\mathcal{H},S],S],
\end{align}  
where the functional $S=\sum_s\int \chi_s\,f_s+\int\bm{\alpha}\cdot\bm{D}\,d^3\bm{X}+\int\bm{\beta}\cdot\bm{B}\,d^3\bm{X}$. Physically, $\delta^2 F$ is the second-order change in the energy functional $\mathcal{H}$ produced by our perturbation. In fact, $\delta^2F$ functions as the (conserved) Hamiltonian of the linearized GVM equations.

We find that $\delta^2 F$ can be written in the form
\begin{align}
\delta^2F=&\sum_s\int\bigg(\frac{1}{2}\omega_s^{\text{gy}}(V^{\text{gy}}_s,\xi_s)\delta f_s+\delta K_s\,\delta f_s\nonumber\\
&+\frac{e_s}{2c}\delta\bm{B}\cdot(V^{\text{gy}}_s)_{\bm{X}}\times(\xi_s)_{\bm{X}}f_s\bigg)\nonumber\\
&+\frac{1}{8\pi}\int\bigg(\delta \bm{D}\cdot\delta \bm{E}+\delta \bm{B}\cdot\delta \bm{H}\bigg)\,d^3\bm{X}.
\end{align}
Here $\bm{X}$ in a subscript denotes the $\bm{X}$-component of a velocity field on phase space. The variations $\delta K_s$, $\delta\bm{E}$, and $\delta\bm{H}$ are given by
\begin{align}
\delta K_s&=\frac{\delta K_s}{\delta \bm{E}}[\delta\bm{E}]+\frac{\delta K_s}{\delta\bm{B}}[\delta\bm{B}]\\
\delta\bm{E}&=\varepsilon^{-1}[\delta \bm{D}]+\eta[\delta \bm{B}]\\
\delta \bm{H}&=\eta^{\dagger}[\delta \bm{D}]+\mu^{-1}[\delta \bm{B}].
\end{align}
where the linear operators $\varepsilon$, $\mu$, and $\eta$ are given by (cf.\ \cite{Morrison_lifting_2013})
\begin{align}
\varepsilon&=1-4\pi\frac{\delta^2\mathcal{K}}{\delta\bm{E}\delta\bm{E}}\\
\mu^{-1}&=1+4\pi\frac{\delta^2\mathcal{K}}{\delta\bm{B}\delta\bm{B}}+(4\pi)^2\frac{\delta^2\mathcal{K}}{\delta\bm{E}\delta\bm{B}}\varepsilon^{-1}\frac{\delta^2\mathcal{K}}{\delta\bm{B}\delta\bm{E}}\\
\eta&=4\pi\varepsilon^{-1}\frac{\delta^2\mathcal{K}}{\delta\bm{B}\delta\bm{E}}.
\end{align}
In principle, an energy principle for electrostatic gyrokinetics analogous to this one could be derived using the Poisson bracket given in \cite{Squire_GK_2013}. However, the authors of that Reference deemed the electrostatic gyrokinetic Poisson bracket too complicated to be practically useful, and so did not attempt deriving an expression for $\delta^2F$.

We have used this expression for $\delta^2F$ to prove that, in the long wavelength limit, the thermal equilibrium state in a uniform background magnetic field is stable. In this case, the gyrocenter kinetic energy is given by
\begin{align}
K=\frac{1}{2}m v_\parallel^2+\omega_c J-\frac{1}{2}mc^2\bigg(\frac{v_\parallel}{c}\frac{\bm{B}_\perp}{B_o}+\frac{\bm{E}\times \hat{b}}{B_o}\bigg)^2,
\end{align}
where $J$ is the gyroaction, $\omega_c$ is the signed gyrofrequency, $\bm{B}_\perp=\bm{B}-\hat{b}\hat{b}\cdot\bm{B}$, and $B_o$ is the magnitude of the background magnetic field. This expression agrees with that given by Krommes in \cite{Krommes_gkh_2013} in the absence of magnetic fluctuations. The linear response functions $\varepsilon^{-1},\mu^{-1},\eta$ are therefore given by the constant matrices
\begin{align}
\varepsilon&=1+\frac{4\pi c^2}{v_A^2}(1-\hat{b}\hat{b})\\
\mu^{-1}&=1-4\pi\beta(1-\hat{b}\hat{b})\\
\eta&=0,
\end{align}
where $\beta=\sum_s\frac{m_sn_s\langle v_\parallel^2\rangle_s}{B_o^2}$ is the plasma $\beta$ and $\langle\cdot\rangle_s$ denotes the velocity space average. Using these expressions and the assumption of thermal equilibrium, a straightforward, but tedious calculation leads to the following form for $\delta^2F$,
\begin{align}
&\delta^2F=\sum_s\int\frac{1}{2T}\bigg(L_{\xi_s}H_{os}-T\frac{\delta B_\parallel}{B_o}\bigg)^2f_s\nonumber\\
&+\frac{1}{8\pi}\int\delta\bm{D}\cdot\varepsilon^{-1}\delta\bm{D}\,d^3\bm{X}\nonumber\\
&+\frac{1}{8\pi}\int\delta\bm{B}_\perp\cdot\mu^{-1}\cdot\delta\bm{B}_\perp d^3\bm{X}\nonumber\\
&+\frac{1}{8\pi}\int(1-4\pi nT/B_o^2)\delta B_\parallel^2\,d^3\bm{X},
\end{align}
where $n=\sum_s n_s$ is the total gyrocenter number density. As long as $4\pi\beta$ and $4\pi nT/B_o^2$ are each less than $1$, a condition that is generally satisfied, $\delta^2F$ is manifestly non-negative, which implies linear stability.

\section{Concluding remarks }The Hamiltonian formulation of the GVM system given in this Chapter is completely determined by two key quantities, the gyrocenter kinetic energy $K_s$, and the guiding center symplectic form $\omega_s^{\text{gc}}$. Suppressing species labels, the gyrocenter kinetic energy is given explicitly to second order in the amplitude of the fluctuating fields, $\epsilon_\delta$, by
\begin{align}
&K(E,B)=\mathcal{H}^{\text{gc}}-\epsilon_\delta \left<\ell\right>+\epsilon_\delta ^2\mathcal{B}^{\text{gy}}(\left<\delta\Xi\right>,\mathbf{d}\left<\ell\right>)\nonumber\\
                &+\frac{1}{2}\epsilon_\delta^2\left<\mathcal{B}_s^{\text{gy}}\bigg(L_R[\delta\tilde{\Xi}-\mathbf{d}I(\tilde{\ell})],[\delta\tilde{\Xi}-\mathbf{d}I(\tilde{\ell})]\bigg)\right>,
\end{align}
where $R$ is the infinitesimal generator of gyrophase rotations times the local gyrofrequency, $I$ is the inverse of the Lie derivative $L_R$, angle brackets denote gyroangle averaging, and $\tilde{Q}=Q-\left<Q\right>$. In standard guiding center coordinates, $L_R=\omega_c\frac{\partial}{\partial \theta}$, where $\theta$ is the gyrophase, which means $I$ amounts to an antiderivative in gyrophase. It can be shown that the second-order gyrocenter kinetic energy has the same general form as Eq.\,(129) in \cite{Brizard_2007}. The relevant correspondences between our symbols and those of \cite{Brizard_2007} are $\ell\leftrightarrow -K_1$, $\mathcal{B}^{\text{gy}ab}\leftrightarrow J_o^{ab}$, $\delta\Xi\leftrightarrow\Delta\Gamma$, and $L_R\delta\Xi\leftrightarrow L_R(\bar{\Gamma}_1+\Gamma_1)$.

From this expression, it is clear that the gyrocenter kinetic energy is determined by the three quantities $\mathcal{H}^{\text{gc}}$, $\ell$, and $\delta\Xi$. $\mathcal{H}^{\text{gc}}$ denotes the guiding center Hamiltonian truncated at some desired order in $\rho/L$. The function $\ell$ and the $1$-form $\delta\Xi$ are defined in terms of any choice of the guiding center Lie generators as follows. Decompose the guiding center transformation $\tau_\text{gc}:TQ\rightarrow TQ$ as $\tau_{\text{gc}}=\tau_2\circ\tau_1$, where
\begin{align}
\tau_1&=\exp(G_1)\\
\tau_2&=\dots\circ\exp(G_3)\circ\exp(G_2)\equiv\exp(\bar{G}_2),
\end{align}
and the $G_k$ are the guiding center Lie generators. The leading-order guiding center transformation, $\tau_1$, must be handled carefully in gyrokinetics because the fluctuating fields are allowed to have short perpendicular wave lengths. The $1$-form
\begin{align}
\delta\Xi=-\frac{e}{c}(\exp(-L_{\tilde{G}_2})\text{i}_{G_1}U(L_{G_1})+\text{i}_{\bar{G}_2}U(L_{\bar{G}_2}) )\bm{B}\cdot dS,
\end{align}
where the function $U(x)=e^{-x/2}\sinh(x/2)/(x/2)$, represents the perturbation to the guiding center Lagrange $1$-form produced by the fluctuating electromagnetic fields. The function
\begin{align}
\delta\mathcal{H}=e(\exp(-L_{\tilde{G}_2})\text{i}_{G_1}U(L_{G_1})+\text{i}_{\bar{G}_2}U(L_{\bar{G}_2}))\bm{E}\cdot dX
\end{align}
represents the perturbation to the guiding center kinetic energy caused by the same fields. The function 
\begin{align}
\ell=\delta\Xi(V_o^{\text{gy}})-\delta H,
\end{align}
where $V_o^{\text{gy}}$ is the unperturbred gyrocenter phase space velocity. 

The Hamiltonian structure of the GVM equations reproduces that of the Vlasov-Maxwell system\,\cite{Morrison_1980,Morrison_divB_1982,Marsden_1982} under the substitutions
\begin{align}
K&\rightarrow\frac{1}{2}m\bm{v}^2\\
\omega^{\text{gc}}&\rightarrow m\, \mathbf{d}x^i\wedge \mathbf{d}v_i.
\end{align}
It is also interesting to compare $[\cdot,\cdot]$ to the bracket given  in \cite{Morrison_lifting_2013}. The only significant difference comes from the manner in which the inductive electric field is built into the kinetic equation. 

Finally, we note two possible directions for future research. (1) It may be useful to identify a Poisson bracket for electromagnetic gyrokinetics in the Darwin approximation, i.e. standard EMGT. The gyrokinetic Vlasov-Darwin equations are somtimes also referred to as the gyrokinetic Vlasov-Poisson-Amp\`ere equations \cite{Sugama_2000}. A Hamiltonian formulation of the non-gyrokinetic Vlasov-Darwin equations has already been given in \cite{Krauss_2007}.  (2) It seems likely that the bracket and Hamiltonian given in this Chapter will provide the Hamiltonian structure for the oscillation center Vlasov-Maxwell equations with appropriate substitutions for $K$ and $\omega^{\text{gc}}$. If this were true, then the benefits that our bracket brings to electromagnetic gyrokinetics could be extended to certain kinds of laser-plasma interactions. 

\section{Acknowledgements} The results presented in this Chapter were obtained in conjunction with Professor Philip Morrison and Professor Alain Brizard. They are also posted on the arXiv at arXiv:1411.1790 and in \cite{Burby_gvm_2015}.

\chapter[Boundary terms and Poisson brackets]{Boundary terms and Poisson brackets}
\label{ch:GKVM}
%
%
%
%
%
%

\section{Introduction}
The purpose of this Chapter is twofold. The first is to describe a systematic procedure to pass from an Euler-Poincar\'e formulation\,\cite{Holm_1998} of a physical system with advected parameters and dynamical fields to a Poisson bracket formulation for that same system. The second is to apply this procedure to the Euler-Poincar\'e formulation of several reduced plasma models, including the gyrokinetic Vlasov-Maxwell system presented in the previous Chapter. The value of a general Eluer-Poincar\'e$\rightarrow$\,Poisson procedure stems from the fact that implementing approximations within the Lagrangian formalism is a well-developed art, whereas finding approximations that respect the Jacobi identity is much more subtle. 

To pass from an Euler-Poincar\'e formulation with a \emph{regular} Lagrangian to a Poisson bracket formulation, all that is necessary is the procedure outlined in Holm's paper on Euler-Poincar\'e theory\,\cite{Holm_1998}, which consists of two steps. First one passes from the parameterized Lagrangian description to a parameterized Hamiltonian description using the Legendre transform. Then one applies the theory developed by Marsden in his paper\,\cite{Marsden_semi_1984} on the Hamiltonian side of semi-direct product theory to ``give life" to the advected parameter and identify a Poisson bracket for the total system.

When the Lagrangian is degenerate, the identification of an appropriate Poisson bracket is more difficult; the conventional Legendre transform-based technique fails. Nevertheless, using the theory of Dirac constraints, Cendra\,\cite{Cendra_1998} and Squire\,\cite{Squire_GK_2013} have derived brackets for specific Euler-Poincar\'e systems with degenerate Lagrangians. Both Cendra and Squire have rederived the Morrison-Marsden-Weinstein bracket. Squire has derived a bracket for the Gyrokinetic Vlasov-Poisson equation. In principle, the methods used by these authors could be adapted to suit our needs, but this is not the path we will follow. Instead we will develop a technique that compliments those of Cendra and Squire. The motivation behind ``reinventing the wheel'' is our desire to simplify the Cendra-Squire approach in such a way that clarifies how advected parameters become dynamical variables and eliminates the need to explicitly introduce the Dirac theory of constraints.

The technique we will develop is insensitive to the degeneracy of the parameter-dependent Lagrangian. In this sense it is similar to the method of Cendra and Squire. However, we will \emph{not} employ the Dirac theory of constraints. Instead, we will apply a technique rooted in a careful analysis of the boundary terms that appear when varying an action functional without keeping endpoints fixed. In detail, our method consists of the following steps.
\\ \\
\emph{step 1---} Identify an Euler-Poincar\'e formulation for the system under consideration. In particular, identify a parameter-dependent Lagrangian, $L_a:TQ\times TG\rightarrow\mathbb{R}$, where $Q$ is the space of dynamical fields, $G$ is a Lie group (usually a diffeomorphism group), and the parameter $a$ is an element of a vector space $V^*$ upon which $G$ acts.
\\ \\
\emph{step 2---}  Eliminate the parameters by introducing a Lagrange multiplier. This method is described in Cendra's \emph{Lagrangian reduction by stages}\,\cite{Cendra_2001}. The result of this simple step will be a Lagrangian $\mathcal{L}:TQ\times TG\times T(V\times V^*)\rightarrow\mathbb{R}$ that embeds the original dynamics in a slightly larger space.
\\ \\
\emph{step 3---} Identify the submanifold $\mathcal{P}_o\subset TQ\times TG\times T(V\times V^*)$ that serves as the augmented system's phase space. Note that this step involves analyzing the initial value problem associated with $\mathcal{L}$'s Euler-Lagrange equations. Possible gauge symmetries and degeneracies of the Lagrangian make this step non-trivial in general.
\\ \\
\emph{step 4---} Identify a Poisson bracket $[\cdot,\cdot]_{\mathcal{P}_o}$ and Hamiltonian $\mathcal{H}_{\mathcal{P}_o}$ on the augmented system's phase space using the boundary symplectic form methodology described in \cite{Marsden_1998}. This bracket is essentially a Pierles bracket.
\\ \\
\emph{step 5---} Observe that the Pierles bracket and Hamiltonian on the augmented phase space are invariant under the action of the semidirect product $S=G\rtimes V$. Perform Poisson reduction using this symmetry, thereby identifying the Poisson bracket on the reduced phase space $\mathcal{P}=\mathcal{P}_o/S$. In this step, the Lagrange multiplier will be eliminated by the Poisson reduction, meaning $\mathcal{P}$ can be thought of as the physical phase space for the system under consideration.
\\ \\
In what follows, we will illustrate this technique by applying it to a number of examples. These include (i) the (generalized) Vlasov-Poisson system, (ii) the Vlasov-Darwin system, (iii) the gyrokinetic Vlasov-Maxwell system, (iv) the neutral Vlasov model\,\cite{Tronci_2015}, and (v) force-free electrodynamics\,\cite{Gralla_2014}.

\section{The generalized Vlasov-Poisson system}
\subsection{Step 1: Euler-Poincar\'e formulation}
Let $P$ be a $2N$-dimensional symplectic manifold with symplectic form $\omega=-\mathbf{d}\vartheta$; note that $P$ is \emph{not} necessarily a cotangent bundle and $\omega$ is \emph{not} necessarily a canonical symplectic form. Let $V^*=\Omega^{2N}(P)$ and $G=\text{Diff}(P)$ denote the space of $2N$-forms on $P$ and the group of diffeomorphisms of $P$, respectively. A typical element of $G$ will be denoted $g\in G$ while a typical element of $V^*$ will be denoted $f\in V^*$.

The generalized Vlasov-Poisson system is defined by the parameter-dependent Lagrangian, $L_{f_o}:TG\rightarrow\mathbb{R}$, given by
\begin{align}
L_{f_o}(g,\dot{g})=\int_P (g_*f_o)\, \vartheta(\dot{g}\circ g^{-1})-\mathcal{H}(g_*f_o),
\end{align}
where $\mathcal{H}:V^*\rightarrow\mathbb{R}$ is the generalized Hamiltonian functional. The generalized Vlasov-Poisson dynamics follow from this Lagrangian by applying Hamilton's principle to the following action functional. Let $\mathfrak{P}(G)$ and $\mathbf{g}\in\mathfrak{P}(G)$ denote the space of paths in $G$ and a typical path in $G$, respectively. The action functional $S_{f_o}:\mathfrak{P}(G)\rightarrow\mathbb{R}$ is given by
\begin{align}
S_{f_o}(\mathbf{g})=\int_{t_1}^{t_2} L_{f_o}(\mathbf{g}(t),\dot{\mathbf{g}}(t))\,dt.
\end{align} 

The Euler-Lagrange equations associated with this action functional can be derived as follows. Let $\mathfrak{g}=\mathfrak{X}(P)$ and $\xi\in\mathfrak{g}$ denote the space of vector fields on $P$ and a typical vector field, respectively. Notice that 
\begin{align}
L_{f_o}(g,\dot{g})=\ell(\dot{g}\circ g^{-1},g_*f_o),
\end{align}
where $\ell:\mathfrak{g}\times V^*\rightarrow\mathbb{R}$ is given by
\begin{align}
\ell(\xi,f)=\int_P f\,\vartheta(\xi)-\mathcal{H}(f).
\end{align}
It follows that the first (fixed-endpoint) variation of the action functional is given by
\begin{align}
\delta S_{f_o}(\mathbf{g})[\delta\mathbf{g}]=&\int_{t_1}^{t_2}\left(\int_P\frac{\delta\ell}{\delta\xi}\left(\dot{\bm{\eta}}(t)+[\bm{\xi}(t),\bm{\eta}(t)]\right)-\frac{\delta \ell}{\delta f}\,L_{\bm{\eta}(t)}\bm{f}(t)\right)\,dt\nonumber\\
=&-\int_{t_1}^{t_2}\int_P\left(\frac{\mathrm{d}}{\mathrm{d}t}\frac{\delta\ell}{\delta\xi}+L_{\bm{\xi}(t)}\frac{\delta\ell}{\delta\xi}-\mathbf{d}\frac{\delta\ell}{\delta f}\otimes \bm{f}(t)\right)\cdot\bm{\eta}(t)\,dt\nonumber\\
=&-\int_{t_1}^{t_2}\int_P\left(\vartheta\otimes\dot{\bm{f}}(t)+L_{\bm{\xi}(t)}(\vartheta\otimes\bm{f}(t))-\mathbf{d}\left(\vartheta(\bm{\xi}(t))-\frac{\delta\mathcal{H}}{\delta f}\right)\otimes\bm{f}(t)\right)\cdot\bm{\eta}(t)\,dt\nonumber\\
=&\int_{t_1}^{t_2}\int_P\bm{f}(t)\,\left(\text{i}_{\bm{\xi}(t)}\omega-\mathbf{d}\frac{\delta\mathcal{H}}{\delta f}\right)\cdot\bm{\eta}(t)\,dt,
\end{align}
where
\begin{align}
\bm{\xi}(t)&=\dot{\mathbf{g}}(t)\circ \mathbf{g}(t)^{-1}\\
\bm{\eta}(t)&=\delta\mathbf{g}(t)\circ \mathbf{g}(t)^{-1}\\
\bm{f}(t)&=\mathbf{g}(t)_*f_o,
\end{align}
and the functional derivatives are evaluated at $(\bm{\xi}(t),\bm{f}(t))$. The Euler-Poincar\'e equations are therefore
\begin{align}
\text{i}_{\bm{\xi}(t)}\omega=\mathbf{d}\frac{\delta\mathcal{H}}{\delta f},
\end{align}
which should be augmented with the equation
\begin{align}
\bm{f}(t)=\mathbf{g}(t)_*f_o.
\end{align}
Note that the last equation implies $\dot{\bm{f}}(t)=-L_{\bm{\xi}(t)}\bm{f}(t)$.
\subsection{Step 2: Introduction of the Lagrange multiplier}
Note that the equation $\dot{\bm{f}}(t)=-L_{\bm{\xi}(t)}\bm{f}(t)$ from the previous section does not follow from the Euler-Lagrange equations associated with the parameter-dependent Lagrangian $L_{f_o}$ because $f_o$ is treated as merely a parameter in Hamilton's principle. We can formally remedy this issue as follows.

Let $V=V^{**}$ and $\chi\in V$ be the dual to $V^*$ and a typical element of $V$, respectively. The space $V$ is properly the space of continuous linear functionals on $V^*$, which is naturally the set of distributional functions on $P$. Define the parameter-independent Lagrangian $\mathcal{L}:TG\times T(V\times V^*)\rightarrow\mathbb{R}$ by
\begin{align}
\mathcal{L}(g,\dot{g},\chi,f_o,\dot{\chi},\dot{f}_o)=L_{f_o}(g,\dot{g})+\int_P\chi\,\dot{f}_o.
\end{align}
When Hamilton's principle is applied to the \emph{augmented action functional}, $\mathcal{S}:\mathfrak{P}(G\times V\times V^*)\rightarrow\mathbb{R}$, given by
\begin{align}
\mathcal{S}(\mathbf{g},\bm{\chi},\bm{f}_o)=\int_{t_1}^{t_2}\mathcal{L}(\mathbf{g}(t),\dot{\mathbf{g}}(t),\bm{\chi}(t),\bm{f}_o(t),\dot{\bm{\chi}}(t),\dot{\bm{f}}_o(t))\,dt,
\end{align} 
the resulting Euler-Lagrange equations are given by
\begin{align}
\dot{\bm{f}}_o(t)&=0\label{aug_ep_f}\\
\dot{\bm{\chi}}(t)&=\mathbf{g}(t)^*\left(\vartheta(\bm{\xi}(t))-\frac{\delta\mathcal{H}}{\delta f}\right)\label{aug_ep_chi}\\
\text{i}_{\bm{\xi}(t)}\omega&=\mathbf{d}\frac{\delta\mathcal{H}}{\delta f}\label{aug_ep_xi},
\end{align}
where the functional derivatives are evaluated at $\mathbf{g}(t)_*\bm{f}_o(t)$. We have thus succeeded in \emph{embedding} the Euler-Poincar\'e equations along with the advection equation into a larger system. We will refer to the system defined by Eqs.\,(\ref{aug_ep_f}), (\ref{aug_ep_chi}), and (\ref{aug_ep_xi}) as \emph{the augmented generalized Vlasov-Poisson equations} (AGVP equations, for short).

\subsection{Step 3: identification of the augmented phase space}
We will now study the initial value problem associated with the augmented generalized Vlasov-Poisson equations. In particular, we would like to identify a submanifold, $\mathcal{P}_o\subset TG\times T(V\times V^*)$, such that the AGVP equations define a first-order (infinite-dimensional) ODE on $\mathcal{P}_o$. In order to accomplish this task, we will merely rearrange the AGVP equations given in the previous section into the form of a first-order equation, and then deduce the allowed set of initial data.

As they were written in the previous section, the AGPV equations are nearly expressed as a first-order system. In order to achieve the desired form, we re-write Eq.\,(\ref{aug_ep_xi}) in terms of $\mathbf{g}(t)$ and substitute Eq.\,(\ref{aug_ep_xi}) into Eq.\,(\ref{aug_ep_xi}), giving
\begin{align}
\dot{\bm{f}}_o(t)&=0\label{one}\\
\dot{\bm{\chi}}(t)&=\mathbf{g}(t)^*\left(\vartheta(X_{\delta\mathcal{H}/\delta f})-\frac{\delta\mathcal{H}}{\delta f}\right)\label{two}\\
\dot{\mathbf{g}}(t)&=X_{\frac{\delta\mathcal{H}}{\delta f}}\circ\mathbf{g}(t),\label{invariant}
\end{align}
where the functional derivatives are evaluated at $\mathbf{g}(t)_*\bm{f}_o(t)$, which is clearly a first-order system of equations in the variables $(\bm{f}_o,\bm{\chi},\mathbf{g})$. That is, there is a vector field $Y$ on $G\times V\times V^*$ such that
\begin{align}
(\mathbf{g}(t),\dot{\mathbf{g}}(t),\bm{\chi}(t),\bm{f}_o(t),\dot{\bm{\chi}}(t),\dot{\bm{f}}_o(t))=Y(\mathbf{g}(t),\bm{\chi}(t),\bm{f}_o(t)).
\end{align}

The triple $(f_o,\chi,g)$ belongs to the set $G\times V\times V^*$, which can naturally be identified with the graph of the vector field $Y$ on $G\times V\times V^*$ that is defined by Eqs.\,(\ref{one}), (\ref{two}), and (\ref{invariant}). Therefore the AGVP equations naturally define a first-order system of ODEs on the submanifold
\begin{align}
\mathcal{P}_o=\{(g,\dot{g},\chi,f_o,\dot{\chi},\dot{f}_o)\in TG\times T(V\times V^*)|(g,\dot{g},\chi,f_o,\dot{\chi},\dot{f}_o)=Y(g,\chi,f_o)\}\approx G\times V\times V^*.
\end{align}
Interestingly, Eq.\,(\ref{invariant}) implies that there is an invariant subset of $\mathcal{P}_o$ given by
\begin{align}
\bar{\mathcal{P}}_o=\{(g,\dot{g},\chi,f_o,\dot{\chi},\dot{f}_o)\in \mathcal{P}_o|g\in\text{Diff}_\omega(P)\}\approx\text{Diff}_\omega(P)\times V\times V^*,
\end{align}
where $\text{Diff}_\omega(P)$ is the set of symplectic diffeomorphisms of $P$. However, $\bar{\mathcal{P}}_o$ is not in one-to-one correspondence with all solutions of the AGVP equations, whereas $\mathcal{P}_o$ is. Therefore we will regard $\mathcal{P}_o$ as the phase space for the AGVP equations.

\subsection{Step 4: derivation of the boundary symplectic form on $\mathcal{P}_o$}
Because $\mathcal{P}_o$ is a valid phase space for the AGVP equations, the AGVP dynamics formally define a time-independent flow map $\mathfrak{F}_t:\mathcal{P}_o=G\times V\times V^*\rightarrow\mathcal{P}_o=G\times V\times V^*$, which is characterized by the relations
\begin{align}
\mathfrak{F}_0&=\text{id}_{\mathcal{P}_o}\\
\frac{\mathrm{d}}{\mathrm{d}t}\mathfrak{F}_t( g,\chi,f_o)&=Y(\mathfrak{F}_t( g,\chi,f_o)),
\end{align}
where $Y$ is the vector field on $G\times V\times V^*$ defined by Eqs.\,(\ref{one}), (\ref{two}), and $(\ref{invariant})$. We can therefore define a mapping $\text{Sol}:\mathcal{P}_o\rightarrow \mathfrak{P}(G\times V\times V^*)$ given by
\begin{align}
\text{Sol}( g,\chi,f_o)(t)=\mathfrak{F}_{t-t_1}( g,\chi,f_o).
\end{align}

The mapping $\text{Sol}$ can be used to pull back the augmented action functional $\mathcal{S}$ to the augmented phase space $\mathcal{P}_o$, thereby defining the \emph{restricted augmented action}
\begin{align}
\mathcal{S}_{\mathcal{P}_o}=\text{Sol}^*\mathcal{S}.
\end{align}
By examining the exterior derivative of the restricted augmented action, we can identify a symplectic form, and therefore Poisson brackets, on the augmented phase space $\mathcal{P}_o$. This can be seen using the following formal manipulation.

Let $( g,\chi,f_o)\in\mathcal{P}_o$ be an arbitrary point in the augmented phase space. Let 
\begin{align}
(\bm{ g}(t),\bm{\chi}(t),\bm{f}_o(t))=\mathfrak{F}_{t-t_1}( g,\chi,f_o).
\end{align}
Because the AGVP dynamical equations imply $\bm{f}_o(t)=f_o$, the restricted augmented action evaluated at $( g,\chi,f_o)$ is given by
\begin{align}
\mathcal{S}_{\mathcal{P}_o}( g,\chi,f_o)=\int_{t_1}^{t_2}L_{f_o}(\bm{ g}(t),\dot{\bm{ g}}(t))\,dt=\int_{t_1}^{t_2}\ell( \bm{\xi}(t),\bm{ g}(t)_*f_o),
\end{align}
where $ \bm{\xi}(t)=\dot{\bm{ g}}(t)\circ\bm{ g}(t)^{-1}$. Using the AGVP equations of motion, we can therefore write
\begin{align}\label{boundary_one_form}
\mathbf{d}S_{\mathcal{P}_o}=\mathfrak{F}_{t_2-t_1}^*\Xi-\Xi,
\end{align}
where $\Xi$ is a one-form on $\mathcal{P}_o$ given by
\begin{align}
\Xi( g,\chi,f_o)[\delta g,\delta\chi,\delta f_o]=\int_P\chi\,\delta f_o+\vartheta(\delta g\circ g^{-1})\, g_*f_o.
\end{align}
Differentiating Eq.\,(\ref{boundary_one_form}) in $t_2$ (and remembering that $\mathcal{S}_{\mathcal{P}_o}$ depends on $t_2$), we obtain
\begin{align}\label{augmented_hamiltons_equations}
\mathbf{d}\dot{\mathcal{S}}_{\mathcal{P}_o}=L_Y\Xi\Rightarrow \text{i}_Y\mathbf{d}\Xi=-\mathbf{d}\left(\Xi(Y)-\dot{\mathcal{S}}_{\mathcal{P}_o}\right),
\end{align}
where the functional $\dot{\mathcal{S}}_{\mathcal{P}_o}:\mathcal{P}_o\rightarrow\mathbb{R}$ is given by
\begin{align}
\dot{\mathcal{S}}_{\mathcal{P}_o}( g,\chi,f_o)=\ell\left(X_{\delta\mathcal{H}/\delta f( g_*f_o)}, g_*f_o\right).
\end{align}
Equation\,(\ref{augmented_hamiltons_equations}) immediately implies that the $2$-form $\omega_{\mathcal{P}_o}=-\mathbf{d}\Xi$ is preserved by the AGVP flow, $\mathfrak{F}_t$. Moreover, because it is not hard to show that $\omega_{\mathcal{P}_o}$ is non-degenerate, the AGVP equations can be written in Poisson bracket form, i.e. given a functional $\mathcal{F}:\mathcal{P}_o\rightarrow\mathbb{R}$,
\begin{align}
\dot{\mathcal{F}}=[\mathcal{F},\mathcal{H}_{\mathcal{P}_o}]_{\mathcal{P}_o},
\end{align}
where 
\begin{align}
\mathcal{H}_{\mathcal{P}_o}( g,\chi,f_o)&=\Xi(Y)-\dot{\mathcal{S}}_{\mathcal{P}_o}\nonumber\\
&=\mathcal{H}( g_*f_o)
\end{align}
is the augmented system's energy functional and $[\cdot,\cdot]_{\mathcal{P}_o}$ is the Poisson bracket obtained by inverting the two-form $\omega_{\mathcal{P}_o}$.

We will conclude this section by deriving an explicit expression for the bracket $[\cdot,\cdot]_{\mathcal{P}_o}$. First observe that if $\mathcal{F}:\mathcal{P}_o\rightarrow\mathbb{R}$ is a functional, then the associated Hamiltonian vector field $Y_\mathcal{F}$, i.e. the vector field on $\mathcal{P}_o$ that satisfies
\begin{align}
\text{i}_{Y_{\mathcal{F}}}\omega_{\mathcal{P}_o}=\mathbf{d}\mathcal{F},
\end{align}
is related to the Poisson bracket according to the formula
\begin{align}\label{trick}
L_{X_{\mathcal{F}}}\mathcal{G}=[\mathcal{G},\mathcal{F}]_{\mathcal{P}_o}.
\end{align}
Because the Lie derivative $L_{X_{\mathcal{F}}}\mathcal{G}$ would be easy to calculate if $X_{\mathcal{F}}$ were known, we will compute a general expression for $X_{\mathcal{F}}$. To this end, it is useful to observe that any integral curve of the vector field $X_{\mathcal{F}}$ is a critical point of a phase space variational principle. That is, an integral curve of $Y_{\mathcal{F}}$ is automatically a critical point of action functional $\mathcal{A}_{\mathcal{F}}:\mathfrak{P}(G\times V\times V^*)\rightarrow\mathbb{R}$ given by
\begin{align}\label{phase_space_action}
\mathcal{A}_{\mathcal{F}}(\bm{ g},\bm{\chi},\bm{f}_o)=\int_{t_1}^{t_2}\left(\Xi_{(\bm{ g}(t),\bm{\chi}(t),\bm{f}_o(t))}[(\dot{\bm{ g}}(t),\dot{\bm{\chi}}(t),\dot{\bm{f}}_o(t))]-\mathcal{F}(\bm{ g}(t),\bm{\chi}(t),\bm{f}_o(t))\right)\,dt.
\end{align}
Therefore we can derive an expression for $Y_\mathcal{F}$ by varying the action given in Eq.\,(\ref{phase_space_action}). For the sake of varying $\mathcal{F}$ w.r.t. $ g$, we introduce the convention that if $H$ is a functional on $\mathcal{P}_o$, then $\delta H/\delta  g( g,\chi,f_o)$ is the unique $1$-form on $P$ that satisfies
\begin{align}
\mathbf{d}H( g,\chi,f_o)[\delta g,0,0]=\int_P \left(\frac{\delta H}{\delta g}( g,\chi,f_o)\otimes  g_*f_o\right)\cdot (\delta g\circ g^{-1}).
\end{align}
The first (fixed-endpoint) variation of $\mathcal{A}_\mathcal{F}$ is given by
\begin{align}
&\delta\mathcal{A}_{\mathcal{F}}(\bm{ g},\bm{\chi},\bm{f}_o)[\delta\bm{ g},\delta\bm{\chi},\delta\bm{f}_o]=\nonumber\\
&\int_{t_1}^{t_2}\int_P\bigg[\left(\dot{\bm{f}}_o(t)-\frac{\delta\mathcal{F}}{\delta \chi}\right)\,\delta\bm{\chi}+\left(\bm{ g}(t)^*(\vartheta( \bm{\xi}(t)))-\frac{\delta\mathcal{F}}{\delta f_o}-\dot{\bm{\chi}}(t)\right)\,\delta \bm{f}_o(t)\nonumber\\
&+\left(\text{i}_{ \bm{\xi}(t)}\omega-\frac{\mathbf{g}(t)_*\dot{\bm{f}}_o}{\mathbf{g}(t)_*\bm{f}_o}\vartheta-\frac{\delta\mathcal{F}}{\delta  g}\right)\otimes\left(\bm{ g}(t)_*\bm{f}_o(t)\right)\cdot\bm{\eta}(t)\bigg]\,dt,
\end{align}
where
\begin{align}
\bm{\xi}(t)=\dot{\mathbf{g}}(t)\circ\mathbf{g}(t)^{-1}\\
\bm{\eta}(t)=\delta\mathbf{g}(t)\circ\mathbf{g}(t)^{-1},
\end{align}
and functional derivatives are evaluated at $(\mathbf{g}(t),\bm{\chi}(t),\bm{f}_o(t))$. It follows that $Y_\mathcal{F}$ is specified by the relations
\begin{align}
\dot{\mathbf{g}}(t)&=\omega^{-1}\left(\delta\mathcal{F}/\delta g+\frac{\mathbf{g}(t)_*\delta\mathcal{F}/\delta\chi}{\mathbf{g}(t)_*\bm{f}_o}\vartheta\right)\circ\mathbf{g}(t)\\
\dot{\bm{\chi}}(t)&=\mathbf{g}(t)^*\left(\vartheta\left(\omega^{-1}\left(\delta\mathcal{F}/\delta g+\frac{\mathbf{g}(t)_*\delta\mathcal{F}/\delta\chi}{\mathbf{g}(t)_*\bm{f}_o}\vartheta\right)\right)\right)-\delta\mathcal{F}/\delta f_o\\
\dot{\bm{f}}_o(t)&=\delta\mathcal{F}/\delta\chi,
\end{align}
where $\omega^{-1}$ denotes the inverse of the linear map $X\rightarrow\text{i}_X\omega$. This formula for $Y_\mathcal{F}$ proves that the two-form $-\mathbf{d}\Xi$ is non-degenerate. Moreover, we can now write down the Poisson bracket $[\cdot,\cdot]_{\mathcal{P}_o}$ using Eq.\,(\ref{trick}). The result is
\begin{align}\label{augmented_bracket}
[\mathcal{F},\mathcal{G}]_{\mathcal{P}_o}=&\int_P B\left(\frac{\delta\mathcal{F}}{\delta g},\frac{\delta\mathcal{G}}{\delta g}\right)\,g_*f_o+\frac{\delta\mathcal{F}}{\delta f_o}\frac{\delta\mathcal{G}}{\delta\chi}-\frac{\delta\mathcal{F}}{\delta\chi}\frac{\delta\mathcal{G}}{\delta f_o}\nonumber\\
+&\int_Pg_*\left(\frac{\delta \mathcal{G}}{\delta\chi}\right)B\left(\frac{\delta\mathcal{F}}{\delta g},\vartheta\right)-g_*\left(\frac{\delta\mathcal{F}}{\delta \chi}\right)B\left(\frac{\delta\mathcal{G}}{\delta g},\vartheta\right),
\end{align}
where $B$ denotes the Poisson tensor assocaited with the symplectic form $\omega$, i.e. given $1$-forms on $P$, $\alpha$ and $\beta$,
\begin{align}
B(\alpha,\beta)=\omega(\omega^{-1}(\alpha),\omega^{-1}(\beta)).
\end{align}
\subsection{Step 5: perform Poisson reduction to obtain bracket on physical phase space}
We have now identified the Hamiltonian, 
\begin{align}
\mathcal{H}_{\mathcal{P}_o}(g,\chi,f_o)=\mathcal{H}(g_*f_o),
\end{align}
and Poisson bracket for the \emph{augmented} generalized Vlasov-Poisson system. The Jacobi identity is satisfied because the bracket has been obtained by inverting the symplectic form $-\mathbf{d}\Xi$. The modifier ``augmented" is appropriate because the dynamical variable $\chi$ has no direct physical meaning. On the other hand, the variables $g$ and $f_o$ together comprise an element of the physical \emph{Lagrangian} (as opposed to Eulerian) phase space; $g$ gives the configuration of particles in the single-particle phase space $P$ and $f_o$ gives the reference phase space density. The purpose of the additional variable $\chi$ is to extend the Lagrangian phase space just enough to allow for a non-degenerate Poisson bracket.

The appearance of the variable $\chi$ perhaps seems awkward at this stage. However, observe the following. The set $V$ is a Lie group under addition that is a symmetry group for the AGVP equations. Specifically, for each $\delta\chi\in V$, we can define a mapping $T_{\delta\chi}:\mathcal{P}_o\rightarrow\mathcal{P}_o$ given by
\begin{align}
T_{\delta\chi}(g,\chi,f_o)=(g,\chi+\delta\chi,f_o),
\end{align}
which clearly satisfies the defining properties of a group action,
\begin{align}
T_{\delta\chi_1+\delta\chi_2}&=T_{\delta\chi_1}\circ T_{\delta\chi_2}\\
T_{0}&=\text{id}_{\mathcal{P}_o}.
\end{align}
This group action leaves the augmented Hamiltonian $\mathcal{H}_{\mathcal{P}_o}$ and the augmented Poisson bracket $[\cdot,\cdot]_{\mathcal{P}_o}$ invariant in the sense that
\begin{align}
T_{\delta\chi}^*\mathcal{H}_{\mathcal{P}_o}&=\mathcal{H}_{\mathcal{P}_o}\label{invariance_lagrangian_hamiltonian}\\
T_{\delta\chi}^*[\mathcal{F},\mathcal{G}]_{\mathcal{P}_o}&=[T_{\delta\chi}^*\mathcal{F},T_{\delta\chi}^*\mathcal{G}]_{\mathcal{P}_o}\label{invariance_lagrangian_bracket},
\end{align}
for arbitrary $\delta\chi\in V$ and functionals $\mathcal{F},\mathcal{G}$. These properties are quick to verify.
Therefore we can define a Hamiltonian and Poisson bracket on the first reduced phase space $\mathcal{P}_L=\mathcal{P}_o/V$, which is naturally the physical Lagrangian phase space, $G\times V^*$.
The Lagrangian Hamiltonian, $\mathcal{H}_{\mathcal{P}_L}:\mathcal{P}_L\rightarrow\mathbb{R}$, is simply given by
\begin{align}
\mathcal{H}_{\mathcal{P}_L}(g,f_o)=\mathcal{H}(g_*f_o).
\end{align}
The Lagrangian Poisson bracket, $[\cdot,\cdot]_{\mathcal{P}_L}$, is defined in terms of the projection map $\pi_L:\mathcal{P}_o\rightarrow\mathcal{P}_L$ given by 
\begin{align}\label{pil}
\pi_L(g,\chi,f_o)=(g,f_o).
\end{align}
We have
\begin{align}\label{reduced_bracket_relation}
\pi_L^*[\mathfrak{F},\mathfrak{G}]_{\mathcal{P}_L}=[\pi_L^*\mathfrak{F},\pi_L^*\mathfrak{G}]_{\mathcal{P}_o},
\end{align}
which defines $[\cdot,\cdot]_{\mathcal{P}_L}$ unambiguously because $T_{\delta\chi}$ leaves the Poisson bracket invariant. In particular,
\begin{align}
[\mathfrak{F},\mathfrak{G}]_{\mathcal{P}_L}(g,f_o)&=[\pi_L^*\mathfrak{F},\pi_L^*\mathfrak{G}]_{\mathcal{P}_o}(g,\tilde{\chi},f_o)\\
&=\int_P B\left(\frac{\delta\mathfrak{F}}{\delta g},\frac{\delta\mathfrak{G}}{\delta g}\right)\,g_*f_o,
\end{align}
where $\tilde{\chi}$ is arbitrary and the functional derivatives are evaluated at $(g,f_o)$. It does not matter which $\tilde{\chi}$ is chosen because
\begin{align}
[\pi_L^*\mathfrak{F},\pi_L^*\mathfrak{G}]_{\mathcal{P}_o}(g,\tilde{\chi}+\delta\chi,f_o)&=T_{\delta\chi}^*([\pi_L^*\mathfrak{F},\pi_L^*\mathfrak{G}]_{\mathcal{P}_o})(g,\tilde{\chi},f_o)\nonumber\\
&=[T_{\delta\chi}^*\pi_L^*\mathfrak{F},T_{\delta\chi}^*\pi_L^*\mathfrak{G}]_{\mathcal{P}_o}(g,\tilde{\chi},f_o)\text{ (by Eq.\,(\ref{invariance_lagrangian_bracket}))}\nonumber\\
&=[\pi_L^*\mathfrak{F},\pi_L^*\mathfrak{G}]_{\mathcal{P}_o}(g,\tilde{\chi},f_o)\text{ (by Eq.\,(\ref{pil}))}.
\end{align}
The relation given in Eq.\,(\ref{reduced_bracket_relation}) shows that the Lagrangian bracket automatically satisfies the Jacobi identity. Thus, the awkwardness introduced by the additional variable $\chi$ is only apparent; we have obtained a physical Hamiltonian formulation for the generalized Vlasov-Poisson in Lagrangian labeling by recognizing that the augmented Hamiltonian and bracket are independent of $\chi$. This Hamiltonian formulation is ``physical" in the sense that all dynamical variables are physically significant.

The set $G$ is a symmetry group of the generalized Vlasov-Poisson system in Lagrangian labeling. Specifically, for each $h\in G$, we can define a mapping $R_h:\mathcal{P}_L\rightarrow\mathcal{P}_L$ given by
\begin{align}
R_h(g,f_o)=(g\circ h,h^*f_o),
\end{align}
that satisfies the properties of a \emph{right} group action, namely
\begin{align}
R_{h_1\circ h_2}&=R_{h_2}\circ R_{h_1}\\
R_{\text{id}_P}&=\text{id}_{\mathcal{P}_L}.
\end{align}
The Lagrangian Hamiltonian and Poisson bracket are each invariant under this group action, in the sense that
\begin{align}
R_{h}^*\mathcal{H}_{\mathcal{P}_L}&=\mathcal{H}_{\mathcal{P}_L}\\
R_{h}^*[\mathcal{F},\mathcal{G}]_{\mathcal{P}_L}&=[R_{h}^*\mathcal{F},R_{h}^*\mathcal{G}]_{\mathcal{P}_L}.
\end{align}
The invariance of the Lagrangian Hamiltonian is quick to verify. The invariance of the Poisson bracket follows from the identity
\begin{align}
\left(\frac{\delta}{\delta g}R^*_h\mathfrak{F}\right)(g,f_o)=\frac{\delta\mathfrak{F}}{\delta g}(g\circ h,h^*f_o).
\end{align}
Therefore, the Lagrangian Hamiltonian and Poisson bracket define corresponding quantities on the \emph{Eulerian} phase space, $\mathcal{P}_E=\mathcal{P}_L/G\approx V^*$. The Eulerian Hamiltonian is given by
\begin{align}
\mathcal{H}_{\mathcal{P}_E}(f)=\mathcal{H}(f).
\end{align}
The Eulerian Poisson bracket is defined in terms of the Eulerian projection map, $\pi_E:\mathcal{P}_{L}\rightarrow\mathcal{P}_E$, given by
\begin{align}
\pi_E(g,f_o)=g_*f_o,
\end{align}
which satisfies the important property $\pi_E\circ R_h=\pi_E$. We have for functionals $\mathsf{F},\mathsf{G}:\mathcal{P}_E\rightarrow\mathbb{R}$,
\begin{align}
[\mathsf{F},\mathsf{G}]_{\mathcal{P}_E}(f)&=[\pi_E^*\mathsf{F},\pi_E^*\mathsf{G}]_{\mathcal{P}_L}(\tilde{g},\tilde{f}_o)\nonumber\\
&=\int_P B\left(\mathbf{d}\frac{\delta\mathsf{F}}{\delta f},\mathbf{d}\frac{\delta\mathsf{G}}{\delta f}\right)\,f\nonumber\\
&=\int_P\left\{\frac{\delta \mathsf{F}}{\delta f},\frac{\delta\mathsf{G}}{\delta f}\right\}f,
\end{align}
where $\tilde{g}$ and $\tilde{f}_o$ are any group element and $6$-form that satisfy $f=\tilde{g}_*\tilde{f}_o$, and $\{\cdot,\cdot\}$ is the Poisson bracket associated with the symplectic form $\omega$. Just as earlier, this bracket automatically satisfies the Jacobi identity. However, while this fact for the Lagrangian bracket is perhaps not well-known, here it comes as no surprise; $[\cdot,\cdot]_{\mathcal{P}_E}$ is none other than a Lie-Poisson bracket. 

\section{The Vlasov-Darwin system}

\subsection{Step 1: Euler-Poincar\'e formulation}
We will use the following notation. 
\\ \\
\emph{Particle configuration space ---} Let $Q =\mathbb{R}^3$ be the single-particle configuration space with metric tensor $\left<\cdot,\cdot\right>$ and associated hodge star $*$. Typical elements of $Q$ will be denoted $q\in Q$. The codifferential on $Q$ will be denoted $\bm{\delta}$, the Laplace-deRham operator will be denoted $\bm{\Delta}=\bm{\delta}\mathbf{d}+\mathbf{d}\bm{\delta}$, and the Laplace-deRham Green operator will be denoted $\bm{G}$. We will also make use of the transverse and longitudinal projection operators $\Pi_T=\bm{\delta}\bm{G}\mathbf{d}$ and $\Pi_L=\mathbf{d}\bm{G}\bm{\delta}$. 
\\ \\
\emph{Particle phase space ---} The set $TQ$ will serve as the single-particle velocity phase space. A typical element of $TQ$ will be denoted $v_q\in T_qQ$. The map $\pi:TQ\rightarrow Q$ will denote the tangent bundle projection. Let $\mathbb{F}:TQ\rightarrow T^*Q$ be the diffeomorphism given by $v_q\mapsto\left<v_q,\cdot\right>$. The symbol $\vartheta$ will stand for $\mathbb{F}^*\theta_o$, where $\theta_o$ is the canonical $1$-form on $T^*Q$. 
\\ \\
\emph{Function spaces ---} The sets $\text{Diff}(TQ)$, $C^\infty(TQ)$, and $\text{den}(TQ)$ denote the diffeomorphism group of $TQ$, the space of smooth functions on $TQ$, and the the space of distributional densities on $TQ$. Typical elements of these  spaces will be denoted $g\in \text{Diff}(TQ)$, $\chi\in C^\infty(TQ)$, and $f\in\text{den}(TQ)$. The sets $\Omega^k(Q)$ for integer $k\geq 0$ are the $k$-forms on $Q$. For our purposes, the $0$-forms, $1$-forms, and $N$-forms are the most important. Typical elements of the latter will be denoted $\phi\in\Omega^0(Q)$, $A\in\Omega^1(Q)$, and $\lambda\in\Omega^{N}(Q)$. The sets $\mathfrak{X}(TQ)$ and $\mathfrak{X}(TQ)^*$ are the vector fields and $1$-form densities on $TQ$. If $\mathfrak{S}$ is any space, $\mathfrak{P}(\mathfrak{S})$ will denote the space of paths in $\mathfrak{S}$ parameterized by the time interval $[t_1,t_2]$. If the symbol $s$ is used to denote a typical element of $\mathfrak{S}$, we will use a bold version of the same symbol to denote a typical path in $\mathfrak{S}$, i.e. $\bm{s}\in\mathfrak{P}(\mathfrak{S})$. 
\\ \\
\emph{Multi-species objects ---}Let $N_s$ be the number of plasma species. Set $G=\text{Diff}(TQ)^{N_s}$, $V=(C^\infty(TQ))^{N_s}$, and $V^*=\text{den}(TQ)^{N_s}$. We will denote typical elements of these spaces with $\tilde{g}=(g_1,...,g_{N_s})\in G$, $\tilde{\chi}=(\chi_1,...,\chi_{N_s})\in V$, and $\tilde{f}_o=(f_{o,1},...,f_{o,N_s})\in V^*$. Set $\mathfrak{g}=\mathfrak{X}(TQ)^{N_s}$. A typical element of $\mathfrak{g}$ will be denoted $\tilde{\xi}\in\mathfrak{g}$.
\\ \\
Sugama\,\cite{Sugama_2013} gives the following Euler-Poincar\'e formulation of the Vlasov-Darwin system (which he calls the Vlasov-Poisson-Amp\`ere system). 
Define the parameter-dependent Lagrangian, $L_{\tilde{f}_o}^S:TG\times T(\Omega^0(Q)\times\Omega^1(Q)\times\Omega^N(Q))\rightarrow\mathbb{R}$, given by
\begin{align}
L^S_{\tilde{f}_o}(\tilde{g},\dot{\tilde{g}},\phi,A,\lambda,\dot{\phi},\dot{A},\dot{\lambda})=\sum_{s=1}^{N_s}&\int_{TQ} g_{s*}f_{o,s}\left(m_s\vartheta+\frac{e_s}{c}\pi^*A\right)(\dot{g}_s\circ g_s^{-1})-g_{s*}f_{o,s}\left(K_s+e_s\pi^*\phi\right)\nonumber\\
&+\int_Q\frac{1}{8\pi}\left(\mathbf{d}\phi\wedge*\mathbf{d}\phi-\mathbf{d}A\wedge*\mathbf{d}A\right)+\frac{1}{4\pi c}\lambda\,\bm{\delta}A,
\end{align}
where $K_s(v_q)=m_s\left<v_q,v_q\right>/2$.  The Vlasov-Darwin system of equations in Lagrangian labeling then follow from Hamilton's principle applied to the action functional $\mathcal{S}^{S}:\mathfrak{P}(G\times\Omega^0(Q)\times\Omega^1(Q)\times\Omega^N(Q))\rightarrow\mathbb{R}$ given by
\begin{align}
\mathcal{S}^{S}_{\tilde{f}_o}(\tilde{\mathbf{g}},\bm{\phi},\bm{A},\bm{\lambda})=\int_{t_1}^{t_2}L^S_{\tilde{f}_o}(\tilde{\mathbf{g}}(t),\dot{\tilde{\mathbf{g}}}(t),\bm{\phi}(t),\bm{A}(t),\bm{\lambda}(t),\dot{\bm{\phi}}(t),\dot{\bm{A}}(t),\dot{\bm{\lambda}}(t))\,dt.
\end{align}

The Euler-Lagrange equations associated with Sugama's Lagrangian are given by
\begin{align}
\label{e_p_equation}\text{i}_{\bm{\xi}_s(t)}\mathbf{d}\theta_s&=-\dot{\theta}_s-\mathbf{d}H_s\\
\bm{\delta}\mathbf{d}\bm{A}(t)&=\frac{4\pi}{c}\sum_{s=1}^{N_s}e_s *u(\bm{\xi}_s(t),\bm{f}_s(t))+\frac{1}{c}\mathbf{d}*\bm{\lambda}(t)\\
\bm{\delta}\mathbf{d}\bm{\phi}(t)&=4\pi\sum_{s=1}^{N_s}e_s *n(\bm{f}_s(t))\\
\bm{\delta}\bm{A}(t)&=0,
\end{align}
where $\bm{\xi}_s(t)=\dot{\mathbf{g}}_s(t)\circ \mathbf{g}_s(t)^{-1}$, $\bm{f}_{s}(t)=\mathbf{g}_s(t)_*f_{o,s}$, the quantities
\begin{align}
\theta_s&=m_s\vartheta +\frac{e_s}{c}\pi^*\bm{A}(t)\\
H_s&=K_s+e_s\pi^*\bm{\phi}(t),
\end{align}
and the operators $u:\mathfrak{X}(TQ)\times\text{den}(TQ)\rightarrow \Omega^2(Q)$ and $n:\text{den}(TQ)\rightarrow \Omega^3(Q)$ are given by the fiber integrals
\begin{align}
u(\xi,f)(q)&=\int_{\pi^{-1}(q)}\text{i}_\xi f\\
n(f)(q)&=\int_{\pi^{-1}(q)}f.
\end{align}
We will refer to $u(\xi,f)$ as the particle flux $2$-form and $n(f)$ as the particle spatial density $3$-form.

We will not use Sugama's variational formulation as our initial Euler-Poincar\'e formulation for the Vlasov-Darwin system. Instead we will insert the elliptic equations for the potentials back into Sugama's Lagrangian, thereby obtaining a new Lagrangian $L_{\tilde{f}_o}:TG\rightarrow\mathbb{R}$. A straightforward calculation shows that $L_{\tilde{f}_o}$ has the simple expression 
\begin{align}
L_{\tilde{f}_o}(\tilde{g},\dot{\tilde{g}})&=\ell(\dot{\tilde{g}}\circ\tilde{g}^{-1},\tilde{g}_*\tilde{f}_o),
\end{align}
where
\begin{align}
\ell(\tilde{\xi},\tilde{f})&=\left(\sum_{s=1}^{N_s}\int_{TQ} f_s\left(m_s\vartheta+\frac{e_s}{2c}\pi^*\mathcal{A}(\tilde{\xi},\tilde{f})\right)(\xi_s)\right)-\mathcal{H}(\tilde{f}).
\end{align}
Here we have introduced the Hamiltonian functional $\mathcal{H}:V^*\rightarrow\mathbb{R}$
\begin{align}
\mathcal{H}(\tilde{f})=\sum_{s=1}^{N_s}\int_{TQ}f_s\left(K_s+\frac{1}{2}e_s\pi^*\Phi(\tilde{f})\right),
\end{align}
the potential operators $\mathcal{A}:\mathfrak{g}\times V^*\rightarrow \Omega^1(Q)$ and $\Phi:V^*\rightarrow \Omega^0(Q)$,
\begin{align}
\mathcal{A}(\tilde{\xi},\tilde{f})&=\frac{4\pi}{c}\bm{G}\Pi_TJ(\tilde{\xi},\tilde{f})\\
\Phi(\tilde{f})&=4\pi\bm{G}\rho(\tilde{f}),
\end{align}
and the charge and current density operators $\rho:V^*\rightarrow\Omega^0(Q)$ and $J:\mathfrak{g}\times V^*\rightarrow\Omega^{1}(Q)$,
\begin{align}
\rho(\tilde{f})&=\sum_{s=1}^{N_s}e_s* n(f_s)\\
J(\tilde{\xi},\tilde{f})&=\sum_{s=1}^{N_s}e_s* u(\xi_s,f_s).
\end{align}

We will now verify directly that our parameter-dependent Lagrangian $L_{\tilde{f}_o}$ reproduces the Vlasov-Darwin equations. Because the equations for the potentials are satisfied by construction, we will merely verify that the Euler-Poincar\'e equation associated with $L_{\tilde{f}_o}$ reproduces Eq.\,(\ref{e_p_equation}). The Euler-Poincar\'e equation follows from Hamilton's principle applied to the action functional $\mathcal{S}_{\tilde{f}_o}:\mathfrak{P}(G)\rightarrow\mathbb{R}$ given by
\begin{align}
\mathcal{S}_{\tilde{f}_o}(\tilde{\mathbf{g}})=\int_{t_1}^{t_2} L_{\tilde{f}_o}(\tilde{\mathbf{g}}(t),\dot{\tilde{\mathbf{g}}}(t))\,dt.
\end{align}
Varying this action, we obtain the general Euler-Poincar\'e equation given originally by Holm,
\begin{align}\label{e_p_general}
\frac{\mathrm{d}}{\mathrm{d}t}\frac{\delta\ell}{\delta\xi_s}+L_{\bm{\xi}_s(t)}\frac{\delta\ell}{\delta\xi_s}=\mathbf{d}\frac{\delta\ell}{\delta f_s}\otimes\bm{f}_s(t).
\end{align}
In order to calculate the functional derivatives appearing in this expression, we will first express the reduced Lagrangian $\ell$ in the form
\begin{align}
\ell(\tilde{\xi},\tilde{f})=\sum_{s=1}^{N_s}\int_{TQ}m_s\vartheta(\xi_s)\,f_s+\frac{1}{2 c}\left<\mathcal{A}(\tilde{\xi},\tilde{f}),J(\tilde{\xi},\tilde{f})\right>-\mathcal{H}(\tilde{f}),
\end{align}
where $\left<\cdot,\cdot\right>$ denotes the natural integration pairing of differential forms 
\begin{align}
\left<\alpha,\beta\right>=\int_Q\alpha\wedge*\beta.
\end{align}
Next we introduce the linear operators $J^{\tilde{f}}:\mathfrak{g}\rightarrow\Omega^{1}(Q)$ and $J_{\tilde{\xi}}:V^*\rightarrow\Omega^1(Q)$ given by
\begin{align}
J^{\tilde{f}}(\tilde{\xi})&=J(\tilde{\xi},\tilde{f})\\
J_{\tilde{\xi}}(\tilde{f})&=J(\tilde{\xi},\tilde{f}).
\end{align}
The adjoint operators $(J_{\tilde{\xi}})^\dagger_s:\Omega^1(Q)\rightarrow C^\infty(TQ)$ and $(J^{\tilde{f}})^\dagger_s:\Omega^1(Q)\rightarrow\mathfrak{X}(TQ)^*$ defined by the relations
\begin{align}
\left<\alpha,J_{\tilde{\xi}}(\delta\tilde{f})\right>&=\sum_{s=1}^{N_s} \int_{TQ}(J_{\tilde{\xi}})^\dagger_s(\alpha)\,\delta f_s\\
\left<\alpha,J^{\tilde{f}}(\delta\tilde{\xi})\right>&=\sum_{s=1}^{N_s}\int_{TQ}(J^{\tilde{f}})^\dagger_s(\alpha)\cdot\delta\xi_s,
\end{align}
are readily found to be given by the formulae
\begin{align}
(J_{\tilde{\xi}})^\dagger_s(\alpha)&=e_s\pi^*\alpha(\xi_s)\\
(J^{\tilde{f}})^\dagger_s(\alpha)&=e_s\pi^*\alpha\otimes f_s.
\end{align}
Finally, we compute the Fr\'echet derivative of $\ell$,
\begin{align}
&D\ell(\tilde{\xi},\tilde{f})[\delta\tilde{\xi},\delta\tilde{f}]=\nonumber\\
&\sum_{s=1}^{N_s}\int_{TQ}(m_s\vartheta\otimes f_s)\cdot\delta\xi_s+\left(m_s\vartheta(\xi_s) -\frac{\delta\mathcal{H}}{\delta f}\right)\,\delta f+\frac{1}{c}\left<\mathcal{A}(\tilde{\xi},\tilde{f}),J^{\tilde{f}}(\delta\tilde{\xi})\right>+\frac{1}{c}\left<\mathcal{A}(\tilde{\xi},\tilde{f}),J_{\tilde{\xi}}(\delta\tilde{f})\right>\nonumber\\
=&\sum_{s=1}^{N_s}\int_{TQ}\bigg(m_s\vartheta\otimes f_s+\frac{1}{c}(J^{\tilde{f}})^\dagger_s(\mathcal{A}(\tilde{\xi},\tilde{f}))\bigg)\cdot\delta\xi_s+\left(m_s\vartheta(\xi_s)+\frac{1}{c}(J_{\tilde{\xi}})^\dagger_s(\mathcal{A}(\tilde{\xi},\tilde{f})) -\frac{\delta\mathcal{H}}{\delta f}\right)\,\delta f\nonumber\\
=&\sum_{s=1}^{N_s}\int_{TQ}\bigg(m_s\vartheta\otimes f_s+\frac{e_s}{c}\pi^*\mathcal{A}(\tilde{\xi},\tilde{f})\otimes f_s\bigg)\cdot\delta\xi_s+\left(m_s\vartheta(\xi_s)+\frac{e_s}{c}\pi^*\mathcal{A}(\tilde{\xi},\tilde{f})(\xi_s) -\frac{\delta\mathcal{H}}{\delta f}\right)\,\delta f,
\end{align}
from which the functional derivatives can be quickly extracted, giving
\begin{align}
\frac{\delta\ell}{\delta\xi_s}&=m_s\vartheta\otimes f_s+\frac{e_s}{c}\pi^*\mathcal{A}(\tilde{\xi},\tilde{f})\otimes f_s\\
\frac{\delta\ell}{\delta f_s}&=m_s\vartheta(\xi_s)+\frac{e_s}{c}\pi^*\mathcal{A}(\tilde{\xi},\tilde{f})(\xi_s) -\frac{\delta\mathcal{H}}{\delta f}.
\end{align}
Note that in computing the Fr\'echet derivative, we have made use of the fact that the operator $\bm{G}\Pi_T$ is self-adjoint. Equation\,(\ref{e_p_general}) therefore reduces to
\begin{align}
\text{i}_{\bm{\xi}_s(t)}\mathbf{d}\left(m_s\vartheta+\frac{e_s}{c}\pi^*\mathcal{A}(\tilde{\bm{\xi}}(t),\tilde{\bm{f}}(t))\right)+\frac{\rm{d}}{\rm{d}t}\left(m_s\vartheta+\frac{e_s}{c}\pi^*\mathcal{A}(\tilde{\bm{\xi}}(t),\tilde{\bm{f}}(t))\right)+\mathbf{d}\frac{\delta\mathcal{H}}{\delta f_s}=0,
\end{align}
which is readily verified to be equivalent to Eq.\,(\ref{e_p_equation}).
\subsection{Step 2: Introduction of the Lagrange multiplier}
We now define the parameter-independent Lagrangian $\mathcal{L}:TG\times T(V\times V^*)\rightarrow\mathbb{R}$ by
\begin{align}
\mathcal{L}(\tilde{g},\dot{\tilde{g}},\tilde{\chi},\tilde{f}_o,\dot{\tilde{\chi}},\dot{\tilde{f}}_o)=L_{\tilde{f}_o}(\tilde{g},\dot{\tilde{g}})+\sum_{s=1}^{N_s}\int_{TQ}\chi_s\,\dot{f}_{o,s}.
\end{align}
When Hamilton's principle is applied to the \emph{augmented action functional}, $\mathcal{S}:\mathfrak{P}(G\times V\times V^*)\rightarrow\mathbb{R}$, given by
\begin{align}
\mathcal{S}(\tilde{\mathbf{g}},\tilde{\bm{\chi}},\tilde{\bm{f}}_o)=\int_{t_1}^{t_2}\mathcal{L}(\tilde{\mathbf{g}}(t),\dot{\tilde{\mathbf{g}}}(t),\tilde{\bm{\chi}}(t),\tilde{\bm{f}}_o(t),\dot{\tilde{\bm{\chi}}}(t),\dot{\tilde{\bm{f}}}_o(t))\,dt,
\end{align} 
the resulting Euler-Lagrange equations are given by
\begin{align}
\dot{\bm{f}}_{o,s}(t)&=0\label{aug_f}\\
\dot{\bm{\chi}}_s(t)&=\mathbf{g}_s(t)^*\frac{\delta\ell}{\delta f_s}=\mathbf{g}_s(t)^*\left(\theta_s(\bm{\xi}_s(t))-\frac{\delta\mathcal{H}}{\delta f_s}\right)\label{aug_chi}\\
\text{i}_{\bm{\xi}_s(t)}\mathbf{d}\theta_s&=-\dot{\theta}_s-\mathbf{d}\frac{\delta\mathcal{H}}{\delta f_s}\label{aug_xi},
\end{align}
where
\begin{align}
\theta_s=m_s\vartheta+\frac{e_s}{c}\pi^*\mathcal{A}(\tilde{\bm{\xi}}(t),\tilde{\bm{f}}(t)),
\end{align}
and the functional derivatives are evaluated at $\tilde{\mathbf{g}}(t)_*\tilde{\bm{f}}_o(t)$. The parameter-independent Lagrangian $\mathcal{L}$ therefore succeeds at \emph{embedding} the Vlasov-Darwin dynamics, including those of the distribution function, into a larger system. We will refer to the system defined by Eqs.\,(\ref{aug_f}), (\ref{aug_chi}), and (\ref{aug_xi}) as \emph{the augmented Vlasov-Darwin equations} (AVD equations, for short).

\subsection{Step 3: Identification of the augmented phase space}
In order to identify a phase space for the AVD equations, we must identify a submanifold of $TG$ upon which the Euler-Lagrange equations associated with the augmented Lagrangian $\mathcal{L}$ define a system of equations that are first order in time. As the Euler-Lagrange equations are written in the previous section, the AVD equations are \emph{not} written as a first order system, and so there is some work to do.

The most non-trivial AVD Euler-Lagrange equation is the one that takes the form of a time-dependent Hamilton equation:
\begin{align}\label{non_trivial}
\text{i}_{\bm{\xi}_s(t)}\mathbf{d}\left(m_s\vartheta+\frac{e_s}{c}\pi^*\mathcal{A}(\tilde{\bm{\xi}}(t),\tilde{\bm{f}}(t))\right)+\frac{\rm{d}}{\rm{d}t}\left(m_s\vartheta+\frac{e_s}{c}\pi^*\mathcal{A}(\tilde{\bm{\xi}}(t),\tilde{\bm{f}}(t))\right)+\mathbf{d}\frac{\delta\mathcal{H}}{\delta f_s}=0.
\end{align}
This equation appears to implicitly relate the Eulerian velocity field $\tilde{\bm{\xi}}(t)$ with its time derivative $\dot{\tilde{\bm{\xi}}}(t)$ and the distribution function $\tilde{\bm{f}}(t)$; the time derivative of $\tilde{\bm{\xi}}(t)$ appears as a result of the identity
\begin{align}
\frac{\rm{d}}{\rm{d}t}\mathcal{A}(\tilde{\bm{\xi}}(t),\tilde{\bm{f}}(t))=\mathcal{A}(\dot{\tilde{\bm{\xi}}}(t),\tilde{\bm{f}}(t))-\mathcal{A}(\tilde{\bm{\xi}}(t),L_{\tilde{\bm{\xi}}(t)}\tilde{\bm{f}}(t)).
\end{align}
Fortunately, this implicit relationship is not quite as complicated as it seems. It turns out that Eq.\,(\ref{non_trivial}) implies $\bm{\xi}_s(t)$ must be a second-order vector field for each $t$. Therefore $\mathcal{A}(\tilde{\bm{\xi}}(t),\tilde{\bm{f}}(t))$ and $u(\bm{\xi}_s(t),\bm{f}_s(t))$ can be expressed in terms of the free-streaming vector field $X_o$, i.e. the unique vector field that satisfies $\text{i}_{X_o}\mathbf{d}m_s\vartheta=-\mathbf{d}K_s$. We have
\begin{align}\label{second_order_simplification}
u(\bm{\xi}_s(t),\bm{f}_s(t))&=u(X_o,\bm{f}_s(t))\\
\mathcal{A}(\tilde{\bm{\xi}}(t),\tilde{\bm{f}}(t))&=\mathcal{A}(\tilde{X}_o,\tilde{\bm{f}}(t)),
\end{align}
which implies that the time derivative of the vector potential simplifies to 
\begin{align}
\frac{\rm{d}}{\rm{d}t}\mathcal{A}(\tilde{\bm{\xi}}(t),\tilde{\bm{f}}(t))=-\mathcal{A}(\tilde{X}_o,L_{\tilde{\bm{\xi}}(t)}\tilde{\bm{f}}(t)).
\end{align}
Thus, the time-dependent Hamilton equation simplifies to 
\begin{align}
\text{i}_{\bm{\xi}_s(t)}\mathbf{d}\left(m_s\vartheta+\frac{e_s}{c}\pi^*\mathcal{A}(\tilde{X}_o,\tilde{\bm{f}}(t))\right)-\frac{e_s}{c}\pi^*\mathcal{A}(\tilde{X}_o,L_{\tilde{\bm{\xi}}(t)}\tilde{\bm{f}}(t))+\mathbf{d}\frac{\delta\mathcal{H}}{\delta f_s}=0,
\end{align}
which is merely a relationship between $\tilde{\bm{\xi}}(t)$ and $\tilde{\bm{f}}(t)$. 

The relationship between $\tilde{\bm{\xi}}(t)$ and $\tilde{\bm{f}}(t)$ can be resolved explicitly as follows. By applying fiber integrals to the Vlasov equation, $\dot{\bm{f}}_s(t)=-L_{\bm{\xi}_s(t)}\bm{f}_s(t)$, we obtain the fluid equation for the particle flux $\bm{u}_s(t)=u(\bm{\xi}_s(t),\bm{f}_s(t))$,
\begin{align}\label{fluid_equation}
\frac{\rm{d}}{\rm{d}t}*\bm{u}_s(t)=-\text{div}(\bm{T}_s(t))^{\flat}-\frac{e_s}{m_s}(*\bm{n}_s)\frac{1}{c}\dot{\mathbf{A}}(t)-\frac{e_s}{m_s}(*\bm{n}_s(t))\mathbf{d}\bm{\phi}(t)+\frac{e_s}{m_s}*(*\bm{u}_s(t)\wedge*\mathbf{d}\mathbf{A}(t)),
\end{align}
where
\begin{align}
\bm{n}_s(t)&=n(\bm{f}_s(t))\\
\mathbf{A}(t)&=\mathcal{A}(\tilde{\bm{\xi}}(t),\tilde{\bm{f}}(t))\\
\bm{\phi}(t)&=\Phi(\tilde{\bm{f}}(t))\\
\bm{T}_s(t)&=T(\bm{f}_s(t))
\end{align}
and $T$ is the stress tensor operator. Given a pair of $1$-forms $\alpha,\beta\in\Omega^1(Q)$, the defining relation for the stress tensor operator is
\begin{align}
T(f)(\alpha,\beta)=*\left(\int_{\pi^{-1}}\pi^*\alpha(X_o)\pi^*\beta(X_o)f\right).
\end{align}
By multiplying the fluid equation for $\bm{u}_s(t)$ by $e_s$ and then summing over species, 
we then obtain
\begin{align}\label{current_density_equation}
&\frac{\rm{d}}{\rm{d}t}*J(\tilde{\bm{\xi}}(t),\tilde{\bm{f}}(t))=\nonumber\\
&-\sum_{s=1}^{N_s}e_s\text{div}(\bm{T}_s(t))^{\flat}-\frac{1}{4\pi c}\bm{\omega}_p^2(t)\dot{\mathbf{A}}(t)-\frac{1}{4\pi}\bm{\omega}_p^2(t)\mathbf{d}\bm{\phi}(t)+\sum_{s=1}^{N_s}\frac{e_s^2}{m_s}*(*\bm{u}_s(t)\wedge*\mathbf{d}\mathbf{A}(t))
\end{align}
as an equation for the time-derivative of the current density. Here we have introduced the local plasma frequency $\bm{\omega}_p^2(t)=\omega_p^2(\tilde{\bm{f}}(t))$, where
\begin{align}
\omega_p^2(\tilde{f})=\sum_{s=1}^{N_s}\frac{4\pi e_s^2}{m_s}*n(f_s).
\end{align}
Finally, by applying the operator $-\frac{4\pi}{c^2}\bm{G}\Pi_T$ to both sides of Eq.\,(\ref{current_density_equation}), we obtain a linear operator $E_I:V^*\rightarrow\Omega^1(Q)$ that gives the inductive electric field in terms of the distribution function,
\begin{align}
E_L(\tilde{f})=\left[1+\frac{1}{c^2}\bm{G}\Pi_T\hat{\omega}_p^2(\tilde{f})\right]^{-1}\frac{1}{c^2}\bm{G}\Pi_T\hat{\omega}_p^2(\tilde{f})[\bm{O}(\tilde{f})],
\end{align}
where $\bm{O}:V^*\rightarrow\Omega^1(Q)$ is given by
\begin{align}
\bm{O}(\tilde{f})=\mathbf{d}\Phi(\tilde{f})+\frac{4\pi}{\omega_p^2(\tilde{f})}\sum_{s=1}^{N_s}\left(e_s\text{div}(T(\tilde{f}))^\flat-\frac{e_s^2}{m_s}*[*u(X_o,f_s)\wedge*\mathbf{d}\mathcal{A}(\tilde{X}_o,\tilde{f})]\right).
\end{align}
The operator $E_L$ is precisely what is necessary to express the Eulerian phase space velocity $\bm{\xi}_s(t)$ in terms of the distribution function. Indeed, we have
\begin{align}
\bm{\xi}_s(t)=\omega_s(\tilde{\bm{f}}(t))^{-1}\left(\mathbf{d}\frac{\delta\mathcal{H}}{\delta f_s}-e_s \pi^*E_I(\tilde{\bm{f}}(t))\right),
\end{align}
where, for each $\tilde{f}\in V^*$, $\omega_s(\tilde{f})$ is the symplectic form on $TQ$ given by
\begin{align}
\omega_s(\tilde{f})=-\mathbf{d}\left(m_s\vartheta+\frac{e_s}{c}\pi^*\mathcal{A}(\tilde{X}_o,\tilde{f})\right).
\end{align}
Note that $\omega_s(\tilde{f})=-\mathbf{d}\theta_s(\tilde{f})$, where
\begin{align}
\theta_s(\tilde{f})=m_s\vartheta+\frac{e_s}{c}\pi^*\mathcal{A}(\tilde{X}_o,\tilde{f}).
\end{align}

With the relationship between $\bm{\xi}_s(t)$ and $\tilde{\bm{f}}(t)$ resolved, we can now substitute it into the AVD Euler-Lagrange equations. When this substitution is performed, the AVD equations become a first-order ODE in the variables $(\tilde{\mathbf{g}}(t),\tilde{\bm{\chi}}(t),\tilde{\bm{f}}_o(t))\in G\times V\times V^*$. Explicitly, we have
\begin{align}
\dot{\mathbf{g}}_s(t)&=\omega_s^{-1}\left(\mathbf{d}\frac{\delta\mathcal{H}}{\delta f_s}-e_s \pi^*E_I\right)\circ\mathbf{g}_s(t)\\
\dot{\bm{\chi}}_s(t)&=\mathbf{g}_s(t)^*\left(\theta_s\left(X_o\right)-\frac{\delta\mathcal{H}}{\delta f_s}\right)\\
\dot{\bm{f}}_{o,s}(t)&=0,
\end{align}
where the operators $\omega_s$, $\theta_s$, $E_I$, and the functional derivatives $\delta\mathcal{H}/\delta f_s$ are evaluated at $\tilde{\mathbf{g}}(t)_*\tilde{\bm{f}}_o(t)\in V^*$. In writing the equation for $\dot{\bm{\chi}}_s(t)$, we have made use of the fact that $\omega_s^{-1}\left(\mathbf{d}\frac{\delta\mathcal{H}}{\delta f_s}-e_s \pi^*E_I\right)$ is a second order vector field. 

As is true of first-order ODEs in general, this first-order ODE for the variables $(\tilde{\mathbf{g}}(t),\tilde{\bm{\chi}}(t),\tilde{\bm{f}}_o(t))\in G\times V\times V^*$ is identifiable with a vector field $Y$ on $G\times V\times V^*$. Setting $\bm{Z}(t)=(\tilde{\mathbf{g}}(t),\tilde{\bm{\chi}}(t),\tilde{\bm{f}}_o(t))$, $Y$ is defined by the relation
\begin{align}
\frac{\rm{d}}{\rm{d}t}\bm{Z}(t)=Y(\bm{Z}(t)).
\end{align}  
Because the AVD equations can be written in this form, it follows that the submanifold $\mathcal{P}_o\subset T(G\times V\times V^*)$ that serves as the AVD phase space is given by
\begin{align}
\mathcal{P}_o=\{(Z,\dot{Z})\in T(G\times V\times V^*)\mid \dot{Z}=Y(Z)\}\approx G\times V\times V^*.
\end{align}

\subsection{Step 4: Derivation of boundary symplectic form on $\mathcal{P}_o$}
We will denote points in $G\times V\times V^*$ with the letter $Z$, i.e. $Z=(\tilde{g},\tilde{\chi},\tilde{f}_o)$. The vector field $Y$ defines a time-independent flow map $\mathfrak{F}_t:\mathcal{P}_o=G\times V\times V^*\rightarrow\mathcal{P}_o=G\times V\times V^*$, which is characterized by the relations
\begin{align}
\mathfrak{F}_0&=\text{id}_{\mathcal{P}_o}\\
\frac{\mathrm{d}}{\mathrm{d}t}\mathfrak{F}_t(Z)&=Y(Z).
\end{align}
We can therefore define a mapping $\text{Sol}:\mathcal{P}_o\rightarrow \mathfrak{P}(G\times V\times V^*)$, given by
\begin{align}
\text{Sol}(Z)(t)=\mathfrak{F}_{t-t_1}(Z),
\end{align}
that sends initial conditions (at $t=t_1$) to their corresponding solution path in $\mathfrak{P}(G\times V\times V^*)$.

The mapping $\text{Sol}$ can be used to pull back the augmented action functional $\mathcal{S}$ to the augmented phase space $\mathcal{P}_o$, thereby defining the \emph{restricted augmented action}
\begin{align}
\mathcal{S}_{\mathcal{P}_o}=\text{Sol}^*\mathcal{S}.
\end{align}
By examining the exterior derivative of the restricted augmented action, we can identify a symplectic form, and therefore Poisson brackets, for the AVD equations.

The exterior derivative of $\mathcal{S}_{\mathcal{P}_o}$ is proportional to the \emph{free-endpoint} variation of the augmented action functional $\mathcal{S}$. Because this variation will be evaluated at a curve in $G\times V\times V^*$ that satisfies the AVD Euler-Lagrange equations, only the endpoint contributions to the free-endpoint variation will appear. Specifically we have
\begin{align}\label{fundamental_exterior_derivative}
\mathbf{d}\mathcal{S}_{\mathcal{P}_o}=\mathfrak{F}_{t_2-t_1}^*\Theta-\Theta,
\end{align}
where $\Theta$ is the $1$-form on $\mathcal{P}_o$ given by
\begin{align}
\Theta(Z)[\delta Z]=\sum_{s=1}^{N_S}\int_{TQ}\left(\theta_s(\tilde{g}_*\tilde{f}_o)\otimes (g_{s})_*f_{o,s}\right)\cdot\eta_s+\chi_s\,\delta f_{o,s}.
\end{align}
Here, $\eta_s=\delta g_s\circ g_s^{-1}$. If we now differentiate Eq.\,(\ref{fundamental_exterior_derivative}) in the variable $t_2$, keeping in mind that $\mathcal{S}_{\mathcal{P}_o}$ depends on $t_2$ \emph{via} the upper limit of time integration, we obtain
\begin{align}\label{intermediate_hamilton_equation}
\mathbf{d}\mathfrak{F}_{t_2-t_1}^*\mathcal{L}_{\mathcal{P}_o}=\mathfrak{F}_{t_2-t_1}^*L_Y\Theta,
\end{align}
where $\mathcal{L}_{\mathcal{P}_o}$ is the augmented Lagrangian pulled back to the augmented phase space via $Y:\mathcal{P}_o\rightarrow T(G\times V\times V^*)$, i.e.
\begin{align}
\mathcal{L}_{\mathcal{P}_o}(Z)&=(Y^*\mathcal{L})(Z)\nonumber\\
&=\mathcal{L}(Y(Z))\nonumber\\
&=\sum_{s=1}^{N_s}\int_P K_s\,(g_{s})_*f_{o,s}+\frac{1}{2 c}\left<\mathcal{A}(\tilde{X}_o,\tilde{g}_*\tilde{f}_o),J(\tilde{X}_o,\tilde{g}_*\tilde{f}_o)\right>-\frac{1}{2}\left<\rho(\tilde{g}_*\tilde{f}_o),\Phi(\tilde{g}_*\tilde{f}_o)\right>.
\end{align}
By applying the identity $L_Y\Theta=\text{i}_{Y}\mathbf{d}\Theta+\mathbf{d}\text{i}_{Y}\Theta$, Eq.\,(\ref{intermediate_hamilton_equation}) can be re-written as
\begin{align}\label{augmented_hamilton_equation}
\text{i}_Y\mathbf{d}\Theta=-\mathbf{d}\mathcal{H}_{\mathcal{P}_o},
\end{align}
where
\begin{align}
\mathcal{H}_{\mathcal{P}_o}(Z)&=(\Theta(Y)-\mathcal{L}_{\mathcal{P}_o})(Z)\nonumber\\
&=\sum_{s=1}^{N_s}\int_P K_s\,(g_{s})_*f_{o,s}+\frac{1}{2 c}\left<\mathcal{A}(\tilde{X}_o,\tilde{g}_*\tilde{f}_o),J(\tilde{X}_o,\tilde{g}_*\tilde{f}_o)\right>+\frac{1}{2}\left<\rho(\tilde{g}_*\tilde{f}_o),\Phi(\tilde{g}_*\tilde{f}_o)\right>
\end{align}
is the augmented Hamiltonian functional. Because the $2$-form $-\mathbf{d}\Theta$ turns out to be non-degenerate, Eq.\,(\ref{augmented_hamilton_equation}) shows that the AVD equations are an infinite-dimensional Hamiltonian system with symplectic form $-\mathbf{d}\Theta$ and Hamiltonian $\mathcal{H}_{\mathcal{P}_o}$.

We will now calculate the Poisson bracket defined by the symplectic form $-\mathbf{d}\Theta$. We will proceed in two steps. First, we will find an expression for an arbitrary Hamiltonian vector field on $\mathcal{P}_o$ by solving the equation
\begin{align}\label{hamiltonian_vector_field_augmented}
\text{i}_{Y_{\mathsf{G}}}\mathbf{d}\Theta=-\mathbf{d}\mathsf{G},
\end{align}
for $Y_{\mathsf{G}}$ given an arbitrary functional $\mathsf{G}:\mathcal{P}_o\rightarrow\mathbb{R}$. Next we will identify an explicit expression for the Poisson bracket associated with $-\mathbf{d}\Theta$  using the formula
\begin{align}\label{lie_poisson_identity}
[\mathsf{F},\mathsf{G}]_{\mathcal{P}_o}=L_{Y_{\mathsf{G}}}\mathsf{F}.
\end{align}
Here, $[\cdot,\cdot]_{\mathcal{P}_o}$ is the augmented system's Poisson bracket. Because calculating the Lie derivative in the last expression is simple, all of the nontrivial work will be done in the first step.

A helpful tool for solving Eq.\,(\ref{hamiltonian_vector_field_augmented}) is the phase space variational principle. This variational principle states that (fixed-endpoint) variations of the action functional $\mathfrak{S}_{\mathsf{G}}:\mathfrak{P}(\mathcal{P}_o)\rightarrow\mathbb{R}$ given by 
\begin{align}
\mathfrak{S}_{\mathsf{G}}(\bm{Z})=\int_{t_1}^{t_2}\bigg(\Theta[\dot{\bm{Z}}(t)]-\mathsf{G}(\bm{Z}(t))\bigg)\,dt
\end{align}
are zero if and only if $\bm{Z}$ is a solution of the equation
\begin{align}
\frac{\rm{d}}{\rm{d}t}\bm{Z}(t)=Y_{\mathsf{G}}(\bm{Z}(t)).
\end{align}
Thus, we know \emph{a priori} that the Euler-Lagrange equations associated with the action $\mathfrak{S}_{\mathsf{G}}$ are Eq.\,(\ref{hamiltonian_vector_field_augmented}). On the other hand, we can compute the fixed-endpoint variation of $\mathfrak{S}_{\mathsf{G}}$ directly, giving
\begin{align}
&\delta\mathfrak{S}_{\mathsf{G}}(\bm{Z})[\delta\bm{Z}]=\nonumber\\
&\sum_{s=1}^{N_s}\int_{t_1}^{t_2}\int_{TQ}\bigg[\mathbf{d}\left(\frac{e_s}{c}\pi^*\mathcal{A}(\tilde{\bm{\xi}}(t),\tilde{\bm{f}}(t))(X_o)\right)-\frac{\rm{d}}{\rm{d}t}\theta_s-\frac{\bm{g}_s(t)_* \dot{\bm{f}}_{o,s}(t)}{\bm{f}_s(t)}\theta_s-\text{i}_{\bm{\xi}_s(t)}\mathbf{d}\theta_s-\frac{\delta\mathsf{G}}{\delta g_s}\bigg]\otimes\bm{f}_s(t)\cdot\bm{\eta}_s(t)\,dt\nonumber\\
+&\sum_{s=1}^{N_s}\int_{t_1}^{t_2}\int_{TQ}\bigg[\mathbf{g}_s(t)^*\left(\theta_s(\bm{\xi}_s(t))+\frac{e_s}{c}\pi^*\mathcal{A}(\tilde{\bm{\xi}}(t),\tilde{\bm{f}}(t))(X_o)\right)-\dot{\bm{\chi}}_s(t)-\frac{\delta\mathsf{G}}{\delta f_{o,s}}\bigg]\,\delta\bm{f}_{o,s}(t)\,dt\nonumber\\
+&\sum_{s=1}^{N_s}\int_{t_1}^{t_2}\int_{TQ}\bigg[\dot{\bm{f}}_{o,s}(t)-\frac{\delta\mathsf{G}}{\delta \chi_s}\bigg]\delta\bm{\chi}_s(t)\,dt,
\end{align}
where $\bm{f}_s(t)=\mathbf{g}_s(t)_*\bm{f}_{o,s}(t)$, $\bm{\eta}_s(t)=\delta\mathbf{g}(t)\circ\mathbf{g}(t)^{-1}$, and $\theta_s$ is evaluated at $\tilde{\bm{f}}(t)$. By the phase space variational principle, if we set this variation equal to zero and then solve for $\dot{\bm{Z}}(t)$ in terms of the functional derivatives of $\mathsf{G}$, the result will be the solution to Eq.\,(\ref{hamiltonian_vector_field_augmented}). We now turn to performing this task.

Just as when we formulated the AVD equations as a first-order ODE, solving for $\dot{\bm{Z}}(t)$ involves dealing with an implicit (linear) equation for $\tilde{\bm{\xi}}(t)$. Indeed, the Euler-Lagrange equation given by
\begin{align}\label{phase_space_e_l_xi_vd}
0=\mathbf{d}\left(\frac{e_s}{c}\pi^*\mathcal{A}(\tilde{\bm{\xi}}(t),\tilde{\bm{f}}(t))(X_o)\right)-\frac{\rm{d}}{\rm{d}t}\theta_s-\frac{\bm{g}_s(t)_* \dot{\bm{f}}_{o,s}(t)}{\bm{f}_s(t)}\theta_s-\text{i}_{\bm{\xi}_s(t)}\mathbf{d}\theta_s-\frac{\delta\mathsf{G}}{\delta g_s},
\end{align}
is an implicit equation for $\bm{\xi}_s(t)$, because $\tilde{\bm{\xi}}(t)$ appears in the first, second, and fourth terms. We will solve this implicit equation by working in the tangent lift of an arbitrary coordinate system $q^i$ on $Q$. First we will express $\mathcal{A}(\tilde{\bm{\xi}}(t),\tilde{\bm{f}}(t))$, which appears in the first term in Eq.\,(\ref{phase_space_e_l_xi_vd}), in terms of $\bm{Z}(t)$. Then we will express $\mathcal{A}(\tilde{X}_o,L_{\tilde{\bm{\xi}}(t)}\tilde{\bm{f}}(t))$, which appears in the second term of Eq.\,(\ref{phase_space_e_l_xi_vd}), in terms of $\bm{Z}(t)$.
\\ \\
\emph{Solving for $\mathcal{A}(\tilde{\bm{\xi}}(t),\tilde{\bm{f}}(t))$ ---} Let $g_{ij}$ denote the components of the metric tensor in our chosen coordinate system. The determinant of this matrix of components will be denoted $|g|$. Set
\begin{align}
\bm{\xi}_s(t)&=u_s^i\frac{\partial}{\partial q^i}+a_s^i\frac{\partial}{\partial \dot{q}^i}\\
\bm{f}_s(t)&=F_s\frac{1}{6}\mathbf{d}\vartheta\wedge\mathbf{d}\vartheta\wedge\mathbf{d}\vartheta\\
\mathcal{A}(\tilde{\bm{\xi}}(t),\tilde{\bm{f}}(t))&=A_i\,\mathbf{d}q^i\\
\frac{\delta\mathsf{G}}{\delta g_s}&=Q_i\,\mathbf{d}q^i+\dot{Q}_i\,\mathbf{d}\dot{q}^i.
\end{align} 
Note that 
\begin{align}
\frac{1}{6}\mathbf{d}\vartheta\wedge\mathbf{d}\vartheta\wedge\mathbf{d}\vartheta=|g|\,\mathbf{d}q^1\wedge\mathbf{d}q^2\wedge\mathbf{d}q^3\wedge\mathbf{d}\dot{q}^1\wedge\mathbf{d}\dot{q}^2\wedge\mathbf{d}\dot{q}^3.
\end{align}

In order to compute $\mathcal{A}(\tilde{\bm{\xi}}(t),\tilde{\bm{f}}(t))$, we will draw upon coordinate expressions for the current density operator $J(\tilde{\bm{\xi}}(t),\tilde{\bm{f}}(t))=\sum_{s}e_s*u(\bm{\xi}_s(t),\bm{f}_s(t))$. As is readily verified, the fiber integrals that give the particle flux $1$-forms, $*u$, can be written as
\begin{align}\label{useful_fiber_integral}
*\int_{\pi^{-1}}\text{i}_{\bm{\xi}_s(t)}\bm{f}_s(t)=\left(\int F_s u_s^ig_{ij}\sqrt{|g|}\,d\dot{q}\right)\,\mathbf{d}q^j,
\end{align}
where $d\dot{q}=d\dot{q}^1d\dot{q}^2d\dot{q}^3$ denotes the ``bare'' measure on $\dot{q}$ space. The current density $1$-form is therefore given by
\begin{align}
J(\tilde{\bm{\xi}}(t),\tilde{\bm{f}}(t))=\sum_{s=1}^{N_s}e_s\left(\int F_s u_s^ig_{ij}\sqrt{|g|}\,d\dot{q}\right)\,\mathbf{d}q^j.
\end{align}
We will also draw upon the expression for $u_s^i$ that is implied by the $\mathbf{d}\dot{q}^i$-component of Eq.\,(\ref{phase_space_e_l_xi_vd}), namely
\begin{align}
g_{ij}u_s^j=-\frac{e_s}{m_s}\frac{1}{c}A_i+\frac{1}{m_s}\dot{Q}_i.
\end{align}

By definition, the $1$-form $\mathcal{A}(\tilde{\bm{\xi}}(t),\tilde{\bm{f}}(t))$ is given by
\begin{align}
\mathcal{A}(\tilde{\bm{\xi}}(t),\tilde{\bm{f}}(t))=\frac{4\pi}{c}\bm{G}\Pi_T\bigg(\sum_{s=1}^{N_s}e_s\left(\int F_s u_s^ig_{ij}\sqrt{|g|}\,d\dot{q}\right)\,\mathbf{d}q^j\bigg),
\end{align}
where we have used the fiber integral identity given above. Both sides of this equation depend on $\tilde{\bm{\xi}}(t)$. However, upon inserting the expression for $u_s^i$ given above, we obtain
\begin{align}
\mathcal{A}(\tilde{\bm{\xi}}(t),\tilde{\bm{f}}(t))&=-\frac{1}{c^2}\bm{G}\Pi_T\bigg(\omega_p^2\,\mathcal{A}(\tilde{\bm{\xi}}(t),\tilde{\bm{f}}(t))\bigg)+\frac{4\pi}{c}\bm{G}\Pi_T\bigg(\sum_{s=1}^{N_s}e_s*u(\omega_s^{-1}[\delta \mathsf{G}/\delta g_s],\bm{f}_s(t))\bigg)\\
&=-\frac{1}{c^2}\bm{G}\Pi_T\bigg(\omega_p^2\,\mathcal{A}(\tilde{\bm{\xi}}(t),\tilde{\bm{f}}(t))\bigg)+\mathcal{A}(\omega^{-1}[\delta\mathsf{G}/\delta\tilde{g}],\tilde{\bm{f}}(t)),
\end{align}
where $\omega^{-1}[\delta\mathsf{G}/\delta\tilde{g}]$ is the element of $\mathfrak{g}$ given by $(\omega^{-1}[\delta\mathsf{G}/\delta\tilde{g}])_s=\omega_s^{-1}[\delta \mathsf{G}/\delta g_s]$, and we have used the identity
\begin{align}
*n(\bm{f}_s(t))=\int F_s\,\sqrt{|g|}\,d\dot{q}. 
\end{align} 
We therefore arrive at the simple conclusion
\begin{align}
\mathcal{A}(\tilde{\bm{\xi}}(t),\tilde{\bm{f}}(t))&=\left[1+\frac{1}{c^2}\bm{G}\Pi_T\hat{\omega}_p^2\right]^{-1}\mathcal{A}(\omega^{-1}[\delta\mathsf{G}/\delta\tilde{g}],\tilde{\bm{f}}(t))\\
&=\bm{n}^2(\tilde{\bm{f}}(t))\left[\mathcal{A}(\omega^{-1}[\delta\mathsf{G}/\delta\tilde{g}],\tilde{\bm{f}}(t))\right],
\end{align}
where we have introduced the squared refractive index operator
\begin{align}
\bm{n}^2(\tilde{f})=\left[1+\frac{1}{c^2}\bm{G}\Pi_T\hat{\omega}_p^2(\tilde{f})\right]^{-1}.
\end{align}
\\ \\
\emph{Solving for $\mathcal{A}(\tilde{X}_o,L_{\tilde{\bm{\xi}}(t)}\tilde{\bm{f}}(t))$ ---} In order to solve for $\mathcal{A}(\tilde{X}_o,L_{\tilde{\bm{\xi}}(t)}\tilde{\bm{f}}(t))$ in terms of $\bm{Z}(t)$, we will make use of coordinate expressions for the $2$-form $\mathbf{d}\theta_s$ and the $1$-form $*u(X_o,L_{\bm{\xi}_s(t)}\bm{f}_s(t))$. We have
\begin{align}
\mathbf{d}\theta_s&=m_s g_{ij}\,\mathbf{d}\dot{q}^i\wedge\mathbf{d}q^j+\bigg(m_s\Gamma_{jki} \dot{q}^i+\frac{1}{2}\frac{e_s}{c}B_{kj}\bigg)\,\mathbf{d}q^k\wedge\mathbf{d}q^j\\
*u(X_o,L_{\bm{\xi}_s(t)}\bm{f}_s(t))&=\left(\frac{1}{\sqrt{|g|}}\frac{\partial}{\partial q^k}\left(\sqrt{|g|}T^{ki}\right)-\int a_s^i F_s\sqrt{|g|}d\dot{q}\right)\,g_{ij}\mathbf{d}q^j,
\end{align}
where
\begin{align}
\Gamma_{jki}=\frac{1}{2}\left(g_{jk,i}+g_{ji,k}-g_{ki,j}\right)
\end{align}
are the Christoffel symbols of the first kind,
\begin{align}
\frac{1}{2}B_{kj}\mathbf{d}q^k\wedge\mathbf{d}q^j=\mathbf{d}\mathcal{A}(\tilde{X}_o,\tilde{\bm{f}}(t)),
\end{align}
and
\begin{align}
T^{ki}=&\int u_s^k\dot{q}^i\,F_s\sqrt{|g|}\,d\dot{q}.
\end{align}
We will also employ a coordinate expression for the inverse of the $2$-form $\omega_s=-\mathbf{d}\theta_s$. Given a $1$-form on $TQ$, $\alpha=\alpha_j\,\mathbf{d}q^j+\dot{\alpha}_j\,\mathbf{d}\dot{q}^j$,
\begin{align}
\omega_s^{-1}(\alpha)=\bigg(\frac{1}{m_s}\dot{\alpha}^l\bigg)\frac{\partial}{\partial q^l}-\bigg(\frac{e_s}{m_s^2 c}\dot{\alpha}^k B^*_{kj}g^{jl}+\frac{1}{m_s}\alpha^l\bigg)\frac{\partial}{\partial \dot{q}^l},
\end{align}
where 
\begin{align}
B^*_{kj}=\frac{cm_s}{e_s}\dot{q}^i[\Gamma_{jki}-\Gamma_{kji}]+B_{kj}.
\end{align}

Set $\mathcal{A}(t)=\mathcal{A}(\tilde{X}_o,L_{\tilde{\bm{\xi}}(t)}\tilde{\bm{f}}(t))$. By definition, the $1$-form $\mathcal{A}(t)$ is given by
\begin{align}
\mathcal{A}(t)=\frac{4\pi}{c}\bm{G}\Pi_T\left(\sum_{s=1}^{N_s}e_s*u(X_o,L_{\bm{\xi}_s(t)}\bm{f}_s(t))\right).
\end{align}
If we decompose $\bm{\xi}_s(t)$ as
\begin{align}
\bm{\xi}_s(t)=\bm{\nu}_s(t)-\frac{e_s}{c}\omega_s^{-1}\left(\mathcal{A}(t)\right),
\end{align}
where $\bm{\nu}_s(t)$ is the vector field on phase space defined by
\begin{align}
\text{i}_{\bm{\nu}_s(t)}\omega_s=&\frac{\delta\mathsf{G}}{\delta g_s}+\frac{\mathbf{g}_s(t)_*(\delta\mathsf{G}/\delta\chi_s)}{\bm{f}_s(t)}\theta_s+\frac{e_s}{c}\pi^*\mathcal{A}(\tilde{X}_o,\tilde{\mathbf{g}}(t)_*(\delta\mathsf{G}/\delta\tilde{\chi}))\nonumber\\
&-\mathbf{d}\left(\frac{e_s}{c}\pi^*\bm{n}^2(\tilde{\bm{f}}(t))\left[\mathcal{A}(\omega^{-1}[\delta\mathsf{G}/\delta\tilde{g}],\tilde{\bm{f}}(t))\right](X_o)\right),
\end{align}
then we can decompose $\mathcal{A}(\tilde{X}_o,L_{\tilde{\bm{\xi}}(t)}\tilde{\bm{f}}(t))$ into two pieces,
\begin{align}\label{solving_for_calA}
\mathcal{A}(t)=\frac{4\pi}{c}\bm{G}\Pi_T\left(\sum_{s=1}^{N_s}e_s*u(X_o,L_{\bm{\nu}_s(t)}\bm{f}_s(t))\right)-\bm{G}\Pi_T\left(\sum_{s=1}^{N_s}\frac{4\pi e_s^2}{c^2}*u(X_o,L_{\omega_s^{-1}(\mathcal{A}(t))}\bm{f}_s(t))\right).
\end{align}
The first term on the right-hand-side of this expression is given entirely in terms of $\bm{Z}(t)$ because the vector field $\bm{\nu}_s(t)$ only depends on $\bm{Z}(t)$. The second term on the right-hand-side involves the quantity we are trying to solve for, $\mathcal{A}(t)$. Using the coordinate identities given earlier, the $1$-form $*u(X_o,L_{\omega_s^{-1}(\mathcal{A}(t))}\bm{f}_s(t))$ can be expressed as
\begin{align}
*u(X_o,L_{\omega_s^{-1}(\mathcal{A}(t))}\bm{f}_s(t))=\frac{1}{m_s}*n(\bm{f}_s(t))\mathcal{A}(t).
\end{align}
Therefore Eq.\,(\ref{solving_for_calA}) simplifies to
\begin{align}
\mathcal{A}(t)=\frac{4\pi}{c}\bm{G}\Pi_T\left(\sum_{s=1}^{N_s}e_s*u(X_o,L_{\bm{\nu}_s(t)}\bm{f}_s(t))\right)-\frac{1}{c^2}\bm{G}\Pi_T\left(\omega_p^2\mathcal{A}(t)\right),
\end{align}
which provides us with the expression for $\mathcal{A}(t)=\mathcal{A}(\tilde{X}_o,L_{\tilde{\bm{\xi}}(t)}\tilde{\bm{f}}(t))$ we have sought after:
\begin{align}
\mathcal{A}(\tilde{X}_o,L_{\tilde{\bm{\xi}}(t)}\tilde{\bm{f}}(t))=\bm{n}^2(\tilde{\bm{f}}(t))\left[\mathcal{A}(\tilde{X}_o,L_{\tilde{\bm{\nu}}(t)}\tilde{\bm{f}}(t))\right].
\end{align}
\\ \\
With these expressions for $\mathcal{A}(\tilde{X}_o,L_{\tilde{\bm{\xi}}(t)}\tilde{\bm{f}}(t))$ and $\mathcal{A}(\tilde{\bm{\xi}}(t),\tilde{\bm{f}}(t))$ in hand, we can finally write down an explicit expression for $Y_{\mathsf{G}}$. Set $Y(\tilde{g},\tilde{\chi},\tilde{f}_o)=(\tilde{g},\tilde{\chi},\tilde{f}_o,\dot{\tilde{g}}_{\mathsf{G}},\dot{\tilde{\chi}}_{\mathsf{G}},(\dot{\tilde{f}}_o)_{\mathsf{G}})$. We have
\begin{align}
(\dot{g}_s)_{\mathsf{G}}\circ g_s^{-1}&=\nu_s\left(\tilde{g}_*\tilde{f}_o;\frac{\delta\mathsf{G}}{\delta\tilde{g}};\tilde{g}_*\frac{\delta\mathsf{G}}{\delta\tilde{\chi}}\right)-\frac{e_s}{c}\omega_s^{-1}\left(\pi^*\bm{n}^2(\tilde{g}_*\tilde{f}_o)\left[\mathcal{A}(\tilde{X}_o,L_{\tilde{\nu}\left(\tilde{g}_*\tilde{f}_o;\frac{\delta\mathsf{G}}{\delta\tilde{g}};\tilde{g}_*\frac{\delta\mathsf{G}}{\delta\tilde{\chi}}\right)}\tilde{g}_*\tilde{f}_o)\right]\right)\\
(\dot{\chi}_s)_{\mathsf{G}}&=-\frac{\delta\mathsf{G}}{\delta f_{o,s}}\nonumber\\
&~~~~~+g_s^*\left(\theta_s\left(\nu_s\left(\tilde{g}_*\tilde{f}_o;\frac{\delta\mathsf{G}}{\delta\tilde{g}};\tilde{g}_*\frac{\delta\mathsf{G}}{\delta\tilde{\chi}}\right)\right)+\frac{e_s}{c}\pi^*\bm{n}^2(\tilde{g}_*\tilde{f}_o)\left[\mathcal{A}(\omega^{-1}[\delta\mathsf{G}/\delta\tilde{g}],\tilde{g}_*\tilde{f}_o)\right](X_o)\right)\\
(\dot{f}_{o,s})_{\mathsf{G}}&=\frac{\delta\mathsf{G}}{\delta\chi_s},
\end{align}
where the operator $\nu_s:V^*\times\mathfrak{g}^*\times V^*\rightarrow \mathfrak{X}(TQ)$ is given by
\begin{align}
\nu_s\left(\tilde{f};\tilde{\alpha};\tilde{h}\right)=\omega_s^{-1}\left(\alpha_s+\frac{h_s}{f_s}\theta_s+\frac{e_s}{c}\pi^*\mathcal{A}(\tilde{X}_o,\tilde{h})-\mathbf{d}\left(\frac{e_s}{c}\pi^*\bm{n}^2(\tilde{f}(t))\left[\mathcal{A}(\omega^{-1}[\tilde{\alpha}],\tilde{f}(t))\right](X_o)\right)\right).
\end{align}

It follows from these expressions together with Eq.\,(\ref{lie_poisson_identity}) that the Poisson bracket on the augmented Vlasov-Darwin phase space can be written down immediately. However, the most obvious form of the bracket is not manifestly antisymmetric. After some toil, an antisymmetric expression for the bracket can be found. It is given by
\begin{align}
&[\mathsf{F},\mathsf{G}]_{\mathcal{P}_o}(Z)=\nonumber\\
&\sum_{s=1}^{N_s}\int_{TQ} B_s\left[\frac{\delta\mathsf{F}}{\delta g_s}-\mathbf{d}\mathfrak{L}_s\left(\omega^{-1}\left(\frac{\delta\mathsf{F}}{\delta\tilde{g}}\right),\tilde{g}_*\tilde{f}_o\right),\frac{\delta\mathsf{G}}{\delta g_s}-\mathbf{d}\mathfrak{L}_s\left(\omega^{-1}\left(\frac{\delta\mathsf{G}}{\delta\tilde{g}}\right),\tilde{g}_*\tilde{f}_o\right)\right]\,g_{s*}f_{os}\nonumber\\
&+ \sum_{s=1}^{N_s}\int_{TQ}B_s\left[\frac{\delta \mathsf{F}}{\delta g_s}-\mathbf{d}\mathfrak{L}_s\left(\omega^{-1}\left(\frac{\delta\mathsf{F}}{\delta\tilde{g}}\right),\tilde{g}_*\tilde{f}_o\right),\theta_s\right]\left(g_{s*}\frac{\delta\mathsf{G}}{\delta\chi_s}\right)\nonumber\\
&-\sum_{s=1}^{N_s}\int_{TQ}B_s\left[\frac{\delta \mathsf{G}}{\delta g_s}-\mathbf{d}\mathfrak{L}_s\left(\omega^{-1}\left(\frac{\delta\mathsf{G}}{\delta\tilde{g}}\right),\tilde{g}_*\tilde{f}_o\right),\theta_s\right]\left(g_{s*}\frac{\delta\mathsf{F}}{\delta\chi_s}\right)\nonumber\\
&-\sum_{s=1}^{N_s}\int_{TQ}\mathfrak{L}_s\left(\omega^{-1}\left(\frac{\delta\mathsf{F}}{\delta\tilde{g}}\right),\tilde{g}_*\tilde{f}_o\right)\left(g_{s*}\frac{\delta\mathsf{G}}{\delta\chi_s}\right)-\mathfrak{L}_s\left(\omega^{-1}\left(\frac{\delta\mathsf{G}}{\delta\tilde{g}}\right),\tilde{g}_*\tilde{f}_o\right)\left(g_{s*}\frac{\delta\mathsf{F}}{\delta\chi_s}\right)\nonumber\\
&+\sum_{s=1}^{N_s}\int_{TQ}\frac{\delta\mathsf{F}}{\delta f_{os}}\frac{\delta\mathsf{G}}{\delta\chi_s}-\frac{\delta\mathsf{G}}{\delta f_{os}}\frac{\delta\mathsf{F}}{\delta\chi_s},
\end{align}
where $\mathfrak{L}_s:\mathfrak{g}\times V^*\rightarrow C^\infty(TQ)$ is a non-linear operator given by
\begin{align}
\mathfrak{L}_s(\tilde{\xi},\tilde{f})=\frac{e_s}{c}\pi^*\bigg(\bm{n}^2(\tilde{f})\left[\mathcal{A}(\tilde{\xi},\tilde{f})\right]\bigg)(X_o),
\end{align}
and $\omega^{-1}\left(\frac{\delta\mathsf{F}}{\delta\tilde{g}}\right)\in\mathfrak{g}$ is given by
\begin{align}
\left[\omega^{-1}\left(\frac{\delta\mathsf{F}}{\delta\tilde{g}}\right)\right]_s=\omega_s^{-1}\left(\frac{\delta\mathsf{F}}{\delta g_s}\right).
\end{align}
In deriving this expression for the augmented system's Poisson bracket, we have made use of the fact that the operator $\bm{n}^2(\tilde{f})\bm{G}\Pi_T$ is self-adjoint. See the next brief subsection for a proof of the self-adjoint property. 

\subsubsection{Properties of the squared refractive index operator}
The squared refractive index operator is defined by
\begin{align}
\bm{n}^2(f)=\left[1+\frac{1}{c^2}\bm{G}\Pi_T\hat{\omega}_p^2(\tilde{f})\right]^{-1},
\end{align}
where $\hat{\omega}_p^2(\tilde{f})$ is the operator that simply multiplies by the (squared) local plasma frequency. $\bm{n}^2(\tilde{f})$ operates on differential forms over $Q$. This definition may be perplexing because it is not immediately obvious that $A\equiv1+\frac{1}{c^2}\bm{G}\Pi_T\hat{\omega}_p^2(\tilde{f})$ should be an invertible operator. To see that it is, suppose that  $A$ has a non-trivial null eigenvector $\alpha_o$. Then $\alpha_o$ would have to satisfy
\begin{align}
\frac{1}{c^2}\bm{G}\Pi_T\hat{\omega}_p^2(\tilde{f})\alpha_o=-\alpha_o.
\end{align}
In particular, $\alpha_o$ would have to be an eigenvector of the operator $\frac{1}{c^2}\bm{G}\Pi_T\hat{\omega}_p^2(\tilde{f})$ with eigenvalue $-1$. This is impossible for the following reason. Define the weighted inner product
\begin{align}
\left<\alpha,\beta\right>_\omega=\int_Q(\alpha\wedge*\beta) \,\omega_p^2(\tilde{f}).
\end{align}
Because the operator $\bm{G}\Pi_T$ is the product of non-negative definite operators, it is non-negative definite itself. Therefore, for each $\alpha$,
\begin{align}
\left<\alpha,\frac{1}{c^2}\bm{G}\Pi_T\hat{\omega}_p^2(\tilde{f})\alpha\right>_\omega=\left<\hat{\omega}_p^2(\tilde{f})\alpha,\bm{G}\Pi_T\hat{\omega}_p^2(\tilde{f})\alpha\right>\geq 0.
\end{align}
But this contradicts our assumption that there is a non-trivial null eigenvector of $\alpha_o$. Indeed,
\begin{align}
\left<\alpha_o,\frac{1}{c^2}\bm{G}\Pi_T\hat{\omega}_p^2(\tilde{f})\alpha_o\right>_\omega=-\left<\alpha_o,\alpha_o\right><0.
\end{align}
It follows that the operator $A$ is invertible, and that $\bm{n}^2(\tilde{f})$ is well-defined.

The most basic property of $\bm{n}^2(\tilde{f})$ is that it commutes with $E\equiv\frac{1}{c^2}\bm{G}\Pi_T\hat{\omega}_p^2(\tilde{f})$. To see this, set $C=\bm{n}^2(\tilde{f})E-E\bm{n}^2(\tilde{f})$. We have
\begin{align}
&(1+E)C=E-(1+E) E\bm{n}^2(\tilde{f})\nonumber\\
\Rightarrow&(1+E)C(1+E)=E(1+E)-(1+E)E=\hat{0}\nonumber\\
\Rightarrow&C=\hat{0}.
\end{align}

While $\bm{n}^2(\tilde{f})$ is not self-adjoint as on operator on the standard $L^2$ space of differential forms, it is self-adjoint as an operator on the weighted $L^2_\omega$ space defined by the weighted inner product introduced earlier. To see this, let $\alpha$ and $\beta$ be arbitrary $k$-forms in $L^2_\omega$. Set $\alpha_1=\bm{n}^2(\tilde{f})\alpha$ and $\beta_1=\bm{n}^2(\tilde{f})\beta$. We have
\begin{align}
\left<\alpha,\bm{n}^2(\tilde{f})\beta\right>_\omega&=\left<\left[1+\frac{1}{c^2}\bm{G}\Pi_T\hat{\omega}_p^2(\tilde{f})\right]\alpha_1,\beta_1\right>_\omega\nonumber\\
&=\left<\hat{\omega}_p^2(\tilde{f})\alpha_1,\beta_1\right>+\left<\frac{1}{c^2}\bm{G}\Pi_T\hat{\omega}_p^2(\tilde{f})\alpha_1,\hat{\omega}_p^2(\tilde{f})\beta_1\right>\nonumber\\
&=\left<\hat{\omega}_p^2(\tilde{f})\alpha_1,\beta_1\right>+\left<\hat{\omega}_p^2(\tilde{f})\alpha_1,\frac{1}{c^2}\bm{G}\Pi_T\hat{\omega}_p^2(\tilde{f})\beta_1\right>\nonumber\\
&=\left<\hat{\omega}_p^2(\tilde{f})\alpha_1,\left[1+\frac{1}{c^2}\bm{G}\Pi_T\hat{\omega}_p^2(\tilde{f})\right]\beta_1\right>\nonumber\\
&=\left<\alpha_1,\left[1+\frac{1}{c^2}\bm{G}\Pi_T\hat{\omega}_p^2(\tilde{f})\right]\beta_1\right>_\omega\nonumber\\
&=\left<\bm{n}^2(\tilde{f})\alpha,\beta\right>_\omega,
\end{align}
where we have used the fact that $\bm{G}\Pi_T$ is a self-adjoint operator on $L^2$. 

When we combine the last two properties of the squared refractive index operator, we obtain the important result that $\bm{n}^2(\tilde{f})\bm{G}\Pi_T$ is self-adjoint on $L^2$. Indeed,
\begin{align}
\left<\alpha,\bm{n}^2(\tilde{f})\bm{G}\Pi_T\beta\right>&=\left<\frac{\alpha}{\omega_p^2(\tilde{f})},\bm{n}^2(\tilde{f})\bm{G}\Pi_T\beta\right>_\omega\nonumber\\
&=\left<\bm{n}^2(\tilde{f})\frac{\alpha}{\omega_p^2(\tilde{f})},\bm{G}\Pi_T\hat{\omega}_p^2(\tilde{f})\frac{\beta}{\omega_p^2(\tilde{f})}\right>_\omega\nonumber\\
&=\left<\bm{G}\Pi_T\hat{\omega}_p^2(\tilde{f})\bm{n}^2(\tilde{f})\frac{\alpha}{\omega_p^2(\tilde{f})},\frac{\beta}{\omega_p^2(\tilde{f})}\right>_\omega\nonumber\\
&=\left<\bm{n}^2(\tilde{f})\bm{G}\Pi_T\alpha,\beta\right>.
\end{align}

\subsection{Step 5: perform Poisson reduction to obtain bracket on physical phase space}
We have now identified the Hamiltonian, 
\begin{align}
\mathcal{H}_{\mathcal{P}_o}(\tilde{g},\tilde{\chi},\tilde{f}_o)=\sum_{s=1}^{N_s}\int_P K_s\,(g_{s})_*f_{o,s}+\frac{1}{2 c}\left<\mathcal{A}(\tilde{X}_o,\tilde{g}_*\tilde{f}_o),J(\tilde{X}_o,\tilde{g}_*\tilde{f}_o)\right>+\frac{1}{2}\left<\rho(\tilde{g}_*\tilde{f}_o),\Phi(\tilde{g}_*\tilde{f}_o)\right>,
\end{align}
and Poisson bracket for the \emph{augmented} Vlasov-Darwin system. The Jacobi identity is satisfied because the bracket has been obtained by inverting the symplectic form $-\mathbf{d}\Xi$. The modifier ``augmented" is appropriate because the dynamical variable $\tilde{\chi}$ has no direct physical meaning. On the other hand, the variables $
\tilde{g}$ and $\tilde{f}_o$ together comprise an element of the physical \emph{Lagrangian} (as opposed to Eulerian) phase space; $\tilde{g}$ gives the configuration of particles in the single-particle phase space $TQ$ and $\tilde{f}_o$ gives the reference phase space density. The purpose of the additional variable $\tilde{\chi}$ is to extend the Lagrangian phase space just enough to allow for a non-degenerate Poisson bracket.

The appearance of the variable $\tilde{\chi}$ perhaps seems awkward at this stage. However, observe the following. The set $V$ is a Lie group under addition that is a symmetry group for the AVD equations. Specifically, for each $\delta\tilde{\chi}\in V$, we can define a mapping $T_{\delta\tilde{\chi}}:\mathcal{P}_o\rightarrow\mathcal{P}_o$ given by
\begin{align}
T_{\delta\tilde{\chi}}(\tilde{g},\tilde{\chi},\tilde{f}_o)=(\tilde{g},\tilde{\chi}+\delta\tilde{\chi},\tilde{f}_o),
\end{align}
which clearly satisfies the defining properties of a group action,
\begin{align}
T_{\delta\tilde{\chi}_1+\delta\tilde{\chi}_2}&=T_{\delta\tilde{\chi}_1}\circ T_{\delta\tilde{\chi}_2}\\
T_{0}&=\text{id}_{\mathcal{P}_o}.
\end{align}
This group action leaves the augmented Hamiltonian $\mathcal{H}_{\mathcal{P}_o}$ and the augmented Poisson bracket $[\cdot,\cdot]_{\mathcal{P}_o}$ invariant in the sense that
\begin{align}
T_{\delta\tilde{\chi}}^*\mathcal{H}_{\mathcal{P}_o}&=\mathcal{H}_{\mathcal{P}_o}\label{invariance_lagrangian_hamiltonian}\\
T_{\delta\tilde{\chi}}^*[\mathsf{F},\mathsf{G}]_{\mathcal{P}_o}&=[T_{\delta\tilde{\chi}}^*\mathsf{F},T_{\delta\tilde{\chi}}^*\mathsf{G}]_{\mathcal{P}_o}\label{invariance_lagrangian_bracket},
\end{align}
for arbitrary $\delta\tilde{\chi}\in V$ and functionals $\mathsf{F},\mathsf{G}$ on the augmented phase space. These properties are quick to verify.
Therefore we can define a Hamiltonian and Poisson bracket on the first reduced phase space $\mathcal{P}_L=\mathcal{P}_o/V$, i.e we can perform \emph{Poisson reduction}. 

The details of applying Poisson reduction to pass from the AVD phase space to the Lagrangian phase space $\mathcal{P}_L$ follow.
The Lagrangian Hamiltonian, $\mathcal{H}_{\mathcal{P}_L}:\mathcal{P}_L\rightarrow\mathbb{R}$, is uniquely determined by requiring
\begin{align}
\mathcal{H}_{\mathcal{P}_L}(\pi_L(\tilde{g},\tilde{\chi},\tilde{f}_o))=\mathcal{H}_{\mathcal{P}_o}(\tilde{g},\tilde{\chi},\tilde{f}_o)
\end{align}
for each $(\tilde{g},\tilde{\chi},\tilde{f}_o)\in\mathcal{P}_o$. Here the projection map $\pi_L:\mathcal{P}_o\rightarrow\mathcal{P}_L$ is given by 
\begin{align}\label{pil_VD}
\pi_L(\tilde{g},\tilde{\chi},\tilde{f}_o)=(\tilde{g},\tilde{f}_o).
\end{align}
The Lagrangian Poisson bracket, $[\cdot,\cdot]_{\mathcal{P}_L}$, is defined by requiring
\begin{align}\label{reduced_bracket_relation_VD}
\pi_L^*[\mathfrak{F},\mathfrak{G}]_{\mathcal{P}_L}=[\pi_L^*\mathfrak{F},\pi_L^*\mathfrak{G}]_{\mathcal{P}_o}.
\end{align}
These definitions make sense because the augmented Hamiltonian and Poisson bracket are invariant under the the action of $V$. We find that the Lagrangian Hamiltonian is given by
\begin{align}
\mathcal{H}_{\mathcal{P}_L}(\tilde{g},\tilde{f}_o)=\sum_{s=1}^{N_s}\int_P K_s\,g_{s*}f_{os}+\frac{1}{2 c}\left<\mathcal{A}(\tilde{X}_o,\tilde{g}_*\tilde{f}_o),J(\tilde{X}_o,\tilde{g}_*\tilde{f}_o)\right>+\frac{1}{2}\left<\rho(\tilde{g}_*\tilde{f}_o),\Phi(\tilde{g}_*\tilde{f}_o)\right>.
\end{align} 
The Lagrangian Poisson bracket is given by
\begin{align}
\left[\mathfrak{F},\mathfrak{G}\right]_{\mathcal{P}_L}=\sum_{s=1}^{N_s}\int_{TQ} B_s\left[\frac{\delta\mathfrak{F}}{\delta g_s}-\mathbf{d}\mathfrak{L}_s\left(\omega^{-1}\left(\frac{\delta\mathfrak{F}}{\delta\tilde{g}}\right),\tilde{g}_*\tilde{f}_o\right),\frac{\delta\mathfrak{G}}{\delta g_s}-\mathbf{d}\mathfrak{L}_s\left(\omega^{-1}\left(\frac{\delta\mathfrak{G}}{\delta\tilde{g}}\right),\tilde{g}_*\tilde{f}_o\right)\right]\,g_{s*}f_{os}.
\end{align}
This bracket and Hamiltonian give a Hamiltonian formulation of the Vlasov-Darwin equations in Lagrangian labeling.

Now we will pass from Lagrangian labeling to Eulerian labeling by applying Poisson reduction a second time. The set $G$ is a symmetry group of the Vlasov-Darwin system in Lagrangian labeling. Specifically, for each $\tilde{h}\in G$, we can define a mapping $R_{\tilde{h}}:\mathcal{P}_L\rightarrow\mathcal{P}_L$ given by
\begin{align}
R_{\tilde{h}}(\tilde{g},\tilde{f}_o)=(\tilde{g}\circ \tilde{h},\tilde{h}^*\tilde{f}_o),
\end{align}
that satisfies the defining properties of a \emph{right} group action, namely
\begin{align}
R_{\tilde{h}_1\circ \tilde{h}_2}&=R_{\tilde{h}_2}\circ R_{\tilde{h}_1}\\
R_{\text{id}_P}&=\text{id}_{\mathcal{P}_L}.
\end{align}
The Lagrangian Hamiltonian and Poisson bracket are each invariant under this group action, in the sense that
\begin{align}
R_{\tilde{h}}^*\mathcal{H}_{\mathcal{P}_L}&=\mathcal{H}_{\mathcal{P}_L}\\
R_{\tilde{h}}^*[\mathcal{F},\mathcal{G}]_{\mathcal{P}_L}&=[R_{\tilde{h}}^*\mathcal{F},R_{\tilde{h}}^*\mathcal{G}]_{\mathcal{P}_L}.
\end{align}
The invariance of the Lagrangian Hamiltonian is quick to verify. The invariance of the Poisson bracket follows from the identity
\begin{align}
\left(\frac{\delta}{\delta \tilde{g}}R^*_{\tilde{h}}\mathfrak{F}\right)(\tilde{g},\tilde{f}_o)=\frac{\delta\mathfrak{F}}{\delta \tilde{g}}(\tilde{g}\circ \tilde{h},\tilde{h}^*
\tilde{f}_o).
\end{align}
Therefore, the Lagrangian Hamiltonian and Poisson bracket define corresponding quantities on the \emph{Eulerian} phase space, $\mathcal{P}_E=\mathcal{P}_L/G\approx V^*$. The Eulerian Hamiltonian is given by
\begin{align}
\mathcal{H}_{\mathcal{P}_E}(\tilde{f})=\sum_{s=1}^{N_s}\int_P K_s\,f_s+\frac{1}{2 c}\left<\mathcal{A}(\tilde{X}_o,\tilde{f}),J(\tilde{X}_o,\tilde{f})\right>+\frac{1}{2}\left<\rho(\tilde{f}),\Phi(\tilde{f})\right>.
\end{align}
The Eulerian Poisson bracket is defined in terms of the Eulerian projection map, $\pi_E:\mathcal{P}_{L}\rightarrow\mathcal{P}_E$, given by
\begin{align}
\pi_E(\tilde{g},\tilde{f}_o)=\tilde{g}_*\tilde{f}_o,
\end{align}
which satisfies the important property $\pi_E\circ R_h=\pi_E$. We have for functionals $\mathcal{F},\mathcal{G}:\mathcal{P}_E\rightarrow\mathbb{R}$,
\begin{align}
[\mathcal{F},\mathcal{G}]_{\mathcal{P}_E}(\tilde{f})&=[\pi_E^*\mathcal{F},\pi_E^*\mathcal{G}]_{\mathcal{P}_L}(\tilde{g},\tilde{f}_o)\nonumber\\
&=\sum_{s=1}^{N_s}\int_{TQ} \left\{\frac{\delta\mathcal{F}}{\delta f_s}-\mathfrak{L}_s\left(\tilde{X}_{\delta\mathcal{F}/\delta \tilde{f}},\tilde{f}\right),\frac{\delta\mathcal{G}}{\delta f_s}-\mathfrak{L}_s\left(\tilde{X}_{\delta\mathcal{G}/\delta\tilde{f}},\tilde{f}\right)\right\}_s\,f_s.
\end{align}
where $\tilde{g}$ and $\tilde{f}_o$ are any group element and $6$-form that satisfy $\tilde{f}=\tilde{g}_*\tilde{f}_o$, $\{\cdot,\cdot\}_s$ is the Poisson bracket associated with the symplectic form $\omega_s$, and $\tilde{X}_{\delta\mathcal{F}/\delta\tilde{f}}\in\mathcal{g}$ is given by
\begin{align}
\left(\tilde{X}_{\delta\mathcal{F}/\delta\tilde{f}}\right)_s=X_{\delta\mathcal{F}/\delta f_s}\equiv\omega_s^{-1}(\mathbf{d}\delta\mathcal{F}/\delta f_s).
\end{align}

\section{The gyrokinetic Vlasov-Maxwell system}
The gyrokinetic Maxwell-Vlasov system is most naturally defined by specifying its action, which is the sum of the net gyrocenter action and the Maxwell action. If a gauge-invariant form of the gyrocenter transformation is employed, we have
\begin{align}
&S_{\tilde{f}_o}(\tilde{\mathbf{g}},\bm{A},\bm{\phi})=\nonumber\\
&\int_{t_1}^{t_2}\bigg(\mathcal{L}_p(\tilde{\bm{\xi}}(t),\tilde{\bm{f}}(t),\bm{E}(t),\bm{B}(t))+\mathcal{L}_{\text{int}}(\tilde{\bm{\xi}}(t),\tilde{\bm{f}}(t),\bm{A}(t),\bm{\phi}(t))+\mathcal{L}_{\text{Max}}(\bm{E}(t),\bm{B}(t))\bigg)\,dt,
\end{align}
where $\tilde{\bm{\xi}}(t)=\dot{\tilde{\mathbf{g}}}\circ\tilde{\mathbf{g}}^{-1}$ is the multi-species Eulerian phase space fluid velocity, $\tilde{\bm{f}}(t)=\tilde{\mathbf{g}}(t)_*\tilde{f}_o$ is the multi-species gyrocenter phase space density (a collection of $6$-forms), $\bm{E}(t)=-\mathbf{d}\bm{\phi}(t)-\dot{\bm{A}}(t)/c$ is the electric field $1$-form, $\bm{B}(t)=\mathbf{d}\bm{A}(t)$ is the magnetic field $2$-form and the various Lagrangian functions are defined as follows. $\mathcal{L}_p$ is the ``free gyrocenter" Lagrangian (in analogy with the notion of a free-particle Lagrangian) given by
\begin{align}
\mathcal{L}_p(\tilde{\xi},\tilde{f},E,B)=\sum_{s=1}^{N_s}\bigg(\left[\int_{TQ}f_s\Xi_s(\xi_s)\right]-\mathcal{K}_s(f_s,E,B)\bigg),
\end{align} 
where $\Xi_s$ is the \emph{guiding} center $1$-form and $\mathcal{K}_s$ is the gyrocenter kinetic energy functional
\begin{align}
\mathcal{K}_s(f,E,B)=\int_{TQ}f K_s(E,B). 
\end{align}
Here $K_s:\Omega^1(Q)\times\mathbf{d}\Omega^1(Q)\rightarrow C^{\infty}(TQ)$ is the gyrocenter kinetic energy function, which has a non-local dependence on the electric and magnetic fields. $\mathcal{L}_{\text{int}}$ is the ``interaction" Lagrangian given by
\begin{align}
\mathcal{L}_{\text{int}}(\tilde{\xi},\tilde{f},A,\phi)=\sum_{s=1}^{N_s}\bigg(\int_{TQ}f_s\left[\frac{e_s}{c}\pi^*A(\xi_s)-e_s\pi^*\phi\right]\bigg)=\frac{1}{c}\left<J(\tilde{\xi},\tilde{f}),A\right>-\left<\rho(\tilde{f}),\phi\right>,
\end{align}
where the current density $1$-form and charge density operators, $J$ and $\rho$, were introduced in the previous section. Finally, $\mathcal{L}_{\text{Max}}$ is the free electromagnetic field action
\begin{align}
\mathcal{L}_{\text{Max}}(E,B)=\frac{1}{8\pi}\left<E,E\right>-\frac{1}{8\pi}\left<B,B\right>.
\end{align}
Explicit expressions for $\Xi$ and $K_s$ will not be necessary; it is enough to know they can be found in principle.

The Euler-Lagrange equations associated with the gyrokinetic action are given by
\begin{align}
\left(\ddt+L_{\bm{\xi}_s(t)}\right)\bigg(\frac{\delta(\mathcal{L}_p+\mathcal{L}_{\text{int}})}{\delta\xi_s}\bigg)&=\B{d}\left(\frac{\delta(\mathcal{L}_p+\mathcal{L}_{\text{int}})}{\delta f_s}\right)\otimes f_s\\
\frac{1}{c}\ddt\frac{\delta(\mathcal{L}_p+\mathcal{L}_{\text{Max}})}{\delta E}+\bm{\delta}\frac{\delta(\mathcal{L}_p+\mathcal{L}_{\text{Max}})}{\delta B}&=-\frac{\delta\mathcal{L}_{\text{int}}}{\delta A}\\
\bm{\delta}\frac{\delta(\mathcal{L}_p+\mathcal{L}_{\text{Max}})}{\delta E}&=\frac{\delta\mathcal{L}_{\text{int}}}{\delta\phi}.
\end{align}
Upon calculating the relevant functional derivatives, these equations reduce to
\begin{align}
\text{i}_{\bm{\xi}_s(t)}\mathbf{d}\Xi_s^{\text{gy}}&=-\dot{\Xi}_s^{\text{gy}}-\mathbf{d}\mathcal{H}_s^{\text{gy}}\label{hamiltons_equations}\\
\frac{1}{c}\ddt\bm{D}(t)-\bm{\delta}\bm{H}(t)&=-\frac{4\pi}{c}J(\T{\bm{\xi}}(t),\T{\bm{f}}(t))\label{ampere}\\
\bm{\delta}\bm{D}(t)&=-4\pi\rho(\T{\bm{f}}(t)),\label{gauss}
\end{align}
where we have introduced the \emph{gyro}center $1$-form and Hamiltonian, 
\begin{align}
\Xi_s^{\text{gy}}&=\Xi_s+\frac{e_s}{c}\pi^*A\\
\mathcal{H}_s^{\text{gy}}&=K_s+e_s\pi^*\phi,
\end{align}
the $1$-form
\begin{align}
\bm{D}(t)=\bm{E}(t)-4\pi\frac{\delta\mathcal{K}}{\delta E}=\hat{D}(\tilde{\bm{f}}(t),\bm{E}(t),\bm{B}(t)),
\end{align}
and the $2$-form
\begin{align}
\bm{H}(t)=\bm{B}(t)+4\pi\frac{\delta\mathcal{K}}{\delta B}=\hat{H}(\tilde{\bm{f}}(t),\bm{E}(t),\bm{B}(t)).
\end{align}
Note that in these expressions $\bm{E}(t)=-\B{d}\bm{\phi}(t)-\dot{\bm{A}}(t)/c$, $\bm{B}(t)=\B{d}\bm{A}(t)$, and $\mathcal{K}=\sum_{s}\mathcal{K}_s$, which implies $\B{d}{\bm{E}}(t)=-\dot{\bm{B}}(t)/c$. Equation \,(\ref{hamiltons_equations}) gives the gyrocenter equations of motion in Hamiltonian form. Equation \,(\ref{ampere}) is the ``macroscopic" Amp\`ere equation (written in terms of forms). Finally, Eq.\,(\ref{gauss}) is the macroscopic Gauss equation.

The reference gyrocenter phase space density $\T{f}_o$ is not a dynamical variable in this formulation of gyrokinetics. However, we can elevate $\T{f_o}$ to the status of a dynamical variable by embedding gyrokinetics within a slightly larger system. In particular, if we define the \emph{augmented} gyrokinetic action
\begin{align}
S(\T{\B{g}},\bm{A},\bm{\phi},\T{\bm{f}}_o,\T{\bm{\chi}})=S_{\T{\bm{f}}_o(t)}(\T{\B{g}},\bm{A},\bm{\phi})+\sum_{s=1}^{N_s}\int_{TQ}\dot{\bm{f}}_{os}(t)\bm{\chi}_s(t),
\end{align}
the resulting Euler-Lagrange equations are given by
\begin{align}
\text{i}_{\bm{\xi}_s(t)}\mathbf{d}\Xi_s^{\text{gy}}&=-\dot{\Xi}_s^{\text{gy}}-\mathbf{d}\mathcal{H}_s^{\text{gy}}\label{hamiltons_equations_aug}\\
\frac{1}{c}\ddt\bm{D}(t)-\bm{\delta}\bm{H}(t)&=-\frac{4\pi}{c}J(\T{\bm{\xi}}(t),\T{\bm{f}}(t))\label{ampere_aug}\\
\bm{\delta}\bm{D}(t)&=-4\pi\rho(\T{\bm{f}}(t)),\label{gauss_aug}\\
\dot{\bm{f}}_{os}(t)&=0\label{f_equation}\\
\dot{\bm{\chi}}_s(t)&=\B{g}_s(t)^*(\Xi_s^{\text{gy}}(\bm{\xi}_s(t))-\mathcal{H}_s^{\text{gy}}).\label{chi_aug}
\end{align}
The first four equations in this set decouple from the fifth and reproduce the gyrokinetic Vlasov-Maxwell dynamics. The fifth equation defines the dynamics of the additional variable $\T{\bm{\chi}}(t)$. We will refer to this larger system of equations as \emph{the augmented} gyrokinetic Vlasov-Maxwell equations, or the AGVM equations for short.

It may seem awkward to introduce the additional dynamical variable $\T{\bm{\chi}}(t)$. However, $\T{\bm{\chi}}(t)$ will be a help rather than a hinderance as we continue our derivation of the gyrokinetic bracket. Moreover, the evolution equation for $\T{\bm{\chi}}(t)$ is very much reminiscent of the evolution equation for the phase of the quantum wave function in the WKB approximation. I don't think this is a coincidence; a semiclassical treatment of gyrokinetics would most likely produce an equation very similar to Eq.\,(\ref{chi_aug}). This is strongly suggested by considering the gauge invariance properties of the augmented gyrokinetic action. Because $\bm{f}_{os}(t)$ depends on time, under the gauge transformation $\bm{A}(t)\rightarrow\bm{A}(t)+\bm{\varphi}(t)$, $\bm{\phi}(t)\rightarrow \bm{\phi}(t)-\dot{\bm{\varphi}}(t)/c$ the augmented gyrokinetic action is not weakly invariant (it changes by more than temporal boundary terms). However, if \emph{we also change} $\bm{\chi}_s(t)$ according to $\bm{\chi}_s(t)\rightarrow\bm{\chi}_s(t)-\frac{e_s}{c}\B{g}_s(t)^*\pi^*\bm{\varphi}(t)$, the gyrokinetic action \emph{is} left weakly invariant. Thus, $\bm{\chi}_s(t)$ changes in the same way as the quantum phase under a gauge transformation.

The AGMV equations determine an evolution equation for the variable $\bm{Z}(t)=(\T{\B{g}}(t),\bm{A}(t),$

\noindent$\bm{D}(t),\T{\bm{f}}_o(t),\T{\bm{\chi}}(t))$ that is first order in time. To see this, first choose the gauge $\bm{\phi}(t)=0$. In this gauge, $\bm{B}(t)=\B{d}\bm{A}(t)$ and $\bm{E}(t)=-\dot{\bm{A}}(t)/c$. Therefore, if we introduce the electric field operator $\hat{E}$, which is defined by the implicit equation
\begin{align}
D=\hat{D}(\tilde{f},\hat{E}(\T{f},D,B),B),
\end{align}
we can write 
\begin{align}
\ddt\bm{A}(t)=-c\hat{E}(\T{\bm{f}}(t),\bm{D}(t),\B{d}\bm{A}(t)),
\end{align}
where $\T{\bm{f}}(t)=\T{\B{g}}(t)_*\T{\bm{f}}_o(t)$, which gives the time derivative of $\bm{A}(t)$ in terms of $\bm{Z}(t)$. For the time derivative of $\T{\B{g}}(t)$, we note that
\begin{align}
\dot{\B{g}}_s(t)=\bm{\xi}_s(t)\circ\B{g}_s(t),
\end{align}
and that $\bm{\xi}_s(t)$ is a functional of $E$ and $B$ according to Hamilton's equations, i.e. $\bm{\xi}_s(t)=X_s^{\text{gy}}(\hat{E}(\T{\bm{f}}(t),\bm{D}(t),\B{d}\bm{A}(t)),\B{d}\bm{A}(t))$ for a functional $X_s^{\text{gy}}:\Omega^1(Q)\times\Omega^2(Q)\rightarrow\mathfrak{X}(TQ)$. Because $\bm{\xi}_s(t)$ can be expressed in terms of $\bm{Z}(t)$, it follows that the macroscopic Amp\`ere equation gives the time derivative of $\bm{D}(t)$ in terms of $\bm{Z}(t)$. For the same reason, the time derivative of $\bm{\chi}_s(t)$ can be written in terms of $\bm{Z}(t)$. Finally, the time derivative of $\bm{f}_{os}(t)$ is trivially a functional of $\bm{Z}(t)$. It follows that the variable $\bm{Z}(t)$ obeys an (infinite-dimensional) autonomous first order ODE
\begin{align}\label{abstract_ode}
\dot{\bm{Z}}(t)=Y(\bm{Z}(t)),
\end{align} 
where $Y$ is a vector field on $\mathcal{P}_o=G\times\Omega^1(Q)\times\Omega^1(Q)\times V^*\times V$. We will refer to the space $\mathcal{P}_o$ as the AGVM phase space. We will use $\mathfrak{F}_t:\mathcal{P}_o\rightarrow\mathcal{P}_o$ to denote the (formal) flow map associated with $Y$. 


By the existence and uniqueness of solutions to first order ODEs, for each point $Z\in\mathcal{P}_o$, there is a unique path $\bm{Z}$ such that $\bm{Z}(t_1)=Z$ and $\dot{\bm{Z}}(t)=Y(\bm{Z}(t))$. In terms of the flow map $\mathfrak{F}_t$, this path is given by
\begin{align}
\bm{Z}(t)=\mathfrak{F}_{t-t_1}(Z)\equiv\text{Sol}(Z)(t),
\end{align}
where we have defined the function $\text{Sol}:\mathcal{P}_o\rightarrow\mathfrak{P}(\mathcal{P}_o)$. The augmented gyrokinetic system therefore gives us a natural way of mapping points in the augmented phase space into a path space. But recall that the augmented gyrokinetic action maps points in a path space into the real numbers. This suggests that we can construct a special real-valued function on $\mathcal{P}_o$ by composing the gyrokinetic action with the function $\text{Sol}$ in some sense. Actually, composition doesn't make literal sense because the augmented gyrokinetic action is a functional defined on a path space that differs from $\mathfrak{P}(\mathcal{P}_o)$; the argument of the augmented gyrokinetic action is a path of the form $(\T{\B{g}},\bm{A},\bm{\phi},\T{\bm{f}}_o,\T{\bm{\chi}})\in\mathfrak{P}(G\times\Omega^1(Q)\times\Omega^0(Q)\times V^*\times V)$. Nevertheless, there is a simple mapping $\Pi:\mathfrak{P}(\mathcal{P}_o)\rightarrow\mathfrak{P}(G\times\Omega^1(Q)\times\Omega^0(Q)\times V^*\times V)$ given by
\begin{align}
\Pi(\T{\B{g}},\bm{A},\bm{D},\T{\bm{f}}_o,\T{\bm{\chi}})=(\T{\B{g}},\bm{A},0,\T{\bm{f}}_o,\T{\bm{\chi}}),
\end{align}
which is motivated by our choice of the gauge $\bm{\phi}(t)=0$. We can still therefore define a special real-valued function on $\mathcal{P}_o$ 
\begin{align}
S_{\mathcal{P}_o}(Z)=S(\Pi(\text{Sol}(Z))).
\end{align}
We will refer to $S_{\mathcal{P}_o}$ as the restricted augmented action. Note that $S_{\mathcal{P}_o}$ implicitly depends on $t_1$ and $t_2$ as these appear in the action integral's limits of integration.

It is extremely interesting to study the (infinite-dimensional) exterior derivative of the restricted augmented action. A simple direct calculation shows
\begin{align}\label{source_of_structure}
\mathbf{d}S_{\mathcal{P}_o}(\delta Z)=\mathfrak{F}_{t_2-t_1}^*\Theta_{\mathcal{P}_o}-\Theta_{\mathcal{P}_o},
\end{align} 
where $\Theta_{\mathcal{P}_o}$ is a $1$-form on $\mathcal{P}_o$ given by
\begin{align}
\Theta_{\mathcal{P}_o}(\delta Z)=\sum_{s=1}^{N_s}\bigg(\int_{TQ}(\Xi^{\text{gy}}\otimes g_{s*}f_{os})\cdot \eta_s+\chi_s\delta f_{os}\bigg)-\frac{1}{4\pi c}\left<D,\delta A\right>.
\end{align}
Here $\T{\eta}=\delta\T{g}\circ \T{g}^{-1}$. This identity becomes very interesting indeed when both sides are differentiated with respect to $t_2$. We have
\begin{align}\label{time_deriv}
\mathfrak{F}_{t_2-t_1}^*\mathbf{d}\dot{S}_{\mathcal{P}_o}=\mathfrak{F}_{t_2-t_1}^*L_Y\Theta_{\mathcal{P}_o}=\mathfrak{F}_{t_2-t_1}^*(\text{i}_{Y}\mathbf{d}\Theta_{\mathcal{P}_o}+\mathbf{d}(\Theta_{\mathcal{P}_o}(Y))),
\end{align}
where $\dot{S}_{\mathcal{P}_o}$ is given by
\begin{align}
\dot{S}_{\mathcal{P}_o}(Z)=\mathcal{L}_p(\tilde{X}^{\text{gy}},\tilde{g}_*\tilde{f}_o,\hat{E},\mathbf{d}A)+\mathcal{L}_{\text{int}}(\tilde{X}^{\text{gy}},A,0)+\mathcal{L}_{\text{Max}}(\hat{E},\mathbf{d}A).
\end{align}
Here we are using the short hand notation $\tilde{X}^{\text{gy}}=\tilde{X}^{\text{gy}}(\hat{E}(\T{g}_*\T{f}_o,D,\mathbf{d}A),\mathbf{d}A)$ and $\hat{E}=\hat{E}(\hat{E}(\T{g}_*\T{f}_o,D,\mathbf{d}A)$. Equation\,(\ref{time_deriv}) tells us that the augmented gyrokinetic dynamical vector field $Y$ obeys an infinite-dimensional version of Hamilton's equations,
\begin{align}
\text{i}_Y\Omega_{\mathcal{P}_o}=\mathbf{d}\mathcal{E}_{\mathcal{P}_o},
\end{align}
where the closed (and non-degenerate, as it turns out) $2$-form $\Omega_{\mathcal{P}_o}=-\mathbf{d}\Theta_{\mathcal{P}_o}$ and the {energy functional} $\mathcal{E}_{\mathcal{P}_o}=\Theta_{\mathcal{P}_o}(Y)-\dot{S}_{\mathcal{P}_o}$. It follows that the AGVM equations have a Poisson formulation on $\mathcal{P}_o$ with a Poisson bracket given by inverting the symplectic form $\Omega$ and a Hamiltonian functional given by $\mathcal{E}_{\mathcal{P}_o}$. 

In order to derive an explicit expression for the AGVM Poisson bracket $[\cdot,\cdot]_{\mathcal{P}_o}$, the simplest approach is to first derive an expression for a general Hamiltonian vector field, $X_{\mathsf{G}}$, using the infinite-dimensional phase space variational principle
\begin{align}
\delta \int_{t_1}^{t_2}\bigg(\Theta_{\mathcal{P}_o}(\dot{\bm{Z}}(t))-\mathsf{G}(\bm{Z}(t))\bigg)\,dt=0,
\end{align}
and then calculate the Poisson bracket using the formula
\begin{align}
[\mathsf{F},\mathsf{G}]_{\mathcal{P}_o}=L_{X_{\mathsf{G}}}\mathsf{F}.
\end{align}
We find that $X_{\mathsf{G}}=(\T{\xi}_{\mathsf{G}}\circ\T{g},\dot{A}_{\mathsf{G}},\dot{D}_{\mathsf{G}},(\dot{\T{f}}_{o})_{\mathsf{G}},\dot{\T{\chi}}_{\mathsf{G}})$, where
\begin{align}
(\xi_{s})_{\mathsf{G}}&=(\omega_s^{\text{gy}})^{-1}\left(\frac{\delta \mathsf{G}}{\delta g_s}-4\pi e_S\pi^*\frac{\delta\mathsf{G}}{\delta D}+\frac{g_{s*}\delta\mathsf{G}/\delta\chi_s}{g_{s*}f_{os}}\Xi^{\text{gy}}\right)\label{g_dot_vm}\\
\dot{A}_{\mathsf{G}}&=-4\pi c\frac{\delta\mathsf{G}}{\delta D}\label{a_dot_vm}\\
\dot{D}_{\mathsf{G}}&=4\pi c\frac{\delta\mathsf{G}}{\delta A}-4\pi J(\T{\xi}_{\mathsf{G}},\T{g}_*\T{f}_o)\label{d_dot_vm}\\
(\dot{f}_{os})_{\mathsf{G}}&=\frac{\delta\mathsf{G}}{\delta\chi_s}\label{fo_dot_vm}\\
(\dot{\chi}_s)_{\mathsf{G}}&=g_s^*(\Xi_s^{\text{gy}}[(\xi_{s})_{\mathsf{G}}])-\frac{\delta\mathsf{G}}{\delta f_{os}}.
\end{align}
Therefore the AGVM Poisson bracket is given by
\begin{align}
[\mathsf{F},\mathsf{G}]_{\mathcal{P}_o}&=\sum_{s=1}^{N_s}\int_{TQ}B_s^{\text{gy}}\left(\frac{\delta\mathsf{F}}{\delta g_s}-4\pi e_s \pi^*\frac{\delta\mathsf{F}}{\delta D},\frac{\delta\mathsf{G}}{\delta g_s}-4\pi e_s \pi^*\frac{\delta\mathsf{G}}{\delta D}\right)g_{s*}f_{os}\nonumber\\
&+\sum_{s=1}^{N_s}\int_{TQ}\left(g_{s*}\frac{\delta\mathsf{F}}{\delta\chi_s}\right)B_s^{\text{gy}}\left(\Xi^{\text{gy}},\frac{\delta\mathsf{G}}{\delta g_s}-4\pi e_s \pi^*\frac{\delta\mathsf{G}}{\delta D}\right)-\left(g_{s*}\frac{\delta\mathsf{G}}{\delta\chi_s}\right)B_s^{\text{gy}}\left(\Xi^{\text{gy}},\frac{\delta\mathsf{F}}{\delta g_s}-4\pi e_s \pi^*\frac{\delta\mathsf{F}}{\delta D}\right)\nonumber\\
&+\left(\sum_{s=1}^{N_s}\int_{TQ}\frac{\delta\mathsf{F}}{\delta f_{os}}\frac{\delta\mathsf{G}}{\delta\chi_s}-\frac{\delta\mathsf{G}}{\delta f_{os}}\frac{\delta\mathsf{F}}{\delta\chi_s}\right)+4\pi c\left(\left<\frac{\delta\mathsf{F}}{\delta D},\frac{\delta\mathsf{G}}{\delta A}\right>-\left<\frac{\delta\mathsf{G}}{\delta D},\frac{\delta\mathsf{F}}{\delta A}\right>\right),
\end{align}
where $B_{s}^{\text{gy}}$ is the gyrocenter Poisson tensor.

The AGVM Poisson bracket in conjunction with the energy functional
\begin{align}
\mathcal{E}_{\mathcal{P}_o}(Z)&=(\Theta_{\mathcal{P}_o}(Y)-\dot{S}_{\mathcal{P}_o})(Z)\nonumber\\
&=\mathcal{K}(\T{g}_*\T{f}_o,\hat{E},\mathbf{d}A)+\left<\hat{P},\hat{E}\right>+\frac{1}{8\pi}\left<\hat{E},\hat{E}\right>+\frac{1}{8\pi}\left<\B{d}A,\B{d}A\right>,
\end{align} 
cast the AGVM equations as an infinite dimensional Poisson dynamical system. Here we have introduced the polarization operator
\begin{align}
\hat{P}(\tilde{f},D,B)=\frac{1}{4\pi}(D-\hat{E}(\tilde{f},D,B)).
\end{align}
Note that the AGVM Poisson bracket is derived by inverting a symplectic form, and so it does \emph{not} have casimirs. The main drawback of this Poisson formulation of gyrokinetics is that it contains a superfluous dynamical variable $\T{\chi}$. A secondary drawback is that it uses the gauge-dependent vector potential $A$ as a dynamical variable. A tertiary drawback is that it is expressed in Lagrangian labeling. 

All of the drawbacks of the augmented Poisson formulation of gyrokinetics can be removed using Poisson reduction. To refresh the reader's memory, the basic premise of Poisson reduction is that if there is a Lie group $H$ that acts freely on a Poisson manifold $P$ in a manner that leaves the Poisson bracket and Hamiltonian function invariant, then the bracket and Hamiltonian induce corresponding objects on the quotient $P/H$. The bracket on $P/H$ is known as the reduced bracket and the Hamiltonian on $P/H$ is known as the reduced Hamiltonian. In our case, the poisson manifold will be $P=\mathcal{P}_o$ equipped with the AGVM bracket, and the Hamiltonian function will be $\mathcal{E}_{\mathcal{P}_o}$. The Lie group will be $H=\Omega^0(Q)\times (G\rtimes V)$. The first factor in $H$ corresponds to gauge symmetry, the second to particle-relabeling symmetry, and the third to symmetry with respect to translations in the augmented variable $\T{\chi}$. The quotient $P/H$ will be the gyrokinetic Eulerian phase space, i.e. the space of triples $(\tilde{f},D,B)$. The reduced bracket and reduced Hamiltonian on the Eulerian phase space will provide us with a Poisson formulation of the gyrokinetic Vlasov-Maxwell system without any of the drawbacks of the AGVM Poisson formulation.

Instead of applying Poisson reduction to the entire symmetry group of the AGVM system $H=\Omega^0(Q)\times (G\rtimes V)$, we will apply the Poisson reduction procedure three times, eliminating one factor of $H$ in each step. The general theory behind breaking a symmetry group into subgroups and then applying step-wise Poisson reduction is known as \emph{Poisson reduction by stages}. We will not need to draw upon any of the general results from this theory. First we will factor out the translations in $\T{\chi}$, which are embodied by the additive Lie group $V$. This will give us a Poisson formulation of gyrokinetics on the space $\mathcal{P}_{LA}=\mathcal{P}_o/V$, which consists of tuples $(\T{g},A,D,\T{f}_o)$. $\mathcal{P}_{LA}$ is the gauge-dependent Lagrangian phase space for gyrokinetics. In the second reduction step, we will factor out the gauge symmetry, which is generated by a second additive Lie group $\Omega^0(Q)$. The result will be a Poisson formulation of gyrokinetics on the gauge-independent Lagrangian phase space $\mathcal{P}_L=\mathcal{P}_{LA}/\Omega^0(Q)$. Finally, we will quotient by the particle relabeling symmetry group, $G$, which will provide us with a Poisson formulation of gyrokinetics on the Eulerian phase space $\mathcal{P}_E=\mathcal{P}_{L}/G$.

The Poisson bracket on the first reduced phase space, $\mathcal{P}_{LA}=\mathcal{P}_o/V$, is given by
\begin{align}
[\mathsf{F},\mathsf{G}]_{\mathcal{P}_{LA}}(\T{g},A,D,\T{f}_o)&=\sum_{s=1}^{N_s}\int_{TQ}B_s^{\text{gy}}\left(\frac{\delta\mathsf{F}}{\delta g_s}-4\pi e_s \pi^*\frac{\delta\mathsf{F}}{\delta D},\frac{\delta\mathsf{G}}{\delta g_s}-4\pi e_s \pi^*\frac{\delta\mathsf{G}}{\delta D}\right)g_{s*}f_{os}\nonumber\\
&+4\pi c\left(\left<\frac{\delta\mathsf{F}}{\delta D},\frac{\delta\mathsf{G}}{\delta A}\right>-\left<\frac{\delta\mathsf{G}}{\delta D},\frac{\delta\mathsf{F}}{\delta A}\right>\right).
\end{align}
The reduced Hamiltonian on $\mathcal{P}_{LA}$ is given by
\begin{align}
\mathcal{E}_{\mathcal{P}_{LA}}(\T{g},A,D,\T{f}_o)
&=\mathcal{K}(\T{g}_*\T{f}_o,\hat{E},\mathbf{d}A)+\left<\hat{P},\hat{E}\right>+\frac{1}{8\pi}\left<\hat{E},\hat{E}\right>+\frac{1}{8\pi}\left<\B{d}A,\B{d}A\right>.
\end{align}
Note that this reduced bracket and Hamiltonian are both invariant under time-independent gauge transformations $A\rightarrow A+\B{d}\varphi$, where $\varphi\in\Omega^0(Q)$.

The Poisson bracket on the second reduced phase space, $\mathcal{P}_L=\mathcal{P}_{LA}/\Omega^0(Q)$, which is the gauge-independent Lagrangian phase space, is given by
\begin{align}
[\mathsf{F},\mathsf{G}]_{\mathcal{P}_{L}}(\T{g},D,B,\T{f}_o)&=\sum_{s=1}^{N_s}\int_{TQ}B_s^{\text{gy}}\left(\frac{\delta\mathsf{F}}{\delta g_s}-4\pi e_s \pi^*\frac{\delta\mathsf{F}}{\delta D},\frac{\delta\mathsf{G}}{\delta g_s}-4\pi e_s \pi^*\frac{\delta\mathsf{G}}{\delta D}\right)g_{s*}f_{os}\nonumber\\
&+4\pi c\left(\left<\frac{\delta\mathsf{F}}{\delta D},\bm{\delta}\frac{\delta\mathsf{G}}{\delta B}\right>-\left<\frac{\delta\mathsf{G}}{\delta D},\bm{\delta}\frac{\delta\mathsf{F}}{\delta B}\right>\right).
\end{align}
The Hamiltonian on $\mathcal{P}_L$ is given by
\begin{align}
\mathcal{E}_{\mathcal{P}_o}(\T{g},D,B,\T{f}_o)
&=\mathcal{K}(\T{g}_*\T{f}_o,\hat{E},B)+\left<\hat{P},\hat{E}\right>+\frac{1}{8\pi}\left<\hat{E},\hat{E}\right>+\frac{1}{8\pi}\left<B,B\right>.
\end{align}
Note that the $2$-form $B$ is required to be exact (ignoring possible Homological complications).

Finally, the Poisson bracket on the Eulerian phase space, $\mathcal{P}_E=\mathcal{P}_L/G$ is given by
\begin{align}
[\mathsf{F},\mathsf{G}]_{\mathcal{P}_{E}}(\T{f},D,B)&=\sum_{s=1}^{N_s}\int_{TQ}B_s^{\text{gy}}\left(\mathbf{d}\frac{\delta\mathsf{F}}{\delta f_s}-4\pi e_s \pi^*\frac{\delta\mathsf{F}}{\delta D},\mathbf{d}\frac{\delta\mathsf{G}}{\delta f_s}-4\pi e_s \pi^*\frac{\delta\mathsf{G}}{\delta D}\right)f_s\nonumber\\
&+4\pi c\left(\left<\frac{\delta\mathsf{F}}{\delta D},\bm{\delta}\frac{\delta\mathsf{G}}{\delta B}\right>-\left<\frac{\delta\mathsf{G}}{\delta D},\bm{\delta}\frac{\delta\mathsf{F}}{\delta B}\right>\right),
\end{align}
and the Eulerian Hamiltonian is given by
\begin{align}
\mathcal{E}_{\mathcal{P}_o}(\T{f},D,B)
&=\mathcal{K}(\T{f},\hat{E},B)+\left<\hat{P},\hat{E}\right>+\frac{1}{8\pi}\left<\hat{E},\hat{E}\right>+\frac{1}{8\pi}\left<B,B\right>.
\end{align}
Note that this bracket, and all of the previous brackets satisfy the Jacobi identity by construction.


\subsection{The augmented gyrokinetic Vlasov-Maxwell symmetry group}
The basic symmetry group of the AGVM equations is $H=\Omega^0(Q)\times (G\rtimes V)$, which as a set is simply $\Omega^0(Q)\times G\times V$. For technical reasons, we will regard $\Omega^0(Q)$ as the functions on $Q$ that vanish at infinity. Some basic properties of $H$ are:
\begin{itemize}
\item The group identity $e=(0,\tilde{\text{id}}_{TQ},0)$. The group product of $s_1=(\varphi_1,\tilde{h}_1,\tilde{\tau}_1)$ and $s_1=(\varphi_2,\tilde{h}_2,\tilde{\tau}_2)$ is given by 
\begin{align}
s_1*s_2=(\varphi_1+\varphi_2,\tilde{h}_1\circ\tilde{h}_2,h_2^*\tau_1+\tau_2).
\end{align}
The inverse of $s=(\varphi,\tilde{h},\tilde{\tau})$ is given by
\begin{align}
s^{-1}=(-\varphi,\tilde{h}^{-1},-\tilde{h}_*\tilde{\tau}).
\end{align}

\item $H$'s Lie algebra $\mathfrak{h}=\Omega^0(Q)\times\mathfrak{g}\times V$. We will denote typical elements of $\mathfrak{h}$ with the symbol $x=(\delta\varphi,\tilde{\zeta},\delta\tilde{\chi})$. The adjoint action of $H$ on $\mathfrak{h}$ is given by
\begin{align}
\text{Ad}_sx=(\delta\varphi,\tilde{h}_*\tilde{\zeta},\tilde{h}_*[L_{\tilde{\zeta}}\tilde{\tau}+\delta\tilde{\tau}]).
\end{align} 
The Lie bracket is therefore
\begin{align}
[x_1,x_2]=(0,-[\tilde{\zeta}_1,\tilde{\zeta}_2],L_{\tilde{\zeta}_2}\delta\tilde{\tau}_1-L_{\tilde{\zeta}_1}\delta\tilde{\tau}_2).
\end{align}

\item The dual to $H$'s Lie algebra $\mathfrak{h}^*=\Omega^0(Q)\times\mathfrak{g}^*\times V^*$. We will denote typical elements of $\mathfrak{h}^*$ with the symbol $\mu=(\rho,\tilde{\alpha},\tilde{f})$. The coadjoint action of $H$ on $\mathfrak{h}^*$ is given by
\begin{align}
\text{Ad}^*_s\mu=(\rho,\tilde{h}_*[\tilde{\alpha}-\mathbf{d}\tilde{\tau}\otimes\tilde{f}],\tilde{h}_*\tilde{f}).
\end{align}
The conditions for a functional $A:\mathfrak{h}^*\rightarrow\mathbb{R}$ to be invariant under the coadjoint action are therefore
\begin{align}
L_{\delta A/\delta\tilde{\alpha}}f&=0\\
L_{\delta A/\delta\tilde{\alpha}}\tilde{\alpha}+\mathbf{d}\frac{\delta A}{\delta\tilde{f}}\otimes\tilde{f}&=0.
\end{align}
Note that these conditions can be regarded as first-order functional partial differential equations for the Casimirs of the Lie Poisson bracket on $\mathfrak{h}^*$.

\item There is a \emph{right} $H$-action on $\mathcal{P}_o$ given by
\begin{align}
R_s(\tilde{g},D,A,\tilde{f}_o,\tilde{\chi})=(\tilde{g}\circ\tilde{h},D,A+\mathbf{d}\varphi,\tilde{h}^*\tilde{f}_o,\tilde{h}^*(\tilde{\chi}+(e/c)\tilde{g}^*\pi^*\varphi)+\tau).
\end{align}
The \emph{right} infinitesimal generator is given by
\begin{align}
x_R(Z)=(\tilde{g}_*\tilde{\zeta}\circ \tilde{g},0,\mathbf{d}\delta\varphi,L_{\tilde{\zeta}}\tilde{f}_o,L_{\tilde{\zeta}}\tilde{\chi}+(e/c)\tilde{g}^*\pi^*\delta\varphi+\delta\tilde{\tau}).
\end{align}
Note that the right infinitesimal generator satisfies $R_s^*x_R=(\text{Ad}_sx)_R$ for each $s\in H$.
\end{itemize}

In order to identify the conservation laws associated with this symmetry group, we will now consider the invariance properties of the $1$-form $\Theta_{\mathcal{P}_o}$. A straightforward calculation shows
\begin{align}
R_s^*\Theta_{\mathcal{P}_o}=\Theta_{\mathcal{P}_o}+\mathbf{d}c_s,
\end{align}
where
\begin{align}
c_s(Z)=\sum_{s=1}^{N_s}\int_{TQ}\frac{e_s}{c}\pi^*\varphi\,g_{s*}f_{os}+h_{s*}\tau_s\,f_{os}.
\end{align}
Differentiating this equation in $s$ at $e\in H$, we obtain
\begin{align}
\text{i}_{x_R}\mathbf{d}\Theta=-\mathbf{d}\left(\Theta(x_R)-\delta c(x)\right),
\end{align}
for each $x\in\mathfrak{h}$. Here we have introduced the mapping $\delta c:\mathfrak{h}\rightarrow C^\infty(\mathcal{P}_o)$ given by
\begin{align}
\delta c(x)(Z)&=\frac{\mathrm{d}}{\mathrm{d}\epsilon}\bigg|_0c_{\exp(\epsilon x)}(Z)\\
                    &=\sum_{s=1}^{N_s}\int_{TQ}\delta\tau_sf_{os}+\frac{1}{c}\left<\rho(\tilde{g}_*\tilde{f}_{o}),\delta\varphi\right>.
\end{align}
$\Theta(x_R)$ is given by
\begin{align}
\Theta(x_R)(Z)=\sum_{s=1}^{N_s}\int_{TQ}(g_s^*\Xi_s^{\text{gy}}\otimes f_{os}-\mathbf{d}\chi\otimes f_{os})\cdot\zeta_s-\frac{1}{4\pi c}\left<\bm{\delta}D,\delta\varphi\right>.
\end{align}
It follows that a momentum mapping $m:\mathcal{P}_o\rightarrow\mathfrak{h}^*$ associated with the action $R_s$ is given by
\begin{align}
m(\tilde{g},D,A,\tilde{f}_o,\tilde{\chi})=\bigg(-\frac{1}{4\pi c}\bm{\delta}D-\frac{1}{c}\rho(\tilde{g}_*\tilde{f}_o),[\tilde{g}^*\tilde{\Xi}^{\text{gy}}-\mathbf{d}\tilde{\chi}]\otimes\tilde{f}_o,-\tilde{f}_o\bigg).
\end{align}
It is straightforward to verify that $m$ satisfies the following equivariance property
\begin{align}
m\circ R_s=\text{Ad}^*_{s^{-1}}\circ m,
\end{align}
i.e. translating in $\mathcal{P}_o$ along the $H$-action only changes the value of $m$ by shifting it along a coadjoint orbit in $\mathfrak{h}^*$. Thus, $m:\mathcal{P}_o\rightarrow\mathfrak{h}^*$ is a Poisson map when $\mathfrak{h}^*$ is equipped with its Lie-Poisson bracket.

We can now formally write down many of the Casimirs of the gyrokinetic Vlasov-Maxwell bracket in Eulerian labeling. Because $m$ is a constant of motion for the AGVM system, any function of $m$ is also a constant of motion. In particular, if $A:\mathfrak{h}^*\rightarrow\mathbb{R}$ is an $\text{Ad}^*$-invariant functional on $\mathfrak{g}^*$, $A\circ m$ is a constant of motion. The constant of motion $A\circ m$ satisfies $R^*_s(A\circ m)=A\circ m$ by the equivariance of $m$ and the $\text{Ad}^*$-invariance of $A$. Therefore $A\circ m$ descends to the quotient $\mathcal{P}_o/H=\mathcal{P}_E$, i.e. there is a functional $C_A:\mathcal{P}_E\rightarrow\mathbb{R}$ uniquely characterized by the formula
\begin{align}
\pi_E^*C_A=A\circ m,
\end{align}
where $\pi_E:\mathcal{P}_o\rightarrow\mathcal{P}_o/H=\mathcal{P}_E$ is the quotient map. $C_A$ is a Casimir of the Eulerian gyrokinetic Vlasov-Maxwell bracket $[\cdot,\cdot]_{\mathcal{P}_E}$ because
\begin{align}
\pi_E^*[C_A,\mathsf{F}]_{\mathcal{P}_E}&=[\pi_E^*C_A,\pi_E^*\mathsf{F}]_{\mathcal{P}_o}\\
                                                                 &=[A\circ m,\pi_E^*\mathsf{F}]_{\mathcal{P}_o}\\
                                                                 &=0.
\end{align}

\section{The neutral Vlasov model}
The neutral Vlasov model was introduced by Tronci and Camporeal in \cite{Tronci_2015}. Its defining equations are given by
\begin{align}
\partial_t f_s+\bm{v}\cdot\nabla f_s+&\frac{e_s}{m_s}(\bm{E}+\bm{v}\times\bm{B})\cdot\nabla_{\bm{v}}f_s=0\\
\partial_t\bm{B}&=-\nabla\times\bm{E}\\
\nabla\times\bm{B}=\mu_o\sum_s e_s\int\bm{v}&f_s\,d\bm{v},~~~~~\sum_se_s\int f_s\,d\bm{v}=0.
\end{align}

We will first formulate the quasineutral model as an initial value problem without assuming $\nabla\cdot\bm{E}=0$, which is not obviously implied by the Euler-Lagrange equations. We will work in terms of potentials in the temporal gauge $\varphi=0$. I will also make use of a hodge decomposition of the vector potential
\begin{align}
\bm{A}=\nabla\times\bm{\alpha}+\nabla\lambda,
\end{align}
where $\bm{\alpha}$ satisfies 
\begin{align}
\nabla\cdot\bm{\alpha}&=0
\end{align}
and $\lambda$ is defined modulo the addition of constants (i.e. $\lambda$ is really an equivalence class of functions whose elements differ from one another by constant functions).
Given an $\bm{A}$, there is a unique pair $\bm{\alpha},\lambda$ that satisfy the previous three equations.

\emph{An equation for $\bm{\alpha}$\,---\,} According to the variational formulation of the neutral Vlasov model given in \cite{Tronci_2015}, one of the Euler-Lagrange equations is $\rho=0$, which implies that $\nabla\cdot \bm{J}=0$. This is consistent with another one of the Euler-Lagrange equations,
\begin{align}
\mu_o\bm{J}=\nabla\times \bm{B}=-\nabla\times\Delta\bm{\alpha}.
\end{align}
Thus, we come to the conclusion that $\bm{\alpha}$ is uniquely determined by $\bm{J}$ (and therefore $f$) \emph{via}
\begin{align}\label{alpha}
\bm{\alpha}=\mu_oG[G[\nabla\times\bm{J}]],
\end{align}
where $G$ is the inverse of the Laplacian ($G$ stands for Green operator). Note that $\lambda$ is not yet determined.

\emph{An equation for $\partial_t\lambda$\,---\,}If we take the time derivative of the Amp\`ere equation, we obtain
\begin{align}
\mu_o\dot{\bm{J}}&=-\nabla\times(\nabla\times\bm{E}).
\end{align}
The time derivative of the current density can also be written in terms of the electromagnetic field using the momentum equation
\begin{align}
\partial_t(n_s\bm{u}_s)+\nabla\cdot\mathbb{T}_s=\frac{e_sn_s}{m_s}(\bm{E}+\bm{u}_s\times\bm{B}),
\end{align}
where $\mathbb{T}$ is the stress tensor,
\begin{align}
\mathbb{T}_s=\int \bm{v}\bm{v}f_s\,d\bm{v}.
\end{align}
We have
\begin{align}
\dot{\bm{J}}=-\nabla\cdot\mathbb{Q}+\epsilon_o\omega_p^2\bm{E}+\epsilon_o\omega_p^2\langle\bm{u}\rangle\times\bm{B},
\end{align}
where
\begin{align}
\mathbb{Q}&=\sum_se_s\mathbb{T}_s\\
\omega_p^2&=\sum_s\omega_{ps}^2=\sum_s\frac{n_s e_s^2}{\epsilon_o m_s}\\
\langle\bm{u}\rangle&=\frac{\sum_s\omega_{ps}^2\bm{u}_s}{\sum_s\omega_{ps}^2}.
\end{align}
If we now equate our two expressions for the time derivative of the current density, we obtain an equation that functionally relates the electric field  to the distribution function and the magnetic field,
\begin{align}
\frac{c^2}{\omega_p^2}\nabla\times(\nabla\times\bm{E})+\bm{E}=-\langle\bm{u}\rangle\times\bm{B}+\frac{1}{\epsilon_o\omega_p^2}\nabla\cdot\mathbb{Q}.
\end{align}
We will now use this expression to derive an expression for $\partial_t\lambda$.

Note that the ($f$-dependent) linear operator $\bm{E}\mapsto\mathcal{D}[\bm{E}]$, where
\begin{align}
\mathcal{D}[\bm{E}]=\frac{c^2}{\omega_p^2}\nabla\times(\nabla\times\bm{E})+\bm{E},
\end{align}
does not have non-zero null eigenvectors. Indeed, if $\bm{E}_o$ is a null eigenvector then we have
\begin{align}
0\leq \int \frac{\omega_p^2}{c^2} |\bm{E}_o|^2\,d\bm{x}=-\int\bm{E}_o\cdot\nabla\times(\nabla\times\bm{E}_o)\,d\bm{x}=-\int|\nabla\times\bm{E}_o|^2\,d\bm{x}\leq 0,
\end{align}
which implies that $|\bm{E}_o|=0$ everywhere. Thus, we can use the inverse of $\mathcal{D}$ to define \emph{an electric field functional} $\bm{\mathcal{E}}=\bm{\mathcal{E}}(f,\bm{B})$, where
\begin{align}
\bm{\mathcal{E}}=\mathcal{D}^{-1}\left[-\langle\bm{u}\rangle\times\bm{B}+\frac{1}{\epsilon_o\omega_p^2}\nabla\cdot\mathbb{Q}\right].
\end{align}
Because the vector potential determines the electric field according to the relation $\bm{E}=-\partial_t{\bm{A}}$, the functional $\bm{\mathcal{E}}$ is related to $\partial_t\lambda$ according to
\begin{align}
\nabla\cdot\bm{\mathcal{E}}(f,\bm{B})=-\Delta\partial_t\lambda.
\end{align} 
Because $\bm{B}$ can be regarded as a functional of $f$ according to Eq.\,(\ref{alpha}), we can now express $\partial_t\lambda$ entirely in terms of $f$:
\begin{align}
\partial_t\lambda=-G[\nabla\cdot\bm{\mathcal{E}}(f,\bm{B})].
\end{align}

\emph{The quasineutral model's initial value formulation\,---\,}We can now formulate the initial value problem for the quasineutral model. The state of the system is determined by the pair $(f,\lambda)$, where $f$ is the multi-species distribution function and $\lambda$ appears in the Hodge decomposition of the vector potential as above. The time evolution equation for $(f,\lambda)$ is given by
\begin{align}
\partial_t f_s&=-\bm{v}\cdot\nabla f_s-\frac{e_s}{m_s}(\bm{\mathcal{E}}+\bm{v}\times\bm{B})\cdot\nabla_{\bm{v}}f_s\\
\partial_t\lambda&=-G[\nabla\cdot\bm{\mathcal{E}}(f,\bm{B})],
\end{align}
where
\begin{align}\label{b_eqn}
\bm{B}=-\mu_oG[\nabla\times\bm{J}]
\end{align}
is expressed in terms of $f$. Note that the dynamics of $f$ decouple from the dynamics of $\lambda$.

\emph{Remarks\,---\,}This system of evolution equations admits solutions that do not satisfy the Amp\`ere equation and the neutrality condition. However, using the definition of $\bm{\mathcal{E}}$, it is straightforward to verify that $\nabla\cdot\bm{J}$ is a constant of motion. Thus, if we choose initial conditions with $\nabla\cdot\bm{J}=0$, then this identity will hold for all subsequent times, and, as a consequence, the Amp\`ere equation will be satisfied for all times (this second statement follows from applying the Laplacian to Eq.\,(\ref{b_eqn})). If, in addition to demanding our initial condition satisfies $\nabla\cdot\bm{J}=0$, we demand that our initial condition satisfies $\rho=0$, then $\rho$ will be zero for all times. Indeed,
\begin{align}
\partial_t\rho=-\nabla\cdot\bm{J},
\end{align}
which implies that $\rho$ is constant along solutions that initially satisfy $\nabla\cdot\bm{J}=0$.

\subsection{Phase space Lagrangian formulation of the neutral Vlasov model}
So far we have deduced that the phase space for the neutral Vlasov model is given by the space of pairs $(f,\lambda)$ subject to the holonomic constraint $\nabla\cdot\bm{J}=0$. Because we will need a slightly more precise notation in the discussion that follows, we will write the constraint as $\nabla\cdot\bm{J}(X_o,f)=0$, where the functional $\bm{J}(V,f)$ depends on the multi-species velocity field $V$ and the multi-species distribution function $f$ according to
\begin{align}
\bm{J}(V,f)=\sum_se_s\int\dot{\bm{x}}_sf_s\,d\bm{v},
\end{align}
where $\dot{\bm{x}}_s=\dot{\bm{x}}_s(\bm{x},\bm{v})$ is the spatial component of the phase space velocity $V_s=(\dot{\bm{x}}_s,\dot{\bm{v}}_s)$. The vector field $X_o$ is given by $\dot{\bm{x}}_s=\bm{v}$ and $\dot{\bm{v}}_s=0$, which implies
\begin{align}
\bm{J}(X_o,f)=\sum_se_s\int \bm{v}f_s\,d\bm{v},
\end{align}
as expected. 

We will derive the Poisson bracket on this phase space in this section. We start by writing down a phase space Lagrangian for an augmented system,
\begin{align}
L_{\mathcal{H}}(g,F_o,\chi,\lambda,\dot{g},\dot{F}_o,\dot{\chi},\dot{\lambda})=\Theta(V,\dot{F}_o,\dot{\chi},\dot{\lambda})-\mathcal{H}(g,F_o,\chi,\lambda)
\end{align}
Here $g$ is a multi-species phase space fluid configuration, i.e. an element of $\text{Diff}(TQ)$; $F_o$ is a multi-species reference phase space density related to the reference distribution function $f_o$ by $F_{os}=f_{os}\,\Omega$, where $\Omega=\mathbf{d}x\wedge\mathbf{d}y\wedge\mathbf{d}z\wedge\mathbf{d}v_x\wedge\mathbf{d}v_y\wedge\mathbf{d}v_z$; $\chi$ is a multi-species function on phase space; and $\lambda$ is a function on $Q$ modulo constant functions. The multi-species phase space fluid velocity is related to $\dot{g}$ by $V=\dot{g}\circ g^{-1}$. The quantities $g,F_o$ are constrained to satisfy $\nabla\cdot \bm{J}(X_o,F)=0$, where $F_s=g_{s*}F_{os}$. The Hamiltonian functional is given by
\begin{align}
\mathcal{H}(g,F_o,\chi,\lambda)=\sum_s\langle K_s,F_s \rangle+\frac{1}{2\mu_o}\langle\nabla\times\bm{A}_T(\bm{J}_o),\nabla\times\bm{A}_T(\bm{J}_o)\rangle
\end{align}
where, $\bm{J}_o=\bm{J}(X_o,F)$, $\bm{A}_T(\bm{J}_o)$ is the transverse vector potential given by
\begin{align}
\bm{A}_T(\bm{J}_o)=\mu_o G[G[\nabla\times(\nabla\times\bm{J}_o)]],
\end{align}
and $K_s=\frac{1}{2}m_s\bm{v}\cdot\bm{v}$ is the single-particle kinetic energy. The Lagrange $1$-form is given by
\begin{align}
\Theta(V,\dot{F}_o,\dot{\chi},\dot{\lambda})=\sum_s\langle \vartheta_s(V_s),F_s\rangle+\sum_s\langle \dot{\chi}_s,F_{os}\rangle.
\end{align}
Here the singe-particle Lagrange $1$-form $\vartheta_s$ is given by
\begin{align}
\vartheta_s=(m_s\bm{v}+e_s[\bm{A}_T(\bm{J}_o)+\nabla\lambda])\cdot d\bm{x}.
\end{align}

The Hamiltonian vector field $X_{\mathcal{H}}$ with Hamiltonian $\mathcal{H}$ is determined by varying the action $S_{\mathcal{H}}=\int_{t_1}^{t_2}L_{\mathcal{H}}\,dt$. More generally, the Hamiltonian vector field $X_{\mathcal{F}}$ with Hamiltonian $\mathcal{F}$ is determined by varying the action $S_{\mathcal{F}}=\int_{t_1}^{t_2}L_{\mathcal{F}}\,dt$. The Poisson bracket $[\mathcal{F},\mathcal{G}]$ on the augmented phase space (i.e. $(g,F_o,\chi,\lambda)$-space) is given by
\begin{align}
[\mathcal{F},\mathcal{G}]=\mathbf{d}\mathcal{F}(X_{\mathcal{G}}).
\end{align}
In order to derive the bracket, we will first find an expression for the general Hamiltonian vector field $X_{\mathcal{F}}$ and then deduce the bracket from the previous formula.

We will find $X_{\mathcal{F}}$ by manipulating the Euler-Lagrange equations associated with the Lagrangian $L_{\mathcal{F}}$. In order to derive the Euler-Lagrange equations associated with the phase space Lagrangian $L_{\mathcal{F}}$, it is necessary to account for the constraint $\nabla\cdot\bm{J}(X_o,F)=0$. This holonomic constraint places a an awkward constraint on the variations of the phase space variables $(g,F_o,\chi,\lambda)$. I have found it very convenient to work with unconstrained variations and a Lagrange multiplier. That is, instead of varying the action $S_{\mathcal{F}}=\int_{t_1}^{t_2}L_{\mathcal{F}}\,dt$ while respecting the divergence-free current constraint, it is easier to vary 
\begin{align}
S^\prime_{\mathcal{F}}=\int_{t_1}^{t_2}(L_{\mathcal{F}}+\langle\psi,\nabla\cdot\bm{J}_o\rangle)\,dt
\end{align}
using unconstrained variations. Here the Lagrange multiplier $\psi$ is a time-dependent function on $Q$ modulo constant functions.

The Euler-Lagrange equations associated with the action $S^\prime_{\mathcal{F}}$ are given as follows. 
\\ \\
\emph{Varying $g_s$\,---\,}Varying the multi-species phase space fluid configuration gives
\begin{align}\label{particle_hamilton_equation}
\frac{d}{dt}\vartheta_s+\text{i}_{V_s}\mathbf{d}\vartheta_s+e_s\mathbf{d}(\bm{v}\cdot\nabla \psi-\bm{v}\cdot\bm{A}_T(\bm{J}))+g_{s*}\left(\frac{\dot{F}_{os}}{F_{os}}\right)\vartheta_s+\delta_{g_s}\mathcal{F}=0.
\end{align}
Here $\bm{J}=\bm{J}(V,F)$ (note that $\bm{J}\neq\bm{J}_o$!) and $\delta_{g_s}\mathcal{F}$ is a $1$-form on the single-particle phase space that can be thought of as the functional derivative of $\mathcal{F}$ with respect to $g_s$. To be precise, 
\begin{align}
\frac{d}{d\epsilon}\bigg|_{0}\mathcal{F}(g_{s\epsilon})=\langle \delta_{g_s}\mathcal{F}(\xi_s),F_s \rangle,
\end{align}
where $\xi_s=\delta g_s\circ g_s^{-1}$. In deriving the expression\,(\ref{particle_hamilton_equation}), it is useful to make use of the self-adjoint property of the transverse vector potential, 
\begin{align}
\langle \bm{J}_1,\bm{A}_T(\bm{J}_2)\rangle=\langle\bm{A}_T(\bm{J}_1),\bm{J}_2 \rangle, 
\end{align}
where $\bm{J}_1,\bm{J}_2$ are arbitrary vector fields on $Q$. It is also useful to note that\,\ref{particle_hamilton_equation} implies that the velocity $V_s$ can be decomposed as
\begin{align}
V_s=\nu_{\mathcal{F}s}+\mathbb{P}_s(e_s\dot{\bm{A}}_{\mathcal{F}}\cdot d\bm{x}),
\end{align}
where $\mathbb{P}_s$ is the Poisson tensor associated with the symplectic form $\omega_s=-\mathbf{d}\vartheta_s$ and 
\begin{align}
\nu_{\mathcal{F}s}=\mathbb{P}_s(\delta_{g_s}\mathcal{F}+e_s\mathbf{d}[\bm{v}\cdot\bm{A}_{o\mathcal{F}}]-e_s\bm{A}_T(\bm{J}_{\mathcal{F}_\chi})-g_{s*}\left(\frac{\delta\mathcal{F}/\delta\chi_s}{F_{os}}\right)\vartheta_s)
\end{align}
\\ \\
\emph{Varying $\lambda$\,---\,} Varying the scalar $\lambda$ gives
\begin{align}\label{lamb_var}
\nabla\cdot\bm{J}+\frac{\delta\mathcal{F}}{\delta\lambda}=0.
\end{align}
Note that $\delta{\mathcal{F}}/\delta\lambda$ lives in the dual to the space of scalar functions modulo constant functions, which is precisely the space of functions on $Q$ with vanishing integral. 
\\ \\
\emph{Varying $F_{os}$\,---\,}Varying the reference phase space density $F_{os}$ gives
\begin{align}
\dot{\chi}_s+g_s^*[\vartheta_s(V_s)+e_s\bm{v}\cdot(\bm{A}_T(\bm{J})-\nabla\psi)]-\frac{\delta\mathcal{F}}{\delta F_{os}}=0.
\end{align}
This equation implies that the phase space function $\chi_s$ behaves very much in the same way as the phase of a quantum wave in the WKB approximation.
\\ \\
\emph{Varying $\chi_s$\,---\,} Varying the phase-like function $\chi_s$ gives
\begin{align}\label{fos_equation}
\dot{F}_{os}+\frac{\delta\mathcal{F}}{\delta \chi_s}=0.
\end{align}
\\ \\
\emph{Varying $\psi$\,---\,}As expected, varying the Lagrange multiplier $\psi$ gives
\begin{align}
\nabla\cdot\bm{J}_o=0.
\end{align}
\\ \\
Taken together, these Euler-Lagrange equations comprise a puzzle that must be solved in order to find the components of the Hamiltonian vector field $X_{\mathcal{F}}$. Some parts of the puzzle are simple. For instance, the time derivative of $F_{os}$ is given immediately by Eq.\,(\ref{fos_equation}). However, Eq.\,(\ref{particle_hamilton_equation}) represents a much more serious challenge. There the velocity field $V_s$ is woven into various terms in a rather intricate manner. Complicating matters further is the fact that the Lagrange multiplier $\psi$ must somehow be eliminated from the equations. 

The key to solving this puzzle is obtaining expressions for the quantities
\begin{align}
\dot{\bm{A}}_{\mathcal{F}}=\nabla\dot{\lambda}-\bm{A}_{T}[\bm{J}(X_o,L_VF)]\\
\bm{A}_{o\mathcal{F}}=\nabla\psi-\bm{A}_{T}[\bm{J}(V,F)],
\end{align}
in terms of functional derivatives of $\mathcal{F}$. In principle, this task is not so difficult. However, obtaining the desired expressions in a useful form is a subtle enterprise. It turns out that a particularly useful way of writing the expressions is
\begin{align}
\dot{\bm{A}}_{\mathcal{F}}&=[n-1]\bigg[\frac{\mu_o\bm{J}_{\mathcal{F}_\nu}+\mu_o\mu[\bm{J}_{\mathcal{F}_{\chi}}]}{\omega_p^2/c^2}\bigg]-\mu_oG\Pi_T \bm{J}_{\mathcal{F}_\chi}\label{key1}\\
\bm{A}_{o\mathcal{F}}&=[n-1]\bigg[\frac{\mu_o\bm{J}_{\mathcal{F}}+\mu_o\nabla G\left[\frac{\delta\mathcal{F}}{\delta\lambda}\right]}{\omega_p^2/c^2}\bigg].\label{key2}
\end{align}
Here the various current densities are given by
\begin{align}
\bm{J}_{\mathcal{F}}&=\bm{J}(\mathbb{P}(\delta_g\mathcal{F}),F)\\
\bm{J}_{\mathcal{F}_\nu}&=\bm{J}(\nu_{\mathcal{F}},F)\\
\bm{J}_{\mathcal{F}_\chi}&=\bm{J}(X_o,g_*(\delta\mathcal{F}/\delta\chi))
\end{align}
and the $f$-dependent linear operators $n,\mu$ are given by
\begin{align}
n&=-\mathcal{D}^{-1}\bigg(\frac{c^2}{\omega_p^2}\Delta\Pi_T\bigg)\\
\mu&=1-\frac{\omega_p^2}{c^2}G\Pi_T.
\end{align}
As we will see, the useful property of this representation is that the operator $n(c^2/\omega_p^2)$ is self-adjoint.  The self-adjoint property can be verified by first noting $\mathcal{D}^{\dagger}=\frac{\omega_p^2}{c^2}\mathcal{D}\frac{c^2}{\omega_p^2}$, which implies
\begin{align}
\bigg(n\frac{c^2}{\omega_p^2}\bigg)^\dagger=-\bigg(\frac{c^2}{\omega_p^2}\Delta\Pi_T\bigg)\mathcal{D}^{-1}\frac{c^2}{\omega_p^2}.
\end{align}
Then one uses the fact that $\frac{c^2}{\omega_p^2}\Delta\Pi_T$ commutes with the operator $\mathcal{D}=1-\frac{c^2}{\omega_p^2}\Delta\Pi_T$ to conclude that $(n\frac{c^2}{\omega_p^2})^\dagger=n\frac{c^2}{\omega_p^2}$. It is also useful to be aware of the identity
\begin{align}
n-1&=-\mathcal{D}^{-1}\bigg(\frac{c^2}{\omega_p^2}\Delta\Pi_T+\mathcal{D}\bigg)\nonumber\\
&=-\mathcal{D}^{-1}.
\end{align}

With equations\,(\ref{key1}) and\,(\ref{key2}) in hand, we can now express the components of the Hamiltonian vector field $X_{\mathcal{F}}$ as
\begin{align}
V_s&=\nu_{\mathcal{F}s}+\mathbb{P}_s(e_s\dot{\bm{A}}_{\mathcal{F}}\cdot d\bm{x})\\
\dot{F}_{os}&=-\frac{\delta\mathcal{F}}{\delta\chi_s}\\
\dot{\chi}_s&=\frac{\delta\mathcal{F}}{\delta F_{os}}-g_s^*\big(\vartheta_s(\nu_{\mathcal{F}s})-e_s\bm{v}\cdot\bm{A}_{o\mathcal{F}}\big)\\
\dot{\lambda}&=G[\nabla\cdot\dot{\bm{A}}_{\mathcal{F}}].
\end{align}
Upon substituting these expressions into the identity $[\mathcal{G},\mathcal{F}]=L_{X_{\mathcal{F}}}\mathcal{G}$, we find the following expression for the Poisson bracket,
\begin{align}
[\mathcal{G},\mathcal{F}]&=\sum_s\langle \omega_s(\nu_{\mathcal{G}s},\nu_{\mathcal{F}s}),F_s\rangle+\sum_s\langle\frac{\delta\mathcal{G}}{\delta\chi_s},\frac{\delta\mathcal{F}}{\delta F_{os}}\rangle-\langle\frac{\delta\mathcal{G}}{\delta F_{os}},\frac{\delta\mathcal{F}}{\delta \chi_{s}}\rangle\nonumber\\
&-\bigg\langle\mu_o\mu(\bm{J}_{\mathcal{G}_\chi})\bigg|\mathcal{D}^{-1}\frac{c^2}{\omega_p^2}\bigg|\bm{J}_{\mathcal{F}}+\nabla G[\delta\mathcal{F}/\delta\lambda]\bigg\rangle+\bigg\langle\mu_o\mu(\bm{J}_{\mathcal{F}_\chi})\bigg|\mathcal{D}^{-1}\frac{c^2}{\omega_p^2}\bigg|\bm{J}_{\mathcal{G}}+\nabla G[\delta\mathcal{G}/\delta\lambda]\bigg\rangle.
\end{align}

This Poisson bracket is defined on a space larger than the phase space we are actually interested in. The ``unimportant" variables are the $\chi_s$ and the $g_s$. We can find the bracket on $(F,\lambda)$-space using Poisson reduction. First we will reduce by the abelian group $\sum_sC^\infty(TQ)$, which acts by translation, $\chi_s\mapsto \chi_s+\delta\chi_s$. It is simple to verify that both the bracket and Hamiltonian given earlier are invariant under this group action. The first reduced bracket is therefore
\begin{align}
[\mathcal{G},\mathcal{F}]_{r1}=\sum_s\langle \omega_s(\nu_{\mathcal{G}s},\nu_{\mathcal{F}s}),F_s\rangle,
\end{align}
where $\mathcal{F},\mathcal{G}$ are functionals of $(g,F_o,\lambda)$ and 
\begin{align}
\nu_{\mathcal{F}s}=\mathbb{P}_s(\delta_{g_s}\mathcal{F}+e_s\mathbf{d}[\bm{v}\cdot\bm{A}_{o\mathcal{F}}]).
\end{align}
Next we will reduce by the product of diffeomorphism groups $\Pi_s\text{Diff}(TQ)$ which acts (on the right) according to $(g_s,F_{os})\mapsto ( g_s\circ h_s,h_{s}^*F_{os})$. Again, the bracket and Hamiltonian given earlier are invariant under this group action. The second reduced bracket is
\begin{align}
[\mathcal{G},\mathcal{F}]_{r2}=\sum_s\langle \{\mathcal{G}_{F_s}+e_s\bm{v}\cdot\bm{A}_{o\mathcal{G}},\mathcal{F}_{F_s}+e_s\bm{v}\cdot\bm{A}_{o\mathcal{F}}\},F_s\rangle,
\end{align}
where
\begin{align}
\bm{A}_{o\mathcal{F}}=-\mathcal{D}^{-1}\frac{c^2}{\omega_p^2}\bigg(\mu_o\bm{J}(X_{\mathcal{F}_F},F)+\mu_o\nabla G[\delta\mathcal{F}/\delta\lambda]\bigg),
\end{align}
and $\mathcal{F}_F=\delta\mathcal{F}/\delta F$. Here, $\mathcal{F},\mathcal{G}$ are functionals on $(F,\lambda)$-space.

\section{Force-free electrodynamics}
Force-free electrodynamics is a single-fluid model sometimes used in astrophysics to describe extremely-magnetized relativistic plasmas. As discussed in \cite{Gralla_2014}, the defining equations (which are Lorentz invariant) are given by
\begin{align}
\partial_t\bm{B}&=-c\nabla\times\bm{E}\\
\partial_t \bm{E}&=c\nabla\times\bm{B}-4\pi\bm{J}\\
\bm{J}&=\frac{c}{4\pi B^2}\bigg[(\nabla\cdot\bm{E})\bm{E}\times\bm{B}+(\bm{B}\cdot\nabla\times\bm{B}-\bm{E}\cdot\nabla\times \bm{E})\bm{B}\bigg]\\
\bm{E}\cdot\bm{B}&=0.
\end{align}

\subsection{Phase space variational principle for FFE}
For the sake of deriving the FFE Poisson bracket by reduction of a non-degenerate bracket, it is useful to formulate a phase space variational principle on an augmented phase space $P_{+}$. The space $P_{+}$ is the space of all ordered lists $(E,A,g,\rho_o,\chi)$ where
\begin{itemize}
	\item $E$ is the electric field $1$-form
	\item $A$ is the vector potential $1$-form 
	\item $g$ is a diffeomorphism of $\mathbb{R}^3$ that represents the fluid configuration
	\item $\rho_o$ is the reference charge density $3$-form
	\item $\chi$ is an auxiliary scalar field,
\end{itemize}
and $E$ and $A$ are constrained to satisfy
\begin{align}
E\wedge\mathbf{d}A=0,
\end{align}
which guarantees that the electric and magnetic fields are perpendicular.
The Lagrange $1$-form on this space is given by
\begin{align}
\Theta(\dot{E},\dot{A},\dot{g},\dot{\rho}_o,\dot{\chi})=\langle A\otimes\rho ,u/c\rangle-\frac{1}{4\pi}(E,\dot{A}/c)+\langle \rho_o,\dot{\chi} \rangle.
\end{align}
Here $u=\dot{g}\circ g^{-1}$, angle brackets denote a ``natural pairing", $\rho=g_*\rho_o$, and parentheses denote the standard inner product of differential forms on $\mathbb{R}^3$. The Hamiltonian functional is given by
\begin{align}
\mathcal{H}(E,A,g,\rho_o,\chi)=\frac{1}{8\pi}(E,E)+\frac{1}{8\pi}(\mathbf{d}A,\mathbf{d}A).
\end{align}
The phase space Lagrangian is given by
\begin{align}
L_{\mathcal{H}}=\Theta(\dot{E},\dot{A},\dot{g},\dot{\rho}_o,\dot{\chi})-\mathcal{H}(E,A,g,\rho_o,\chi).
\end{align}

To verify that this Lagrangian reproduces the equations of FEE, it is useful to calculate the first variation of $L_{\mathcal{F}}$, which is given by
\begin{align}
\delta L_{\mathcal{F}}&=\bigg(*[\iota_{u/c}\rho]+\frac{1}{4\pi}\dot{E}/c+\frac{1}{4\pi}\bm{\delta}\left(*\frac{(b\cdot \dot{A}/c)E}{|B|}-B\right),\delta A \bigg)\nonumber\\
&+\bigg(-\frac{1}{4\pi}\dot{A}/c-\frac{1}{4\pi}E,\delta E_\perp\bigg)\nonumber\\
&+\bigg\langle [-\dot{A}/c-\iota_{u/c}B-(A/c)g_*(\dot{\rho}_o/\rho_o)]\otimes\rho,\xi\bigg\rangle\nonumber\\
&+\langle g^*(A(u/c))+\dot{\chi},\delta\rho_o\rangle\nonumber\\
&+\langle-\dot{\rho}_o,\delta\chi\rangle\nonumber\\
&+\frac{d}{dt}\Theta(\delta{E},\delta{A},\delta{g},\delta{\rho}_o,\delta{\chi}).
\end{align}
Here $B=\mathbf{d}A$, $|B|=\sqrt{*(B\wedge*B)}$, $b=*B/|B|$, the dot product between two $k$-forms $\alpha,\beta$ is given by
\begin{align}
\alpha\cdot\beta=*(\alpha\wedge*\beta),
\end{align}
and if $\alpha$ is a $1$-form, $\alpha_\perp=\alpha-(\alpha\cdot b)b$. Note that $\delta E$ is constrained to satisfy
\begin{align}
\delta E\cdot b=-E\cdot *\frac{\delta B}{|B|},
\end{align}
i.e. the parallel variation of $E$ is completely determined by the variation of $B$. By setting the variation of $L_{\mathcal{H}}$ equal to zero, the following equations of motion emerge:
\begin{align}
\dot{\rho}_o&=0\\
\dot{\chi}&=-g^*(A(u/c))\\
\dot{A}/c&=-E\\
\bm{\delta}B&=*[\iota_{u/c}\rho]+\frac{1}{4\pi}\dot{E}/c\\
E&=\iota_{u/c}B,
\end{align} 
which are equivalent to the equations of force free electrodynamics provided $4\pi *\rho=-\bm{\delta}E$ at $t=0$ (note that $4\pi *\rho+\bm{\delta}E$ is a constant of motion).

\subsection{Symmetries of the phase space Lagrangian}
\subsubsection{Symmetry under translations in $\chi$}
Consider the abelian group $G_o=C^\infty(\mathbb{R}^3)$, whose typical element we will denote $\tau$. $G_o$ acts on the augmented FFE phase space according to $\chi\mapsto \chi+\tau$. The phase space Lagrangian is invariant under this symmetry, which implies that $\Theta(0,0,0,0,\tau)=\langle \rho_o,\tau\rangle$ is a constant of motion for each $\tau\in G_o$. Because $\tau$ is an arbitrary smooth function, $\rho_o$ must be independent of time.
\subsection{Symmetry under time-independent gauge transformations}
Consider again the abelian group $G_o$. Let $s\in G_o$ be a typical element. This group acts on $P_+$ in a second way, namely
\begin{align}
A\mapsto A+\mathbf{d}s\\
\chi\mapsto \chi-\frac{1}{c}g^*s.
\end{align}
Note that this transformation preserves the constraint $E\wedge \mathbf{d}A=0$. The phase space Lagrangian is invariant under this symmetry which implies that 
\begin{align}
\mu_s=\Theta(0,\mathbf{d}s,0,0,-g^*s/c)=-\frac{1}{4\pi c}(\bm{\delta}E+4\pi*\rho,s)
\end{align}
is a constant of motion for each $s\in G_o$. Thus, $\bm{\delta}E+4\pi*\rho$ is a constant of motion.

\subsubsection{Symmetry under particle relabling}
Let $G_1=\text{Diff}(\mathbb{R}^3)$ be the non-abelian group of diffeomorphisms of $\mathbb{R}^3$ with typical element $h\in G_1$. There is a (right) $G_1$ action on $P_+$ given by
\begin{align}
g\mapsto g\circ h\\
\rho_o\mapsto h^*\rho_o\\
\chi\mapsto h^*\chi.
\end{align}
The phase space Lagrangian is invariant this symmetry. Thus, the contraction of the Lagrange $1$-form with the infinitesimal generator of this symmetry is a constant of motion. The infinitesimal generator is given by
\begin{align}
\xi_{P_+}=(\delta E,\delta A,\delta g,\delta\rho,\delta\chi)=(0,0,Tg\circ \xi,L_\xi\rho,L_\xi\chi),
\end{align}
where $\xi$ is an arbitrary vector field. This implies that
\begin{align}
\mu_\xi=\Theta(\xi_{P_+})=\langle (g^*A/c+\mathbf{d}\chi)\otimes\rho_o,\xi)\rangle
\end{align}
is a constant of motion for each $\xi$. Because $\xi$ is arbitrary and $\rho_o$ is a constant of motion, the quantity $A_o=g^*A+\mathbf{d}c\chi$ is a constant of motion. To understand this conservation law, consider the exterior derivative of $A_o$,
\begin{align}
\mathbf{d}A_o=g^*B.
\end{align}
By applying the pushforward by $g$ to each side of this expression, and noting that $A_o$ is a constant of motion, we conclude that the magnetic field $2$-form is advected by the fluid velocity $u$,
\begin{align}
B=g_*\mathbf{d}A_o\Rightarrow \dot{B}=-L_u B.
\end{align}
But this implies that the vector potential $A$ is advected modulo an exact $1$-form,
\begin{align}
A=g_*A_o+\mathbf{d}\psi.
\end{align}
Thus, a second conclusion we can draw about the conservation of $\mu_\xi$ is that $\mathbf{d}c\chi$ can be interpreted as the difference between $A$ in the temporal gauge and $A$ in the ``advection gauge", wherein $A$ is advected as a $1$-form. 

\subsubsection{Symmetry under isometries of configuration space}
Let $G_2$ be the non-abelian group of isometries of $\mathbb{R}^3$ with typical element $R$. $G_2$ is naturally a matrix group, and so it is equipped with a natural (left) action on $\mathbb{R}^3$,
\begin{align}
\bm{x}\mapsto R\bm{x}.
\end{align}
Given an element $\nu\in\mathfrak{g}_2$ we can therefore define an infinitesimal generator on $\mathbb{R}^3$,
\begin{align}
\nu_Q=\frac{d}{d\epsilon}\bigg|_0\exp(\epsilon \nu),
\end{align}
where $\exp$ denotes the matrix exponential. The left action of $G_2$ on $\mathbb{R}^3$ lifts to a left action on $P_+$ given by
\begin{align}
A&\mapsto R_*A\\
E&\mapsto R_*E\\
g&\mapsto R\circ g.
\end{align}
The infinitesimal generator of this action is given by
\begin{align}
\nu_{P_+}=(\delta E,\delta A,\delta g,\delta\rho,\delta\chi)=(-L_{\nu_Q}E,-L_{\nu_Q}A,\nu_Q\circ g,0,0).
\end{align} 
The phase space Lagrangian is invariant under this symmetry. Therefore
\begin{align}
\mu_{\nu}=\Theta(\nu_{P_+})=\frac{1}{4\pi}(A(\nu_Q/c),4\pi*\rho+\bm{\delta}E)+\frac{1}{4\pi}(E,\iota_{\nu_Q}B)
\end{align}
is a constant of motion for each $\nu\in \mathfrak{g}_2$. This conservation law is equivalent to the global conservation of linear and angular momentum. Provided initial conditions are chosen so that the Gauss equation is satisfied, $\mu_\nu$ is equal to the total momentum of the electromagnetic field.

\subsubsection{Combining the symmetry groups $G_o,G_o$, and $G_1$.}
There is a ``big group" that encodes three of the symmetries discussed so far: phase translation, gauge transformation, and particle relabling. The big group will be denoted $\mathcal{G}=G_o\times (G_o\rtimes G_1)$, which is suggestive of the fact that $H$ is a direct product of the abelian gauge group $G_o$ with the non-ablian semidirect product $G_o\rtimes G_1$. The identity element is $e=(0,0,\text{id})\in G_o\times G_o\times G_1$. The group product is given by
\begin{align}\label{product}
(s_1,\tau_1,h_1)*(s_2,\tau_2,h_2)=(s_1+s_2,h_2^*\tau_1+\tau_2,h_1\circ h_2),
\end{align}
which can be seen to be associative by direct calculation. The group inverse is given by
\begin{align}
(s,\tau,h)^{-1}=(-s,-h_*\tau,h^{-1}).
\end{align}

While the structure of this group may appear a bit mysterious, it can be uncovered in a straightforward way as follows. By an abuse of notation, denote the right actions of $G_o$, $G_o$, and $G_1$ by $R_s$, $R_\tau$, and $R_h$, respectively. We have
\begin{align}
R_s(E,A,g,\rho_o,\chi)&=(E,A+\mathbf{d}s,g,\rho_o,\chi-g^*s/c)\\
R_\tau(E,A,g,\rho_o,\chi)&=(E,A,g,\rho_o,\chi+\tau)\\
R_{h}(E,A,g,\rho_o,\chi)&=(E,A,g\circ h,h^*\rho_o,h^*\chi).
\end{align}
Note that we have the following commutation relations amongst these maps
\begin{align}
R_s\circ R_\tau&=R_\tau\circ R_s\\
R_s\circ R_h&=R_h\circ R_s\\
R_h\circ R_\tau&=R_{h^*\tau}\circ R_h.
\end{align} 
Therefore we have the identity
\begin{align}\label{nice_identity}
(R_{s_2}\circ R_{\tau_2}\circ R_{h_2})\circ(R_{s_1}\circ R_{\tau_1}\circ R_{h_1})&=R_{s_1+s_2}\circ R_{h_2^*\tau_1+\tau_2}\circ R_{h_1\circ h_2}.
\end{align}
This identity strongly suggests defining the group product given above. Morevoer, by combining this identity with the fact that the product\,(\ref{product}) does indeed satisfy the group axioms, we can quickly deduce the appropriate right action of $\mathcal{G}$ on $P_+$, namely
\begin{align}\label{right_action}
R_{(s,\tau,h)}=R_s\circ R_\tau\circ R_h.
\end{align}
The identity\,(\ref{nice_identity}) can be rewritten as
\begin{align}
R_{(s_2,\tau_2,h_2)}\circ R_{(s_1,\tau_1,h_1)}=R_{(s_1,\tau_1,h_1)*(s_2,\tau_2,h_2)},
\end{align}
which is the most non-trivial property satisfied by a right $\mathcal{G}$ action. The other properties of a right action follow immediately from the definition\,(\ref{right_action}) and the commutation relations.

\subsubsection{The quotient space $P=P_+/\mathcal{G}$}
The Poisson bracket, $\{\cdot,\cdot\}_{P_+}$, on $P_+$ given by inverting the symplectic form $-\mathbf{d}\Theta$ will automatically be symmetric under the group $\mathcal{G}$ because $\Theta$ has the same symmetry. Therefore there is a \emph{reduced} Poisson bracket on $P=P_+/\mathcal{G}$ that is given formally as follows. Let $\pi:P_+\rightarrow P$ be the projection map that sends a point in $P_+$ to its orbit under the action of $\mathcal{G}$. Given a pair of functionals $F,G:P\rightarrow\mathbb{R}$, we can pull them back to $P_+$, thereby obtaining the functionals $\pi^*F,\pi^*G:P_+\rightarrow\mathbb{R}$. Each of these functionals on $P_+$ is automatically $\mathcal{G}$-invariant, which implies that the functional $\{\pi^*F,\pi^*G\}_{P_+}$ is also $\mathcal{G}$-invariant. Because $\pi$ is surjective, there is therefore a unique functional, $\{F,G\}_P:P\rightarrow\mathbb{R}$, that satisfies
\begin{align}\label{reduced_pb}
\pi^*\{F,G\}_P=\{\pi^*F,\pi^*G\}_{P_+}.
\end{align}
The formula\,(\ref{reduced_pb}) defines the Poisson bracket on $P$. The Jacobi and Leibniz identities are straightforward to check.

Because $P$ represents a ``physical" phase space for FFE, the bracket on $P$ is in some ways more desireable than the bracket on $P_+$. It is therefore useful to have a concrete model of the space $P$. The purpose of this subsection is to supply this model. 

It will turn out that $P$ can be represented as the space of triples $(E,B,\rho)$, where $E$ is the electric field $1$-form, $B$ is the (exact) magnetic field $2$-form, $\rho$ is the charge density $3$-form, and the electric and magnetic fields are constrained to satisfy $E\wedge B=0$. To see this, first note that there is a surjective map $\pi:P_+\rightarrow P$ given by
\begin{align}
\pi(E,A,g,\rho_o,\chi)=(E,\mathbf{d}A,g_*\rho_o).
\end{align}
Next suppose $(E,A,g,\rho_o,\chi)$ and $(E^\prime,A^\prime,g^\prime,\rho_o^\prime,\chi^\prime)$ each map to $(E,B,\rho)$ under $\pi$. Immediately we see that $E^\prime=E$ and $\mathbf{d}A^\prime=\mathbf{d}A$, which implies that there is some $s\in G_o$ such that $A^\prime=A+\mathbf{d}s$. If we now set $h=g^{-1}\circ g^\prime$ and $\tau=\chi^\prime-h^*\chi+h^*g^*s/c$, a simple calculation shows that $(E^\prime,A^\prime,g^\prime,\rho_o^\prime,\chi^\prime)=R_{(s,\tau,h)}(E,A,g,\rho_o,\chi)$. This shows that points in $\pi^{-1}(E,B,\rho)$ are all on the same $\mathcal{G}$-orbit. Because it is also true that $\pi\circ R_{(s,\tau,h)}=\pi$ for each $(s,\tau,h)\in\mathcal{G}$, we must therefore conclude that the preimages of points in $P$ along $\pi$ are precisely the $\mathcal{G}$-orbits. This proves that $P_+/\mathcal{G}$ is diffeomorphic to $P$.

\subsection{Derivation of the Poisson bracket on $P$}
In order to compute the Poisson bracket on $P$, we will compute $\{\pi^*F,\pi^*G\}_{P_+}$, where $F,G:P\rightarrow\mathbb{R}$ are functionals on $P$. By the definition of a Hamiltonian vector field, we have
\begin{align}\label{basic_identity}
\{\pi^*F,\pi^*G\}_{P_+}=(\mathbf{d}\pi^*F)(X_{\pi^*G}).
\end{align}
In order to evaluate the right hand side of Eq.\,(\ref{basic_identity}), we must compute the components of the Hamiltonian vector field $X_{\pi^*G}=(\dot{E}_G,\dot{A}_G,\dot{g}_G,\dot{\rho}_{oG},\dot{\chi}_G)$. This computation is most easily done by analyzing the Euler-Lagrange equations associated with the phase space Lagrangian 
\begin{align}
L_G(E,A,g,\rho_o,\chi,\dot{E},\dot{A},\dot{g},\dot{\rho}_o,\dot{\chi})=\Theta(\dot{A},\dot{g},\dot{\rho}_o,\dot{\chi})-\pi^*G(E,A,g,\rho_o,\chi).
\end{align}
The first variation of $L_G$ is given by
\begin{align}
\delta L_G&=\bigg(*[\iota_{u/c}\rho]+\frac{1}{4\pi}\dot{E}/c+\frac{1}{4\pi}\bm{\delta}\left(*\frac{(b\cdot[ \dot{A}/c+4\pi\delta G/\delta E])E}{|B|}-4\pi\frac{\delta G}{\delta B}\right),\delta A \bigg)\nonumber\\
&+\bigg(-\frac{1}{4\pi}\dot{A}/c-\frac{\delta G}{\delta E},\delta E_\perp\bigg)\nonumber\\
&+\bigg\langle [-\dot{A}/c-\iota_{u/c}B-(A/c)g_*(\dot{\rho}_o/\rho_o)-\mathbf{d}\delta G/\delta\rho]\otimes\rho,\xi\bigg\rangle\nonumber\\
&+\langle g^*(A(u/c))+\dot{\chi}-g^*\delta G/\delta\rho,\delta\rho_o\rangle\nonumber\\
&+\langle-\dot{\rho}_o,\delta\chi\rangle\nonumber\\
&+\frac{d}{dt}\Theta(\delta{E},\delta{A},\delta{g},\delta{\rho}_o,\delta{\chi}).
\end{align}
It follows that the components of the Hamiltonian vector field $X_{\pi^*G}$ are given by
\begin{align}
\dot{\rho}_o&=0\\
\frac{1}{c}\dot{A}_G&=-4\pi\bm{\alpha}_{G\perp}-4\pi\mathbf{d}\frac{\delta G}{\delta\rho}\\
\frac{1}{c}j_G&=\frac{4\pi*\rho}{|B|}*(\bm{\alpha}_G\wedge b)-\bigg(\frac{E}{|B|}\cdot\bm{\delta}(*\bm{\alpha}_{G\perp})+b\cdot\bm{\delta}\bigg(\bm{\alpha}_{G\parallel}\frac{*E}{|B|}-\frac{\delta G}{\delta B}\bigg)\bigg)b\\
\frac{1}{c}\dot{E}&=- \frac{4\pi*\rho}{|B|}*(4\pi\bm{\alpha}_G\wedge b)-4\pi\bm{\delta}\bigg(\bm{\alpha}_{G\parallel}\frac{*E}{|B|}-\frac{\delta G}{\delta B}\bigg)_\perp+\bigg(\frac{E}{|B|}\cdot\bm{\delta}(*\bm{\alpha}_{G\perp})\bigg)b\\
\dot{\chi}_G&=g^*\bigg(\frac{\delta G}{\delta\rho}-A(u_G/c)\bigg),
\end{align}
where $\bm{\alpha}_G=\delta G/\delta E-(1/4\pi)\mathbf{d}\delta G/\delta \rho$ and $j_G=*\iota_{u_G}\rho$ is the current density $1$-form.

After some tedious but straightforward calculations involving substituting the previous expressions into the identity\,(\ref{basic_identity}), we find that the Poisson bracket on $P$ is given by
\begin{align}
\{F,G\}_P&=-4\pi c\bigg(\frac{(4\pi*\rho)B}{|B|^2},\bm{\alpha}_F\wedge\bm{\alpha}_G\bigg)\nonumber\\
&+4\pi c\bigg(\bm{\alpha}_{F\perp},\bm{\delta}\bigg(\frac{\delta G}{\delta B}-\bm{\alpha}_{G\parallel}*e\bigg)\bigg)-4\pi c\bigg(\bm{\alpha}_{G\perp},\bm{\delta}\bigg(\frac{\delta F}{\delta B}-\bm{\alpha}_{F\parallel}*e\bigg)\bigg),
\end{align}
where $e=E/|B|$. 

It is not difficult to show that any functional of $\bm{\delta}E+4\pi*\rho$ is a Casimir of the bracket $\{\cdot,\cdot\}_P$. Therefore the submanifold of $P$ defined by $\bm{\delta}E+4\pi*\rho=0$ is a Poisson submanifold that can be parameterized by the space $P_o$ of pairs $(E,B)$ that satisfy $E\wedge B=0$. Being a Poisson submanifold, $P_o$ has a bracket $\{\cdot,\cdot\}_{P_o}$ that is naturally induced by $\{\cdot,\cdot\}_P$. The expression for this bracket on $(E,B)$-space is given by
\begin{align}
\{F,G\}_{P_o}=&4\pi c\bigg(\frac{(\bm{\delta}E)B}{|B|^2},\frac{\delta F}{\delta E}\wedge\frac{\delta G}{\delta E}\bigg)\nonumber\\
+&4\pi c\bigg(\frac{\delta F}{\delta E}_{\perp},\bm{\delta}\bigg(\frac{\delta G}{\delta B}-\frac{\delta G}{\delta E}_{\parallel}*e\bigg)\bigg)\nonumber\\
-&4\pi c\bigg(\frac{\delta G}{\delta E}_{\perp},\bm{\delta}\bigg(\frac{\delta F}{\delta B}-\frac{\delta F}{\delta E}_{\parallel}*e\bigg)\bigg).
\end{align}

\chapter[Energetically-consistent gyrokinetic collision operator]{Energetically-consistent gyrokinetic collision operator}
\label{ch:GK_collision_operator}
%
%
%
%
%
%

\section{Introduction} One of the greatest unsolved problems in the theory of magnetically-confined plasmas is understanding and controlling the turbulent flux of particles and heat into a fusion reactor's wall\,\cite{Kikuchi_2012}. It is believed that the predominant cause of these fluxes is low-frequency fluctuating electromagnetic fields with wavelengths on the order of the gyroradius. While a collisionless gyrokinetic model of these fluctuating fields has been developed that is fully consistent with the First Law of Thermodynamics (for a recent review see \cite{Brizard_2007}), this energetically-consistent model has the serious flaw of ignoring collisions altogether.

In order to accurately describe irreversible plasma transport processes, the effects of collisions must be incorporated into gyrokinetic theory. Previous work on linear gyrokinetic collision operators \cite{Abel_2008,Li_Ernst_2011,Madsen_2013} assumed a strict two-scale separation between a large-scale equilibrium distribution function $F_o$ and a small-scale fluctuating part $\delta F=F-F_o$. Conservation properties of the collision operator in \cite{Abel_2008}, for example, were discussed in the gyroBohm limit. Here, we will focus on nonlinear gyrokinetic collision operators for a global 
full-$F$ approach that do not make this split, and that can thus investigate more completely the possible effects of finite $\epsilon=\rho_i/L$ in experiments, such as corrections to gyroBohm scaling and non-local turbulence spreading (see footnote 5 on p. 427 in \cite{Brizard_2007}.)


When finite-$\epsilon$ effects are accounted for, preserving exact conservation properties, and therefore ensuring consistency with the First Law of Thermodynamics, is a nontrivial unsolved problem. The collision operators in \cite{Abel_2008,Li_Ernst_2011}, for example, were obtained by transforming a particle-space collision operator with exact conservation properties into the lowest-order guiding center coordinates. While this approach guarantees the existence of energy and momentum-like quantities that annihilate the collision operator, these same quantities are not conserved by the full-$F$ collisionless gyrokinetic system, and therefore fail to be conserved by the full-$F$ collisional system. More generally, existing gyrokinetic collision operators are not energetically consistent in a full-$F$ formalism because: (a) the gyrocenter coordinate transformation, and therefore any collision operator transformed into gyrocenter coordinates, is only known as an asymptotic expansion in the gyrokinetic ordering parameter $\epsilon$; and (b) replacing the asymptotic expansion of such an operator with a truncated power series destroys exact conservation laws. The purpose of this Chapter is to present the first collisional formulation of global full-$F$ gyrokinetics with exact conservation laws. 

\section{Electrostatic Model} For the sake of simplicity, our discussion will focus on quasi-neutral electrostatic gyrokinetics (for instance, see \cite{Parra_2011}). However, the ideas behind our discussion apply equally-well to electromagnetic gyrokinetics (for example, see \cite{Sugama_2000}.) Our primary result consists of an expression for the non-linear Landau operator in gyrocenter coordinates that is corrected by small terms to ensure exact energy and momentum conservation [see Eq.\,(\ref{gyro_landau}).] These correction terms are analogous to the $B_\parallel^*$-denominators in the Hamiltonian guiding center theory introduced by Littlejohn \cite{Littlejohn_1981}; they do not increase the theory's order of accuracy, but they are essential to include for the sake of ensuring exact energy and momentum conservation.

As a first step, we review how the energy conservation law is discussed in collisionless kinetic theory. The governing equations of collisionless electrostatic kinetic theory are the Vlasov-Poisson equations,
\begin{align}
\partial_t f_s+\{f_s,H_s\}=0\label{lp_vlasov_eqn}\\
\Delta\varphi=-4\pi\rho(f), \label{Poisson}
\end{align}
where $f_s$ is the species-$s$ distribution function, $\varphi$ is the electrostatic potential, $\rho(f)$ is the charge density, $H_s= {p^2}/{2m_s}+e_s\varphi$, and $\{\cdot,\cdot\}$ is the standard canonical Poisson bracket. Equations (\ref{lp_vlasov_eqn})-(\ref{Poisson}) conserve the total energy
\begin{align}
{\cal E}&=\sum_{s}\int \frac{p^2}{2 m_s}f_s\,dz+\left\langle\varphi,\frac{}{}\rho(f)+\frac{1}{8\pi}\Delta\varphi\right\rangle,
\label{vp_energy}
\end{align}
where $\langle\cdot,\cdot\rangle$ denotes the standard $L^2$-pairing of functions on configuration space and $dz = d\bm{x}\,d\bm{p}$. Because binary collisions conserve energy, Eq.~(\ref{vp_energy}) must also be conserved in collisional kinetic theory. In particular, if the Vlasov-Poisson equations are modified by the addition of a bilinear collision operator,
\begin{align}
\partial_t f_s+\{f_s,H_s\}=\sum_{\bar{s}} C_{s\bar{s}}(f_s,f_{\bar{s}})\label{landau_poisson_a}\\
\Delta\varphi=-4\pi\rho(f)\label{landau_poisson_b},
\end{align}
then $C_{s\bar{s}}$ must be chosen to satisfy the condition
\begin{align}
0=&\frac{d{\cal E}}{dt} = \sum_{s}\int  H_s\,\partial_t f_s\,dz+\left\langle\partial_t \varphi,\rho(f)+\frac{1}{4\pi}\Delta\varphi\right\rangle\nonumber\\
=&\sum_{s,\bar{s}}\int H_sC_{s\bar{s}}(f_s,f_{\bar{s}})\,dz.
\end{align}
Because this identity must hold for an arbitrary multi-species distribution function, the collision operator therefore has to satisfy the well-known identities
\begin{align}
\int H_sC_{s\bar{s}}(f_s,f_{\bar{s}})\,dz + \int H_{\bar{s}}C_{\bar{s}s}(f_{\bar{s}},f_s)\,d\bar{z} = 0,
\label{basic_identities}
\end{align}
which express the fact that the energy gained by species $s$ due to collisions with species $\bar{s}$ is precisely the energy lost by species $\bar{s}$ due to collisions with species $s$. The non-linear Landau operator (summation rule is implied),
\begin{align}
C_{s\bar{s}}(f_s,f_{\bar{s}})=-\frac{\Gamma_{s\bar{s}}}{2}\{x_i,\gamma_{i}^{s\bar{s}}\},\label{particle_C}
\end{align}
satisfies the identities\,(\ref{basic_identities}), and therefore defines an energetically-consistent collisional kinetic theory. Here $\Gamma_{s\bar{s}}=4\pi e_s^2e_{\bar{s}}^2\ln\Lambda$; the $3$-component vector $\bm{\gamma}^{s\bar{s}}$ is
\begin{align}
\gamma_{i}^{s\bar{s}}(z)=\int \delta(\bm{x}-\bar{\bm{x}})\;\mathbb{Q}^{s\bar{s}}(z,\bar{z})\,\bm{A}_{s\bar{s}}(z,\bar{z})\,d\bar{z};
\end{align}
the $3\times 3$ matrix $\mathbb{Q}^{s\bar{s}}$ is given by
\begin{align}
\mathbb{Q}^{s\bar{s}}(z,\bar{z})=\frac{1}{W_{s\bar{s}}(z,\bar{z})}\mathbb{P}[\bm{W}_{s\bar{s}}(z,\bar{z})],
\end{align}
where $\mathbb{P}(\bm{\xi}) \equiv \mathbb{I} - \hat{\bm{\xi}}\hat{\bm{\xi}}$ is the orthogonal projection onto the plane perpendicular to the vector 
$\bm{\xi}$; the velocity difference $\bm{W}_{s\bar{s}}$ is given by
\begin{align}
\bm{W}_{s\bar{s}}(z,\bar{z})=\{\bm{x},H_s\}(z)-\{\bm{x},H_{\bar{s}}\}(\bar{z});
\end{align}
and the vector 
\begin{align}
\bm{A}_{s\bar{s}}(z,\bar{z})=f_{s}(z)\{\bm{x},f_{\bar{s}}\}(\bar{z})-\{\bm{x},f_s\}(z)f_{\bar{s}}(\bar{z}).
\end{align}
When comparing this form of the Landau operator to more conventional expressions, it is useful to note that $\{\bm{x},g\}=\partial_{\bm{p}}g$, where $g$ is any function on phase space, so that the collision operator (\ref{particle_C}) describes collisions in momentum space. Moreover, the identities
(\ref{basic_identities}) follow immediately from the fact that the velocity difference $\bm{W}_{s\bar{s}}$ is a null-eigenvector of the matrix 
$\mathbb{Q}^{s\bar{s}}$.

\section{Electrostatic Gyrokinetic Model} In order to apply this same argument to gyrokinetic theory, we start with the gyrokinetic Vlasov-Poisson system
\begin{align}
\partial_t F_s+\{F_s,H_s^{\text{gy}}\}_s^{\text{gc}}=0\label{gk_vlasov}\\
\nabla\cdot\bm{P}=\rho(F).\label{qn}
\end{align}
Here, $F_s$ is the gyrocenter distribution function; $\varphi$ is the electrostatic potential; $\{\cdot,\cdot\}_s^{\text{gc}}$ is the guiding center Poisson bracket;
\begin{align}
H_s^{\text{gy}}&=H_s^{\text{gc}}+e_s\left<\psi\right>+\frac{e_s^2}{2}\langle\{\tilde{\psi},\tilde{\Psi}\}_s^{\text{gc}}\rangle \equiv K_s(\bm{E})+e_s\varphi
\end{align}
is the gyrocenter Hamiltonian; $\psi(z)=\varphi(\bm{X}+\bm{\rho}_{os})$, where $\bm{\rho}_{os}$ is the lowest-order guiding-center gyroradius; 
$\langle\cdot\rangle$ denotes the gyroaverage; $\tilde{\Psi}$ denotes the gyroangle antiderivative of $\tilde{\psi} \equiv \psi - \langle\psi\rangle$; $K_s(\bm{E})$ is the gyrocenter kinetic energy; $\bm{P}=-\,
\delta\mathcal{K}/\delta\bm{E}$ is the gyrocenter polarization density; $\mathcal{K}=\sum_s\int F_sK_s(\bm{E})\,dz_s^{\text{gc}}$; and $dz_s^{\text{gc}}$ denotes the guiding center Liouville volume element. These equations govern collisionless quasineutral electrostatic gyrokinetic theory in the ``high-flow" regime (see \cite{Krommes_Hammett_2013} and references therein) and they conserve the total energy,
\begin{align}
{\cal E}^{\text{gy}}=\sum_s\int F_s H_s^{\text{gy}}\,dz_s^{\text{gc}},
\end{align}
exactly. Note that the quasineutrality equation\,(\ref{qn}) implies that this system governs plasma dynamics on time scales long compared to the period of plasma oscillations.

The equations governing collisional gyrokinetic theory are given by adding a bilinear collision operator to the gyrokinetic Vlasov-Poisson equations, 
\begin{align}\label{gk_landau_poisson}
\partial_t F_s+\{F_s,H_s^{\text{gy}}\}_s^{\text{gc}}&=\sum_{\bar{s}}C^{\text{gy}}_{s\bar{s}}(F_s,F_{\bar{s}})\\
\nabla\cdot\bm{P}&=\rho(F).
\end{align}
Because the conservation laws of ordinary collisional kinetic theory are consistent with those of collisionless kinetic theory, the gyrokinetic collision operator $C^{\text{gy}}_{s\bar{s}}$ must not alter the conservation of ${\cal E}^{\text{gy}}$. Thus, 
\begin{align}
0=&\frac{d{\cal E}^{\text{gy}}}{dt}=\sum_s\int H_s^{\text{gy}}\partial_tF_s\,dz_s^{\text{gc}}+\left\langle\rho(F)-\nabla\cdot\bm{P},\frac{}{}\partial_t\varphi\right\rangle\nonumber\\
=&\sum_{s,\bar{s}}\int H_s^{\text{gy}}C^{\text{gy}}_{s\bar{s}}(F_s,F_{\bar{s}})\,dz_s^{\text{gc}}.
\end{align}
This identity will be satisfied for a general multi-species gyrocenter distribution function if and only if
\begin{align}
\int H_s^{\text{gy}}C_{s\bar{s}}^{\text{gy}}(F_s,F_{\bar{s}})\,dz_s^{\text{gc}}+\int H_{\bar{s}}^{\text{gy}}C_{\bar{s}s}^{\text{gy}}(F_{\bar{s}},F_s)\,d\bar{z}_{\bar{s}}^{\text{gc}}=0,\label{gk_identity}
\end{align}
which is the gyrokinetic version of Eq.\,(\ref{basic_identities}). The identities\,(\ref{gk_identity}) must be satisfied exactly by any energetically-consistent gyrokinetic collision operator.

\section{An energetically-consistent collision operator} While Eq.\,(\ref{gk_identity}) imposes important qualitative constraints, they cannot determine the form of the gyrokinetic collision operator by themselves.  A quantitative constraint is necessary as well. To this end, it is important that the gyrokinetic collision operator agrees with the the transformation of the particle-space Landau operator \footnote{Necessary conditions for the use of the Landau operator are $\omega_c<\omega_p$ and $(\partial_tF)/(\omega_p F)<1$. When these conditions are not satisfied, our discussion must be modified.} into gyrocenter coordinates, at least up to some desired order in the gyrokinetic ordering parameter $\epsilon$. Is it possible to satisfy these qualitative and quantitative constraints simultaneously? The answer is ``yes". 

We have discovered an accurate gyrokinetic collision operator that is consistent with the conservation laws of collisionless gyrokinetic theory, and therefore the first law of thermodynamics. The form of the operator is suggested by the somewhat-peculiar presentation of the particle-space Landau operator given earlier. Let $\bm{y}_s=\bm{X}+\bm{\rho}_{os}$ and define the gyrocenter velocity difference
\begin{align}\label{gy_vel_dif}
\bm{W}_{s\bar{s}}^{\text{gy}}(z,\bar{z})=\{\bm{y}_s,H_s^{\text{gy}}\}^{\text{gc}}_s(z)-\{\bm{y}_{\bar{s}},H_{\bar{s}}^{\text{gy}}\}^{\text{gc}}_{\bar{s}}(\bar{z}),
\end{align}
the associated $3\times 3$ matrix
\begin{align}
\mathbb{Q}^{s\bar{s}}_{\text{gy}}(z,\bar{z})=\frac{1}{W^{\text{gy}}_{s\bar{s}}(z,\bar{z})}\mathbb{P}[\bm{W}^{\text{gy}}_{s\bar{s}}(z,\bar{z})],
\end{align}
and the vector
\begin{align}
\bm{A}_{s\bar{s}}^{\text{gy}}(z,\bar{z})=F_{s}(z)\{\bm{y}_{\bar{s}},F_{\bar{s}}\}^{\text{gc}}_{\bar{s}}(\bar{z})-\{\bm{y}_s,F_s\}^{\text{gc}}_s(z)F_{\bar{s}}(\bar{z}).
\end{align}
The energetically-consistent gyrokinetic Landau operator is given by
\begin{align}\label{gyro_landau}
C_{s\bar{s}}^{\text{gy}}(F_s,F_{\bar{s}})=-\frac{\Gamma_{s\bar{s}}}{2}\{y_{s\,i},\gamma^{s\bar{s}}_{\text{gy}\,i}\}^{\text{gc}}_s,
\end{align}
where
\begin{align}
\bm{\gamma}^{s\bar{s}}_{\text{gy}}(z)=\int \delta^{\text{gy}}_{s\bar{s}}(z,\bar{z})\mathbb{Q}^{s\bar{s}}_{\text{gy}}(z,\bar{z})
\bm{A}_{s\bar{s}}^{\text{gy}}(z,\bar{z})\,d\bar{z}_{\bar{s}}^{\text{gc}},
\end{align}
and $\delta^{\text{gy}}_{s\bar{s}}(z,\bar{z})=\delta(\bm{y}_s(z)-\bm{y}_{\bar{s}}(\bar{z}))$. Note that this operator depends explicitly on the electric field through the gyrocenter Hamiltonians that appear in Eq.~(\ref{gy_vel_dif}). Using a straightforward, but tedious argument that is not reproduced here, we have shown that this operator agrees with the Landau operator transformed into gyrocenter coordinates with leading-order accuracy. 

Because the proof is simple, we will now show explicitly that the gyrokinetic Landau-Poisson system\,(\ref{gk_landau_poisson}) defined in terms of the collision operator\,(\ref{gyro_landau}) has exact conservation laws for energy and momentum. We hope to convey the similarity of this demonstration with the analogous demonstration for the ordinary Landau-Poisson system\,(\ref{landau_poisson_a})-(\ref{landau_poisson_b}). However, a word of caution is in order here. It is essential that the guiding center Poisson brackets that appear in Eq.~(\ref{gyro_landau}) be genuine Poisson brackets (i.e., the brackets must satisfy the Leibniz and Jacobi identities). Dropping terms from a bracket that satisfies these properties will destroy the gyrokinetic Landau-Poisson system's exact conservation laws.

\section{Energy conservation}Proving that the gyrokinetic Landau operator\,(\ref{gyro_landau}) satisfies the identities\,(\ref{gk_identity}) is very similar to proving that the particle-space Landau operator satisfies the identities\,(\ref{basic_identities}). Setting $\dot{\mathcal{E}}_{s\bar{s}}=\int H_s^{\text{gy}}C_{s\bar{s}}^{\text{gy}}(F_s,F_{\bar{s}})\,dz^{\text{gc}}_s$, it is simple to verify that
\begin{align}
&\dot{\mathcal{E}}_{s\bar{s}}+\dot{\mathcal{E}}_{\bar{s}s}=\frac{\Gamma_{s\bar{s}}}{2}\iint(\bm{W}_{s\bar{s}}^{\text{gy}})^{\dagger}\mathbb{Q}^{s\bar{s}}_{\text{gy}}\bm{A}_{s\bar{s}}^{\text{gy}}\delta_{s\bar{s}}^{\text{gy}}\,d\bar{z}_{\bar{s}}^{\text{gc}}\,dz_s^{\text{gc}},
\end{align}
where all two-point quantities in the integrand are evaluated at $(z,\bar{z})$ and $\cdot^\dagger$ denotes the ordinary matrix transpose. Because $\mathbb{Q}^{s\bar{s}}_{\text{gy}}$ is a symmetric matrix with null eigenvector $\bm{W}_{s\bar{s}}^{\text{gy}}$, the right-hand-side of this equation vanishes exactly. Thus the gyrokinetic Landau operator\,(\ref{gyro_landau}) satisfies the identities\,(\ref{gk_identity}) exactly, and the gyrokinetic Landau-Poisson system\,(\ref{gk_landau_poisson}) has an exact energy conservation law, $d{\cal E}^{\text{gy}}/dt=0$.

\section{Toroidal momentum conservation}We will prove that if the background magnetic field is axisymmetric, then the gyrokinetic Landau-Poisson system conserves the total toroidal momentum
\begin{align}
P_\phi=\sum_s\int p_{\phi s}F_s\,dz_s^{\text{gc}},\label{gy_Pphi}
\end{align}
where $p_{\phi s}$ is the guiding center canonical toroidal momentum \footnote{Rather than give an explicit expression for $p_{\phi s}$, which will depend on ones choice of guiding center representation, it is better to define it operationally \emph{via} the guiding center Poisson bracket: for each phase space function $f$, the canonical toroidal momentum satisfies $\{f,p_{\varphi s}\}_s^{\text{gc}}=\partial_\phi f$, where $\partial_\phi$ is the toroidal angle derivative.}. If the background magnetic field has additional symmetries, a similar proof of the conservation of the corresponding total momentum can easily be constructed. The time derivative of Eq.~(\ref{gy_Pphi}) yields
\begin{align}
\frac{dP_\phi}{dt}&=\sum_{s,\bar{s}}\int p_{\phi s}C_{s\bar{s}}^{\text{gy}}(F_s,F_{\bar{s}})\,dz_s^{\text{gc}}=\sum_{s,\bar{s}}\dot{P}_{\phi s\bar{s}},
\end{align}
where $P_\phi$ is conserved exactly by the gyrokinetic Vlasov-Poisson system. Here, we find
\begin{align}
&\dot{P}_{\phi s\bar{s}}+\dot{P}_{\phi\bar{s}s}=\nonumber\\
&\frac{\Gamma_{s\bar{s}}}{2}\iint(\{\bm{y}_s,p_{\phi s}\}^{\text{gc}}_s-\{\bm{y}_{\bar{s}},p_{\phi\bar{s}}\}^{\text{gc}}_{\bar{s}})^\dagger\mathbb{Q}^{s\bar{s}}_{\text{gy}}\bm{A}_{s\bar{s}}^{\text{gy}}\delta_{s\bar{s}}^{\text{gy}}\,d\bar{z}_{\bar{s}}^{\text{gc}}\,dz_s^{\text{gc}}.
\end{align}
Now using the fact that $p_{\phi s}$ is the generator of infinitesimal toroidal rotations, we can see that $\{\bm{y}_s,p_{\phi s}\}^{\text{gc}}_s=e_z\times\bm{y}_s$, where $e_z$ is the unit vector along the axis of rotation. Therefore the vector quantity $(\{\bm{y}_s,p_{\phi s}\}^{\text{gc}}_s-\{\bm{y}_{\bar{s}},p_{\phi\bar{s}}\}^{\text{gc}}_{\bar{s}})\,\delta_{s\bar{s}}^{\text{gy}}=e_z\times(\bm{y}_s-\bm{y}_{\bar{s}})\,\delta_{s\bar{s}}^{\text{gy}}=0$, which follows from standard $\delta$-function properties. This shows that $\dot{P}_{\phi s\bar{s}}+\dot{P}_{\phi\bar{s}s}=0$, which in turn implies total toroidal momentum conservation $dP_{\phi}/dt=0$.

\section{Entropy production}As we have discussed, these conservation laws ensure that the gyrokinetic Landau-Poisson system is consistent with the the First Law of Thermodynamics. On the other hand, they do not directly imply that the gyrokinetic Landau-Poisson system is consistent with the Second Law of Thermodynamics. To verify that entropy is indeed a non-decreasing function of time, we have computed the time 
derivative of $S=-\sum_s\int F_s\text{ln} F_s\,dz_s^{\text{gc}}$ and found  
\begin{align}
\frac{dS}{dt}=\frac{\Gamma_{s\bar{s}}}{2}\iint \frac{1}{F_sF_{\bar{s}}}(\bm{A}_{s\bar{s}}^{\text{gy}})^\dagger\mathbb{Q}^{s\bar{s}}_{\text{gy}}\bm{A}_{s\bar{s}}^{\text{gy}}\delta_{s\bar{s}}^{\text{gy}}\,d\bar{z}_{\bar{s}}^{\text{gc}}\,dz_s^{\text{gc}}.\label{S_dot}
\end{align} 
Because $\mathbb{Q}^{s\bar{s}}_{\text{gy}}$ is a positive semi-definite matrix and the distribution function is positive \footnote{Positivity of the distribution function is also guaranteed by the positive semi-definiteness of $\mathbb{Q}^{s\bar{s}}_{\text{gy}}$.}, the right-side of Eq.~(\ref{S_dot}) is non-negative, which is the desired result.

Note that this proves one ``half" of a gyrokinetic version of Boltzmann's $H$-theorem. The missing ingredient is a complete characterization of the distributions that satisfy $dS/dt=0$, i.e. the gyrokinetic Maxwellians. Because the guiding center Poisson bracket is rather complicated, we have not yet found a complete characterization. However, we have verified that the distribution
\begin{align}
F_{Ms}=\frac{1}{Z_s}\exp\bigg(-\frac{H_{s}^{\text{gy}}}{T}\bigg),
\end{align}
where $Z_s=\int \exp(-H_s^{\text{gy}}/T)\,dz_s^{\text{gc}}$ is the partition function, maximizes the entropy. We leave the characterization of the most general gyrokinetic Maxwellian, which would be useful for the sake of deriving dissipative gyrofluid models with exact conservation laws \cite{Madsen_gyrofluid_2013}, as a topic for future study.

\section{Gyroaveraging} When the collision frequency is much smaller than the gyrofrequency \cite{Brizard_2004}, the full gyrokinetic Landau operator\,(\ref{gyro_landau}) can be replaced with that operator's gyroaverage, $\langle C_{s\bar{s}}^{\text{gy}}\rangle$. When this is done, the gyrokinetic Landau-Poisson system becomes the gyroaveraged Landau-Poisson system,
\begin{align}
\partial_t F_s+\{F_s,H_s^{\text{gy}}\}_s^{\text{gc}}&=\sum_s\langle C^{\text{gy}}_{s\bar{s}}(F_s,F_{\bar{s}})\rangle\\
\nabla\cdot\bm{P}&=\rho(F),
\end{align} 
where $F_s$ is now interpreted as the gyroaveraged part of the distribution function. Because the functions $H_{s}^{\text{gy}}$ and $p_{\phi s}$ are independent of the gyrophase, the proofs of energy and momentum conservation given earlier work with $C_{s\bar{s}}^{\text{gy}}$ replaced by $\langle C_{s\bar{s}}^{\text{gy}}\rangle$. Thus, the gyroaveraged Landau-Poisson system has exact energy and momentum conservation laws.

\section{Linearization} Closely related to the gyroaveraged Landau-Poisson system is the collisionally-linear gyroaveraged Landau-Poisson system, 
\begin{align}
\partial_t  F_s+\{ F_s,H_s^{\text{gy}}\}_s^{\text{gc}}&=\sum_{\bar{s}}\left(\delta C_{s\bar{s}}^{\text{test}} \;+\frac{}{}
\delta C_{s\bar{s}}^{\text{field}}\right), \\
\nabla\cdot\bm{P}&=\rho(F),
\end{align} 
where the linearized test-particle and field-particle collision operators are
\begin{align}
\delta C_{s\bar{s}}^{\text{test}}(F_s)&=\langle C^{\text{gy}}_{s\bar{s}}(F_s,F_{M\bar{s}})\rangle, \label{deltaC_test} \\
\delta C_{s\bar{s}}^{\text{field}}( F_{\bar{s}})&=\langle C^{\text{gy}}_{s\bar{s}}( F_{Ms}, F_{\bar{s}})\rangle. \label{deltaC_field}
\end{align}
This system of equations is obtained from the gyroaveraged Landau-Poisson system by assuming $F_s=F_{Ms}+\delta F_{s}$ and then dropping the non-linear term in the collision operator, $\langle C_{s\bar{s}}^{\text{gy}}(\delta F_s,\delta F_{\bar{s}})\rangle$. Note that $\langle C_{s\bar{s}}^{\text{gy}}( F_{Ms}, F_{M\bar{s}})\rangle=0$ \footnote{Note that this identity does not contradict the message presented in \cite{Madsen_2013}. In that reference, the gyrokinetic Maxwellian is defined using only the lowest-order gyrocenter Hamiltonian.}.  Because the gyrokinetic Landau operator satisfies the identities\,(\ref{gk_identity}), it is straightforward to prove that these equations have the same conservation laws for energy and momentum as the gyroaveraged Landau-Poisson system.

\section{Concluding remarks} The key to deriving an energetically-consistent formulation of collisional gyrokinetics was first expressing the particle-space Landau operator in terms of Poisson brackets ``as much as possible," which was an idea first championed by Brizard in \cite{Brizard_2004}. In particular, the identity
\begin{align}
\bm{v}-\bar{\bm{v}}=\{\bm{x},H_s\}(z)-\{\bm{x},H_{\bar{s}}\}(\bar{z})
\end{align}
suggests that the appropriate definition of the gyrocenter velocity difference is given by Eq.\,(\ref{gy_vel_dif}). This idea, together with the procedure given earlier for determining the energetic consistency constraints, appears to be appropriate for deriving energetically-consistent collision operators for other reduced plasma models as well. In future work, we will report on the energy-conserving collisional formulations of electromagnetic gyrokinetics and oscillation center theory. 

We note that, although the gyrokinetic Landau operator (\ref{gyro_landau}) and its linearized forms (\ref{deltaC_test})-(\ref{deltaC_field}) may prove difficult to implement numerically, they identify the proper formalism for the inclusion of collisional transport in gyrokinetic theory. Hence, these gyrokinetic collision operators form the basis from which approximations can be implemented for practical applications.

Lastly, by setting $\varphi=0$ in the above formulas, our results reduce to an energy-momentum-conserving guiding center collision operator. This operator would be ideally suited to incorporating collisions into orbit-following codes such as ORBIT\,\cite{White_1984}; see \cite{Hirvijoki_2013} for recent work on the Monte Carlo implementation of a 5D guiding center Fokker-Planck collision operator. All previous guiding center collision operators that have been applied in orbit-following codes either resort to \emph{ad hoc} methods to ensure exact conservation laws \cite{Boozer_collisions_1981}, or else do not fully account for inhomogeneities in the magnetic field\,\cite{Tessarotto_1994}.

\section{Acknowledgements} The results presented in this chapter were obtained in collaboration with Professor Alain Brizard. They can also be found on the arXiv at arXiv:1503.07185.

\chapter[Hamiltonian mechanics of stochastic acceleration]{Hamiltonian mechanics of stochastic acceleration}
\label{ch:stoch_ham}
%
%
%
%
%
%

\section{General theory}
\subsection{Introduction}The term ``stochastic acceleration" refers to the chaotic motion of particles subjected to a prescribed random force.  Such motion occurs in myriad contexts; the turbulent electromagnetic fields present in the interstellar medium and the RF wave fields found in magnetic fusion devices are just two examples. In the astrophysical context, it is thought to be partially responsible for the presence of cosmic rays in our solar system \cite{Fermi_1949}. In the magnetic fusion context, it might explain the presence of certain high-energy tails observed in the National Spherical Torus Experiment when neutral beams are fired into RF-heated plasmas \cite{Liu_2009}. 


Robust modeling of stochastic acceleration requires statistical approaches. The dominant approach is to employ the Fokker-Planck equation \cite{Sturrock_1966,Hall_1967,Barbosa_1979,Petrosian_2004,Hamilton_1992} for the one-particle distribution function. However, when studying Richardson dispersion \cite{Richardson_1926,Jullien_1999}, and more generally any phenomenon governed by the two-particle distribution function \cite{Mukhopadhyay_2012}, the one-particle Fokker-Planck equation is insufficient. This is because spatial correlations in the random force field prevent the two-particle distribution function from factoring as a product of one-particle distribution functions. A superior statistical model when multi-particle statistics are in question would be a Langevin equation for particle trajectories. A wisely-chosen Langevin equation could capture the physics of the one- and two-particle distribution functions while providing an attractive means to perform Monte Carlo simulations of stochastic acceleration. Currently, there are no satisfactory methods for finding such a Langevin equation. 

The purpose of this Chapter is to describe, for the first time, a systematic procedure for passing from a microscopic description of stochastic acceleration in terms of Hamiltonian equations of motion to the physically-correct Langevin equation for particle trajectories in the long-time limit. We will also show that, aside from reproducing the correct multi-particle statistics, this Langevin equation inherits the Hamiltonian structure of the microscopic dynamics. Specifically, we will show that the Langevin equation is a Hamiltonian stochastic differential equation (SDE) \cite{Lazaro-Cami_2008}. Thus, this work proves that symmetries of the macroscopic physical laws governing stochastic acceleration lead to conservation laws. 

We will focus our attention on stochastic acceleration problems similar to those studied in \cite{Sturrock_1966,Hall_1967,Barbosa_1979,Petrosian_2004}. These consist of a collection of non-interacting particles moving through a prescribed Hamiltonian force field. By assumption, the force will consist of a small-amplitude perturbation superimposed over a time-independent background. The perturbed force felt by a particle will be assumed to have a correlation time much shorter than any bounce time associated with the perturbation, zero mean, and temporally homogeneous statistics. These assumptions preclude treating Coulomb collisions because the polarization field produced by a particle cannot be modeled as a prescribed field; the polarization force depends on the history of a particle's orbit. They also preclude the treatment of strong turbulence \cite{DuBois_1978}.  


\subsection{The main idea} Mathematically, this type of problem can be described as follows. Each particle moves through a $2n$-dimensional single-particle phase space $M$ according to a dynamical law given by a time-dependent vector field $X_t$; if $z_t\in M$ denotes the trajectory of a particle in $M$, then 
\begin{align}\label{micro}
\dot{z_t}=X_t(z_t).
\end{align} 
Because the only forces present are Hamiltonian, $X_t$ must be Hamiltonian in the sense that there is some Poisson bracket $\{\cdot,\cdot\}$ and some time-dependent Hamiltonian, $H_t$, such that  $\dot{z}^i=\{z^i,H_t\}$, where $z^i$ denotes an arbitrary coordinate system on $M$ \cite{Grebogi_1979}. By standard mathematical convention, this is written $X_t=X_{H_t}$  \cite{Abraham_2008}. The presumed form of the force then implies $H_t=H_0+\epsilon h_t$, where $\epsilon\ll1$, $H_0$ describes the mean time-independent background, and $h_t$ describes the small-amplitude random perturbation. Moreover, $X_{h_t}$ evaluated on a particle trajectory must have a correlation time $\tau_{\text{ac}}$ much shorter than some constant $\tau$, which, in turn, is much shorter than any bounce time associated with the perturbation $\tau_b$, $\tau_{\text{ac}}\ll\tau\ll\tau_b$.

Our goal in this Chapter is to find the correct coarse-grained version of the microscopic equations of motion, $X_{H_t}$. Specifically, we seek a Langevin equation in the form 
\begin{align}\label{ansatz}
\delta z_t=X_0(z_t)\,\mathrm{d}t+\sum_{k\geq 1}X_k(z_t)\,\delta W^k_t
\end{align}
whose solutions correctly reproduce the late-time statistical behavior of solutions to the microscopic equations of motion.
Here $X_k$ are vector fields on $M$ that must be determined, $W^k$ are independent ordinary Wiener processes, and $\delta$ denotes the Stratonovich differential \cite{Gardiner_2009} (sometimes also written $\circ \mathrm{d}$). We will identify the $X_k$ by demanding that Eq.\,(\ref{ansatz}) possess two properties: it must generate the Fokker-Planck equations for the one- and two-particle distribution functions, $f_t(z)$ and $g_t(z_1,z_2)$. The two-particle distribution function is defined such that the probability particle $1$ is in the region $U_1\subset M$ and particle 2 is in the region $U_2\subset M$ at time $t$ is given by $\int_{U_1}\int_{U_2}g_t\,\mathrm{d}z_1\,\mathrm{d}z_2$, where $\mathrm{d}z$ denotes the Liouville measure \cite{Abraham_2008}. Baxendale \cite{Baxendale_1984} has proven that a Langevin equation is uniquely determined by its one- and two- particle Fokker-Planck equations. Therefore, these conditions uniquely specify the Langevin equation we seek. In particular, the requirement that two-particle statistics be accurately reproduced is critical; Baxendale's work implies that constraining the Langevin equation only to be consistent with the one-particle Fokker-Planck equation would not identify it uniquely.

Physically, the reason that the two-particle Fokker-Planck equation contains more information than the one-particle Fokker-Planck equation can be understood as follows. After a short amount of time $\Delta t$, the displacement of a particle initially located at $z_1$ at time $t$ is given approximately by $\Delta t\, X_t(z_1)$. Similarly, the displacement of a particle initially located at $z_2$ is nearly $\Delta t\, X_t(z_2)$. Because the random force field generally has spatial correlations, $X_t(z_1)$ and $X_t(z_2)$ are not statistically independent. Thus, the probability distribution of $(z_1^\prime,z_2^\prime)$, where $z_i^\prime\approx z_i+ \Delta t \,X_t(z_i)$, will not be given by the product of the distribution of $z_1^\prime$ with that of $z_2^\prime$. This failure-to-factor precludes determining the two-particle distribution function from the mere knowledge of the one-particle distribution function. Note that this is true in spite of the fact that these particles do not \emph{interact}; because the random force is assumed to be prescribed, the time-evolution of $z_1$ is decoupled from the time-evolution of $z_2$.


\subsection{Identifying the Langevin equation} The one-particle Fokker-Planck equation associated with Eq.\,(\ref{ansatz}) is given by \cite{Gardiner_2009,Baxendale_1984}
\begin{align}\label{fok_one_gen}
\frac{\partial f_t}{\partial t}&=-\text{div}(f_tX_0)+\frac{1}{2}\sum_{k\geq 1}\text{div}(\text{div}(f_t X_k)X_k)\nonumber\\
&=A_1f_t,                                     
\end{align}
while the two-particle Fokker-Planck equation \cite{Baxendale_1984,Schmalfuss_2001,Kunita_1987} is given by
\begin{align}\label{fok_two_gen}
\frac{\partial g_t}{\partial t}=&A_1^{(1)}g_t+A_1^{(2)}g_t\nonumber\\
                                           &+\sum_{k\geq 1}\text{div}^{(1)}\text{div}^{(2)}:g_tX_k(z_1)\otimes X_k(z_2).
\end{align}
The divergence operators in these expressions are defined relative to the Liouville volume form and the colon indicates the full contraction of second-rank tensors, $a:b\equiv a^{ij}b_{ij}$. Because these equations follow from Eq.\,(\ref{ansatz}) \emph{via} rigorous mathematics, we will refer to them as the mathematical Fokker-Planck equations.

On the other hand, under our assumption that the correlation time of the perturbed force is much shorter than a bounce time, standard coarse-graining procedures \cite{Risken_1996,Bazant_2006} together with a decomposition theorem for time-ordered exponentials \cite{Lam_1998} lead to the late-time evolution laws for the one- and two-particle distribution functions associated with the microscopic equations of motion, Eq.\,(\ref{micro}). The physical one-particle Fokker-Planck equation is given by 
\begin{align}\label{one_particle_fok}
\frac{\partial f_t}{\partial t}&=-\left\{f_t,H_0+\frac{\epsilon^2}{\tau}\mathbb{E}[s_2]\right\}+\frac{\epsilon^2}{2\tau}\mathbb{E}[\left\{\left\{f_t,s_1\right\},s_1\right\}]\nonumber\\
&=A_1f_t,                     
\end{align}  
while the physical two-particle Fokker-Planck equation (see the supplementary material for a derivation) is given by
\begin{align}\label{two_particle_fok}
\frac{\partial g_t}{\partial t}=A_1^{(1)}g_t+A_1^{(2)}g_t+\frac{\epsilon^2}{\tau}\mathbb{E}[\alpha:\mathbf{d}^{(1)}\mathbf{d}^{(2)}g_t].
\end{align}
The notation introduced in these two equations is defined as follows: $\mathbb{E}$ denotes an expectation value; the functions $s_1,s_2$ are defined by
\begin{subequations}
\label{kick}
\begin{align}
&s_1=\int_0^\tau\!\exp(\lambda X_{H_0})_{*}h_{\tau-\lambda}\mathrm{d}\lambda\label{s1}\\
&s_2=\frac{1}{2}\!\int_0^\tau\!\!\!\!\int_0^a\!\{\exp(bX_{H_0})_{*}h_{\tau-b},\exp(aX_{H_0})_{*}h_{\tau-a}\}\mathrm{d}b\,\mathrm{d}a;\label{s2}
\end{align}
\end{subequations}
$\exp(Y):M\rightarrow M$ denotes the time-one advance map of the dynamical system defined by the vector field $Y$; $(\exp(Y)_{*}h)(z)\equiv h(\exp(-Y)(z))$; the superscripts indicate which argument of $g_t$ that $A_1$ and the exterior derivative $\mathbf{d}$ should be applied to; and $\alpha(z_1,z_2)\equiv\mathbb{E}[X_{s_1}(z_1)\otimes X_{s_1}(z_2)]$ is the two-point covariance tensor. 

The $X_k$ must be chosen so that the mathematical Fokker-Planck equations, Eqs.\,(\ref{fok_one_gen}) and (\ref{fok_two_gen}), are equivalent to the physical Fokker-Planck equations, Eqs.\,(\ref{one_particle_fok}) and (\ref{two_particle_fok}). However, a direct comparison of these two pairs of equations is difficult with Eqs.\,(\ref{one_particle_fok}) and (\ref{two_particle_fok}) in their current form. To eliminate this issue, we will obtain a special decomposition of the two-point covariance tensor $\alpha(z_1,z_2)$. 

As a first step, notice that if we fix a one-form $\xi\in T_{z_1}^*M$, then we can define a vector field $Y_\xi$ on $M$ by contracting $\xi$ with $\alpha$ on the left according to 
\begin{align}\label{ydefined}
Y_\xi(z_2)&=\alpha(z_1,z_2)(\xi,\cdot)\nonumber\\
&=\mathbb{E}[\xi(X_{s_1}(z_1))X_{s_1}(z_2)].
\end{align}
By forming all possible linear combinations of vector fields of this form, we can construct a (potentially infinite dimensional) linear space of vector fields \cite{Aronszajn_1950, Baxendale_1976}, which we will denote $\mathcal{H}$,
\begin{align}\label{defH}
\mathcal{H}=\{\text{linear combinations of }Y_\xi,~\xi\in T^*M\}.
\end{align}
Because each $Y_\xi$ is of the form $Y_\xi(z)=X_{\bar{H}}(z)$ with $\bar{H}(z)=\mathbb{E}[\xi(X_{s_1}(z_o)){s_{1}}(z)]$, and the sum of Hamiltonian vector fields is again Hamiltonian, $\mathcal{H}$ consists entirely of Hamiltonian vector fields. Moreover, following Baxendale \cite{Baxendale_1984, Baxendale_1976}, we see that $\mathcal{H}$ is a real Hilbert space whose inner product is defined by the formula
\begin{align}\label{ip}
\left<Y_\xi,Y_\eta\right>_{\mathcal{H}}&=\alpha(z_1,z_2)(\xi,\eta)\nonumber\\
                                      &=\mathbb{E}[\xi(X_{s_1}(z_1))\eta(X_{s_1}(z_2))],
\end{align}
where $\xi\in T^*_{z_1}M$ and $\eta\in T^*_{z_2} M$. Therefore we may choose an orthonormal basis $\{e_k\}_{k\ge 1}$ for $\mathcal{H}$, where each $e_k$ must be of the form $e_k=X_{H_k}$. A simple calculation then leads to the desired decomposition of $\alpha$:
\begin{align}\label{decomp}
\alpha(z_1,z_2)=\sum_{k\ge 1}X_{H_k}(z_1)\otimes X_{H_k}(z_2).
\end{align}

Using this decomposition of the two-point covariance tensor, it is straightforward to manipulate Eqs.\,(\ref{one_particle_fok}) and (\ref{two_particle_fok}) into the same form as Eqs.\,(\ref{fok_one_gen}) and (\ref{fok_two_gen}). After doing so, it is trivial to identify the correct $X_k$. Indeed, we have found that the physical Langevin equation is given by
\begin{align}\label{result}
\delta z_t=X_{\tilde{H}_0}(z_t)\,\mathrm{d}t+\sum_{k\geq 1}X_{\tilde{H}_k}(z_t)\,\delta W^k_t,
\end{align}
where
\begin{align}
\tilde{H}_0=H_0+\frac{\epsilon^2}{\tau}\mathbb{E}[s_2],~~~\tilde{H}_k&=\frac{\epsilon}{\sqrt{\tau}}H_k\label{hks}
\end{align}
Recall that the $X_{H_k}$ are defined to be an orthonormal basis of the Hilbert space $\mathcal{H}$ defined in Eq.\,(\ref{defH}). Also recall that all of the above manipulations have been performed under the assumption that the correlation time of the perturbed force felt by a particle is much shorter than any bounce time associated with the perturbation.

Because the coefficients in the Langevin equation for stochastic acceleration, Eq.\,(\ref{result}), are all Hamiltonian vector fields, this equation is an example of a \emph{stochastic Hamiltonian system}, the foundations of which are developed in \cite{Lazaro-Cami_2008}. It is in this sense that the Langevin equation for stochastic acceleration inherits the Hamiltonian structure of the microscopic equations. In particular, SDEs of this type are known to arise from a stochastic variational principle for which Noether's theorem applies. Thus, even at the dissipative macroscopic level, symmetries imply the presence of conservation laws.


\subsection{Example 1} We will find the physical Langevin equation for two example stochastic acceleration problems. Generally speaking, finding the coefficients of the physical Langevin equation involves finding an orthonormal basis for the space $\mathcal{H}$, a task which may be analytically intractable. But, by Mercer's theorem \cite{Mercer_1909}, this task can be cast as an eigenvalue problem for which there are existing numerical solution methods. In any case, in these examples, the analytical route is tractable. 

First, consider a single-species, unmagnetized plasma subjected to a random weak electrostatic pulse at $\tau$-second intervals. Assume that the pulses are uniform in space and constant in magnitude, but uniformly and independently distributed in direction. Thus, the $k$'th pulse is generated by a potential of the form $\phi_k(\bm{x},t)=(\bm{z}_k\cdot \bm{x})\phi_o u(t-k\tau)$, where $\bm{z}_k$ is a random vector uniformly distributed over the unit sphere and $u(t)$ is a temporal windowing function localized at $t=\tau/2$.

In order to find the Langevin equation governing the plasma dynamics at times much longer than $\tau$, we must (a) calculate $s_1$ and $s_2$ using Eqs.\,(\ref{s1}) and (\ref{s2}), (b) find an orthonormal basis $\{X_{H_k}\}_{k\geq1}$ for the space $\mathcal{H}$ defined in Eq.\,(\ref{defH}), and (c) write down Eq.\,(\ref{result}) with $\tilde{H}_0$ and $\tilde{H}_k$ calculated using Eq.\,(\ref{hks}). The results of these three steps are as follows.

(a) A quick calculation shows that 
\begin{subequations}
\begin{align}
s_1&=m_o\bm{z}\cdot \bm{x}-m_1\bm{z}\cdot \bm{v}\\ 
s_2&=\text{const}
\end{align}
\end{subequations}
where $m_o=(q/m)\phi_o\int_0^\tau u(s)\mathrm{d}s$, $m_1=(q/m)\phi_o\int_0^\tau(\tau-s)u(s)\mathrm{d}s$, and $q/m$ is the charge-to-mass ratio.

(b) Each $Y_{\xi}$ must be of the form $Y_\xi=X_{g_{\bm{\beta}\bm{\gamma}}}$, where
\begin{align}
g_{\bm{\beta}\bm{\gamma}}(\bm{x},\bm{v})=\frac{1}{3}(m_1\bm{\beta}+m_o\bm{\gamma})\cdot(m_1\bm{v}-m_o\bm{x}),
\end{align}
and $\bm{\beta},\bm{\gamma}$ are arbitrary constant 3-component vectors. Using this expression, it is simple to find an orthonormal basis for $\mathcal{H}$. One is given by $\{X_{\bar{H}_k}\}_{k=1..3}$, with
\begin{align}
H_i(\bm{x},\bm{v})=\frac{1}{\sqrt{3}}e_i\cdot(m_1\bm{v}-m_o\bm{x}),
\end{align}
where $\{e_i\}_{i=1..3}$ is the standard basis for $\mathbb{R}^3$.

(c) Finally, the physical Langevin equation is given by
\begin{subequations}
\label{phys_lang}
\begin{align}
\delta x^i=v^i\,\mathrm{d}t+\frac{1}{\sqrt{3\tau}}m_1\,\delta W^i\\
\delta v^i=\frac{1}{\sqrt{3\tau}}m_o\,\delta W^i,
\end{align}
\end{subequations}
where $i=1,2,3$.

As is readily verified, the one-particle Fokker-Planck equation for this SDE is given by
\begin{align}\label{fpe}
\frac{\partial f_t}{\partial t}+v\cdot\nabla f_t=\frac{1}{6\tau}&(m_1^2\nabla^2f_t+m_om_1\nabla\cdot\nabla_vf_t\nonumber\\
&+m_om_1\nabla_v\cdot\nabla f_t+m_o^2\nabla_v^2f_t).
\end{align}
On the other hand, given an arbitrary function $\phi(\bm{x},\bm{v})$, the  SDE 
\begin{subequations}
\label{counter}
\begin{align}
\delta x^i=&v^i\,dt+\frac{m_1}{\sqrt{3\tau}}\left(\cos(\phi) \,\delta W^{1,i}-\sin(\phi) \,\delta W^{2,i}\right)\\
\delta v^i=&\frac{m_o}{\sqrt{3\tau}}\left(\cos(\phi) \,\delta W^{1,i}-\sin(\phi) \,\delta W^{2,i}\right),
\end{align}
\end{subequations}
where the $W^{1,i},W^{2,j}$ are six independent ordinary Wiener processes, will also generate Eq.\,(\ref{fpe}). However, when $\phi$ is not constant, the two-particle Fokker-Planck equation generated by Eq.\,(\ref{counter}) will differ from the two-point Fokker-Planck equation generated by Eq.\,(\ref{phys_lang}). This can be verified using Eq.\,(\ref{two_particle_fok}). The procedure identified here selects $\phi=0$ as the physical choice. In particular, it shows that a Langevin equation with the correct one-particle Fokker-Planck equation may still incorrectly reproduce the two-particle distribution function.

The inadequacy of Eq.\,(\ref{counter}) can also be understood intuitively as follows. Chaotic motions of any two particles experiencing the electrostatic pulses are ``synchronized" since the pulses are independent of $\bm{x}$ and $\bm{v}$. The Langevin equation (\ref{counter}), on the other hand, desynchronizes particle trajectories by involving additional Wiener processes, in spite of giving the correct one-particle Fokker-Planck equation.


\subsection{Example 2} Next, consider a minority population of magnetized fast ions moving through a plane lower-hybrid wave that propagates perpendicular to the magnetic field. Assume the wave has a high harmonic number and a wavelength small compared to a typical ion gyroradius. Karney \cite{Karney_1979} has shown that the dynamics of the perpendicular velocity of these ions are governed by a canonical time-dependent Hamiltonian system with Hamiltonian
\begin{align}
H_t=I-\epsilon \sin(\sqrt{2I}\sin\theta-\nu t),
\end{align}
where $I$ is the normalized magnetic moment, $t$ the time normalized by the gyroperiod, $\theta$ the gyrophase, $\nu$ the harmonic number, and $\epsilon$ the normalized wave amplitude. Moreover, when $\epsilon$ exceeds a threshold value, an ion's motion becomes chaotic. This chaotic motion comes as the result of the effective randomization of the wave phase felt by an ion after a gyroperiod. Thus, above the threshold for chaos, we can model the wave phase as being randomized every gyroperiod by a random variable $\eta$. That is, we can replace the exact \emph{chaotic} ion motion with a stochastic approximation; see \cite{Chirikov_1979} for Chirikov's application of the same modeling approach to the standard map. This allows us to apply the formalism developed in this Chapter to find the physical Langevin equation describing the stochastic particle trajectories at times much longer than the gyroperiod.

As in the previous example, the first step is to calculate $s_1$ and $s_2$. Set $\tau=2\pi$ and adopt the rough approximation
\begin{align}\label{approx}
\sum_{n=-\infty}^\infty\frac{J_n}{\nu-n}\exp(in\theta)\approx\frac{J_{n_o}}{\delta}\exp(in_o\theta),
\end{align}
where $\nu=n_o+\delta$, $|\delta|<\frac{1}{2}$, and $J_n=J_n(\sqrt{2I})$ denotes the Bessel function of the first kind \cite{Abramowitz_1964}. This approximation amounts to selecting the most slowly varying term in the sum in Eq.\,(\ref{approx}). Then, upon directly evaluating the integrals in Eqs.\,(\ref{s1}) and (\ref{s2}), the resulting expressions for $s_1$ and $\mathbb{E}[s_2]$ are
\begin{subequations}
\begin{align}
s_1&= 2\pi\text{sinc}(\pi\delta)J_{n_o}\sin(n_o\theta+\eta)\label{s1_2}\\
\mathbb{E}[s_2]&=\frac{\pi}{2}\sum_{m=-\infty}^\infty\frac{J_{m+1}^2-J_{m-1}^2}{m-\nu}\nonumber\\
                             &~~~+\frac{\pi}{2}\text{sinc}(2\pi\delta)\frac{J_{n_o+1}^2-J_{n_o-1}^2}{\delta},
\end{align}
\end{subequations}
where $\eta$ is a random variable uniformly distributed over the interval $[0,2\pi]$ and $\text{sinc}(x)=\sin(x)/x$.

Next, the space $\mathcal{H}$ can be constructed using the above expression for $s_1$. In this case, $\mathcal{H}$ is two-dimensional and has a basis $\{X_{H_1},X_{H_2}\}$, where
\begin{subequations}
\begin{align}
H_1(I,\theta)&=\sqrt{2}\pi\text{sinc}(\pi\delta) J_{n_o}(\sqrt{2I})\cos(n_o\theta)\\
H_2(I,\theta)&=\sqrt{2}\pi\text{sinc}(\pi\delta) J_{n_o}(\sqrt{2I})\sin(n_o\theta).
\end{align}
\end{subequations}

Finally, the coefficients for the Langevin equation, Eq.\,(\ref{result}), can be derived using Eq.\,(\ref{hks}). The result is
\begin{subequations}
\label{karney_sde}
\begin{align}
\delta I=&\epsilon\sqrt{\pi}\text{sinc}(\pi\delta)n_oJ_{n_o}(\sqrt{2I})\nonumber\\
&\times\left(\sin(n_o\theta)\delta W^1-\cos(n_o\theta)\delta W^2\right)\\
\delta \theta=&\left(1+\frac{\epsilon^2}{2\pi}\frac{\partial}{\partial I}\mathbb{E}[s_2]\right)\mathrm{d}t\nonumber\\
&+\bigg(\epsilon\sqrt{\frac{\pi}{2I}}\text{sinc}(\pi\delta)J_{n_o}^\prime(\sqrt{2I})\nonumber\\
&~~\times\left(\cos(n_o\theta)\delta W^1+\sin(n_o\theta)\delta W^2\right)\bigg).
\end{align}
\end{subequations}
The diffusion of the magnetic moment $I$ predicted by Eq.\,(\ref{karney_sde}) has already been studied by Karney \cite{Karney_1979}. However, Eq.\,(\ref{karney_sde}) extends and compliments Karney's results by predicting the appropriate diffusion in gyrophase, as well as the correct two-particle statistics.


\subsection{Concluding remarks} We have shown how to derive the physical Langevin equation for particle trajectories undergoing stochastic acceleration. This SDE correctly generates the correct one- and two-particle Fokker-Planck equations and inherits the Hamiltonian structure of the microscopic equations of motion. This inheritance is theoretically satisfying because it is a direct consequence of demanding consistency with the physical one- and two-particle Fokker-Planck equations. It also implies that symmetries of the macroscopic physical laws governing stochastic acceleration imply the presence of conservation laws. While this relationship is well known at the microscopic level, it is a pleasant surprise that it remains intact upon passing to dissipative macroscopic equations.

A Hamiltonian Langevin equation \cite{Lazaro-Cami_2008} is a Stratonovich SDE of the form given in Eq.\,(\ref{result}). If a loop of initial conditions for this SDE evolves under a given realization of the noise, then the action of that loop is constant in time. In addition, these equations arise from a stochastic action principle \cite{Lazaro-Cami_2008} for which Noether's theorem applies. Thus, by showing the physical Langevin equation is Hamiltonian, we have also identified potentially powerful tools for the analysis of stochastic acceleration. In particular, using the methods of Bou-Rabee \cite{Bou-Rabee_2009}, the stochastic action principle can be used to develop variational integrators for Eq.\,(\ref{result}). Because these integrators are known to possess superior long-term statistical fidelity \cite{Bou-Rabee_2010}, this approach may prove to be useful in Monte Carlo simulations of stochastic acceleration.

\subsection{Derivation of the physical two-particle Fokker-Planck equation}
This supplement to the article ``The Hamiltonian mechanics of stochastic acceleration" consists of a derivation of the physical two-particle Fokker-Planck equation. In the main text, the two-particle Fokker-Planck equation is given in Eq.\,(6). The derivation will freely draw upon notation defined in the article. The essential idea behind this derivation is not novel; the same idea is presented in \cite{Bazant_2006} in the simpler context of a one-dimensional random walker.

Let $F_{t,s}$ be the time advance map \cite{Abraham_2008} associated with the dynamical vector field $X_{H_t}$ (Eq.\,(1) in our manuscript); $F_{t,s}(z)$ gives the time $t$ phase space location of a particle located at $z\in M$ at time $s$. Because this time advance map satisfies the identity $F_{t,r}\circ F_{r,s}=F_{t,s}$, where $\circ$ denotes the composition of functions, we have $F_{N\tau,0}=F_{N\tau,(N-1)\tau}\circ  F_{(N-1)\tau,(N-2)\tau}\circ...\circ F_{\tau,0}$. This decomposition of the time advance map provides a stroboscopic description of particle dynamics; as the integer $N$ increases, it tells us the phase space location of a particle at the times $t=0,t=\tau,t=2\tau,...$. 

Because $\tau\ll \tau_b$,  the results in \cite{Lam_1998} may be used to write $F_{(k+1)\tau,k\tau}=\exp(X_{s_{(k+1)\tau}})\circ\exp(\tau X_{H_0})$, where $s_t=\epsilon s_{1,t}+\epsilon^2 s_{2,t}+...$ is given to $O(\epsilon^2)$ by
\begin{align}\label{kick}
&s_{1,t}=\int_0^\tau\!\exp(\lambda X_{H_0})_{*}h_{t-\lambda}\,d\lambda\\
&s_{2,t}=\frac{1}{2}\!\int_0^\tau\!\!\!\!\int_0^a\!\{\exp(bX_{H_0})_{*}h_{t-b},\exp(aX_{H_0})_{*}h_{t-a}\}\,db\,da.\nonumber
\end{align}
Here $\exp(Y):M\rightarrow M$ is the time-one advance map of the dynamical system defined by the vector field $Y$; $\{\cdot,\cdot\}$ denotes the Poisson bracket; and $(\exp(Y)_{*}h)(z)=h(\exp(-Y)(z))$. This means that each $\tau$-second step in the discrete-time dynamics is a deterministic drift, $\exp(\tau X_{H_0})$, followed by a small random kick, $\delta_k\equiv\exp(X_{s_{(k+1)\tau}})$. Moreover, the statistical assumptions on the perturbed force imply that the various $\delta_k$ are independent identically distributed maps. Therefore, each discrete-time particle trajectory, $z_{N\tau}=F_{N\tau,0}(z)$, is a temporally-homogeneous Markov process.

In terms of $F_{t,s}$, the trajectory of a pair of particles at $\bm{z}=(z_1,z_2)\in M\times M$ at time $s$ is given by
\begin{align}
\bm{z}_t=(F_{t,s}(z_1),F_{t,s}(z_2)).
\end{align}
This motivates introducing the two-particle time-advance map, $\mathbf{F}_{s,t}:M\times M\rightarrow M\times M$, which is defined by the formula
\begin{align}
\mathbf{F}_{t,s}(\bm{z})=(F_{t,s}(z_1),F_{t,s}(z_2)).
\end{align}
The two-particle time-advance map inherits many of the qualitative features of the one-particle time-advance map. In particular, $\mathbf{F}_{t,r}\circ\mathbf{F}_{r,s}=\mathbf{F}_{t,s}$, which implies that $\mathbf{F}_{N\tau,0}=\mathbf{F}_{N\tau,(N-1)\tau}\circ  \mathbf{F}_{(N-1)\tau,(N-2)\tau}\circ...\circ \mathbf{F}_{\tau,0}$. Moreover, $\mathbf{F}_{(k+1)\tau,k\tau}$ admits the decomposition $\mathbf{F}_{(k+1)\tau,k\tau}=\bm{\delta}_k\circ\mathbf{exp}(\tau \bar{X}_{H_0})$, where
\begin{align}
\bm{\delta}_k(\bm{z})=(\delta_k(z_1),\delta_k(z_2)),
\end{align}
and $\mathbf{exp}(\tau\bar{X}_{H_0})$ is the time-$\tau$ advance map associated with the vector field on $M\times M$ given by the formula
\begin{align}
\bar{X}_{H_0}(\bm{z})=X_{H_0}(z_1)\oplus X_{H_0}(z_2)\in T_{z_1}M\oplus T_{z_2}M.
\end{align}

The time-homogeneous Markov property implies that the linear operator $P_N$ defined on two-particle observables $\mathcal{Q}:M\times M\rightarrow\mathbb{R}$ by the formula $(P_N\mathcal{Q})(z)=\mathbb{E}[\mathcal{Q}(\mathbf{F}_{N\tau,0}(z))]$ satisfies the semigroup property $P_{N+M}=P_{N}P_M$. Therefore, if we define the time evolution of a two-particle observable as $\mathcal{Q}_{N\tau}=P_N\mathcal{Q}$, then $\frac{1}{\tau}(\mathcal{Q}_{(N+1)\tau}-\mathcal{Q}_{N\tau})=\frac{1}{\tau}(P_1 -1)\mathcal{Q}_{N\tau}$. For times sufficiently large compared with $\tau$, intuition suggests that the left-hand side of this identity approaches the partial time derivative $\partial \mathcal{Q}_t/\partial t$ and that $\frac{1}{\tau}(P_1-1)$ may be approximated by a differential operator. This intuition can be made precise through the use of a Kramers-Moyal expansion \cite{Risken_1996,Bazant_2006}, which we will describe now. Following \cite{Bazant_2006}, we will obtain this limiting partial differential equation by scaling the time variable by the appropriate power of $\epsilon$ and looking for a dominant balance of the equation
\begin{align}\label{perfect}
\frac{1}{\tau}(\mathcal{Q}_{(N+1)\tau}-\mathcal{Q}_{N\tau})=\frac{1}{\tau}(P_1 -1)\mathcal{Q}_{N\tau}
\end{align}
as $\epsilon\rightarrow 0$. 

First we eliminate the mean drift associated with the background fields by working with the function 
\begin{align}
\bar{\mathcal{Q}}_t(\bm{z})=\bigg(\mathbf{exp}(-t\bar{X}_{H_o})^*\mathcal{Q}_t\bigg)(\bm{z})\equiv \mathcal{Q}_t\bigg(\mathbf{exp}(-t \bar{X}_{H_o})(\bm{z})\bigg).
\end{align}
instead of $\mathcal{Q}_t$. In terms of $\bar{\mathcal{Q}}_t$ and $t=N\tau$, Eq.\,(\ref{perfect}) becomes
\begin{align}\label{km}
&\bar{\mathcal{Q}}_t-\mathbf{exp}(-\tau \bar{X}_{H_o})^*\bar{\mathcal{Q}}_t+\sum_{k=1}^\infty \frac{\partial^k \bar{\mathcal{Q}}_t}{\partial t^k}\tau^k=\\
&\bar{\mathcal{Q}}_t-\mathbf{exp}(-\tau \bar{X}_{H_o})^*\bar{\mathcal{Q}}_t+\mathbb{E}\bigg[\epsilon[\bar{\mathcal{Q}}_t,\bar{\bm{s}}_{1}]+\epsilon^2([\bar{\mathcal{Q}}_t,\bar{\bm{s}}_2]+[[\bar{\mathcal{Q}}_t,\bar{\bm{s}}_1],\bar{\bm{s}}_1]/2)\bigg]+O(\epsilon^3),\nonumber
\end{align}
where 
\begin{align}
\bar{\bm{s}}_1&=\mathbf{exp}(-t \bar{X}_{H_o})^*\left(\pi_1^*s_{1,\tau}+\pi_2^*s_{1,\tau}\right)\\
\bar{\bm{s}}_2&=\mathbf{exp}(-t \bar{X}_{H_o})^*\left(\pi_1^*s_{2,\tau}+\pi_2^*s_{2,\tau}\right);
\end{align}
the bracket $[\cdot,\cdot]$ is the Poisson bracket on $M\times M$ defined by the formula
\begin{align}
[f,g](z_1,z_2)=\{f(z_1,\cdot),g(z_1,\cdot)\}(z_2)+\{f(\cdot,z_2),g(\cdot,z_2)\}(z_1);
\end{align} 
and $\pi_1,\pi_2:M\times M\rightarrow M$ are the projection maps onto the first and second factor respectively.  The Taylor expansion in time is the key step here. It is justified by the fact that we will be considering late times when the evolution of $\bar{\mathcal{Q}}_t$ has had time to slow down as a result of diffusion.  Notice that because $\mathbb{E}[s_{1,t}]=0$, $\mathbb{E}[\bar{\bm{s}}_1]=0$ as well. 

Next we set $t=(\tau/\epsilon^2)\lambda$, where $\lambda$ is a renormalized dimensionless time. As $\epsilon\rightarrow 0$, the dominant balance of Eq.\,(\ref{km}) is given by
\begin{align}
\frac{\partial \bar{\mathcal{Q}}_\lambda}{\partial \lambda}=\mathbb{E}\bigg[[\bar{\mathcal{Q}}_\lambda,\bar{\bm{s}}_2]+[[\bar{\mathcal{Q}}_\lambda,\bar{\bm{s}}_1],\bar{\bm{s}}_1]/2\bigg].
\end{align}
Or, in terms of $t$ and $\mathcal{Q}_t=\mathbf{exp}(t \bar{X}_{H_o})^*\bar{\mathcal{Q}}_t$,
\begin{align}
\frac{\partial \mathcal{Q}_t}{\partial t}&=[\mathcal{Q}_t,\pi_1^*H_o+\epsilon^2\mathbb{E}[\pi_1^*s_{2,\tau}]/\tau]+\frac{\epsilon^2}{2\tau}\mathbb{E}\big[[\mathcal{Q}_t,\pi_1^*s_{1,\tau}],\pi_1^*s_{1,\tau}]\big]\nonumber\\
&+[\mathcal{Q}_t,\pi_2^*H_o+\epsilon^2\mathbb{E}[\pi_2^*s_{2,\tau}]/\tau]+\frac{\epsilon^2}{2\tau}\mathbb{E}\big[[\mathcal{Q}_t,\pi_2^*s_{1,\tau}],\pi_2^*s_{1,\tau}]\big]\nonumber\\
&+\frac{\epsilon^2}{\tau}\mathbb{E}\big[[[\mathcal{Q}_t,\pi_1^*s_{1,\tau}],\pi_2^*s_{1,\tau}]\big]\nonumber\\
&=L_{2}\mathcal{Q}_t.
\end{align}

Thus, for late times $\mathcal{Q}_t$ is given formally by
\begin{align}
\mathcal{Q}_t(\bm{z})=\left<\exp(t L_2)\mathcal{Q},\delta_{\bm{z}}\right>=\left<\mathcal{Q},g_{t,\bm{z}}\right>,
\end{align}
where $\left<\cdot,\cdot\right>$ denotes the $L^2$ pairing of functions on $M\times M$ relative to the two-particle Liouville measure $dz_1\,dz_2$, $\delta_{\bm{z}}$ is a delta function concentrated at $\bm{z}$, and $g_{t,\bm{z}}$ is the distribution function of a pair of particles that begin at $\bm{z}\in M\times M$ when $t=0$. Because this identity holds for arbitrary functions $\mathcal{Q}$, it implies that $g_{t,\bm{z}}$ evolves according to
\begin{align}
g_{t,\bm{z}}=\exp(t A_2)\delta_{z},
\end{align}
where $A_2=L_2^*$ is the $L^2$ adjoint of the operator $L_2$. Differentiating this last identity in time and integrating against the initial two-particle distribution function finally leads to the two-particle Fokker-Planck equation
\begin{align}
\frac{\partial g_t}{\partial t}=A_2g_t,
\end{align} 
where $A_2$ is given by
\begin{align}
(A_2g_t)(\bm{z})=A_1^{(1)}g_t+A_1^{(2)}g_t+\frac{\epsilon^2}{\tau}\mathbb{E}[\alpha:\mathbf{d}^{(1)}\mathbf{d}^{(2)}g_t].
\end{align}
The quantities $A_1$ and $\alpha$ are defined in the main text. 
 
\section{The Lorentz plasma}
\subsection{Introduction}
The Lorentz plasma consists of a noninteracting gas of electrons moving through a neutralizing random collection of fixed, infinitely-massive, Debye-screened ions. The typical approach to studying the dynamics of the Lorentz plasma is to derive a Fokker-Planck equation governing the single-electron distribution function on time scales long compared with the plasma period. This approach eliminates the need to resolve the complicated structure of the ionic potential, and thereby greatly reduces the analytical and computational resources required to understand the plasma's behavior. 

The structure of the Fokker-Planck equation, which takes the form of a Vlasov equation corrected by a collision operator, is ultimately determined by the microscopic Hamiltonian equations of motion for a single electron. Therefore qualitative features of the electronic equations of motion ought to have counterparts at the level of the collision operator. For instance, because an electron that passes through the screened potential of an ion suffers no change in its kinetic energy, it would be surprising if the Fokker-Planck equation didn't have a kinetic energy conservation law. Likewise, because there is no mechanism for electron absorption, the collision operator should be consistent with the conservation of electron number. The purpose of this Chapter is to study the implications on the collision operator of a more subtle qualitative feature of the electronic dynamical equations than either energy conservation of particle conservation.

The qualitative feature we will be concerned with is the Hamiltonian nature of single-electron dynamics. Using the technique described in \cite{Burby_2013}, we will show that because the electronic equations of motion are Hamiltonian, there is a Fokker-Planck equation for the Lorentz plasma that is Hamiltonian in a stochastic sense \cite{Lazaro-Cami_2008}. In particular, this Fokker-Planck equation is the  Kolmogorov forward equation associated with a stochastic differential equation that can be derived from a stochastic variational principle. We will then compare and contrast this Hamiltonian Fokker-Planck equation with the classical result
\begin{align}\label{Lorentz_equation}
\partial_t f+v\cdot\nabla f=C_L(f),
\end{align}
where $C_L$ is the Lorentz collision operator.  We will find that the collision operator, $C_{HL}$, in the Hamiltonian Fokker-Planck equation is not identical to $C_L$, but agrees with the latter asymptotically in the limit $\epsilon_o,\epsilon_1\rightarrow 0$, where $\epsilon_o=\tau_{\text{ac}}/\tau$ and $\epsilon_1=v_{th}\tau/L$, $\tau$ is the coarse-graining time step, $\tau_{\text{ac}}\sim \omega_p^{-1}$ is the Lagrangian autocorrelation time, and $L$ is the length scale of the electron distribution function. We will also show that there is tension between the stochastic Hamiltonian property possessed by $C_{HL}$ and the energy-conserving property of $C_L$ in the following sense. The operator $C_{HL}$ slowly produces energy for all non-zero $\epsilon_o,\epsilon_1$, while the limiting energy-conserving operator $C_L$ is provably not Hamiltonian in the sense of \cite{Lazaro-Cami_2008}. Finally, we will prove that any ``reasonable" stochastic Hamiltonian collision operator that is associated with a path-wise energy-conserving stochastic differential equation must be signficantly different from the Lorentz collision operator $C_L$.

Altogether, these results might give the impression that the stochastic Hamiltonian formalism is not appropriate as an underlying mathematical structure for the pitch angle scattering process. After all, sacrificing energy conservation is a hard pill to swallow. However, by carefully analyzing why the Hamiltonian Fokker-Planck equation for the Lorentz plasma slowly produces energy, we uncover a clue as to how stochastic Hamiltonian dynamics and energy conservation might be reconciled. We will argue that by relaxing the Markov assumption, and thereby allowing for some memory effects, the stochastic Hamiltonian approach may be able to accommodate energy conservation. This possibility is interesting on mathematical, as well as physical grounds because there is presently no mathematical formalism for describing stochastic Hamiltonian dynamics with memory.

\subsection{Microscopic Hamiltonian description of the Lorentz plasma}
The set $Q=\mathbb{R}^3$ will serve as the configuration space for the non-interacting electrons. The velocity phase space for a single electron is therefore $M=TQ\approx Q\times\mathbb{R}^3$. If $\phi$ is the potential produced by the background ions, then the dynamics of each electron are governed by the Hamiltonian
\begin{align}
\mathfrak{H}(x,v)=\frac{1}{2}v^2+\frac{q_e}{m_e} \phi(x),
\end{align}
where $q_e$ and $m_e$ are the electron charge and mass, respectively. The relationship between this Hamiltonian and the electron dynamical vector field $X$ is given by Hamilton's equations,
\begin{align}\label{ham_basic}
\text{i}_X\omega_o=\mathbf{d}\mathfrak{H},
\end{align}
where $\omega_o=\mathbf{d}x^i\wedge\mathbf{d}v_i$.

\subsubsection{Structure of the ionic potential}
Let $\lambda_D$ and $b_o$ be the Debye length and the electron distance of closest approach, respectively. The plasma parameter $\Lambda=\lambda_D/b_o$. The structure of the electrostatic potential produced by each background ion is determined by $\lambda_D$ and $b_0$ in the following manner. The charge density produced by an ion centered at $x_j$ is given by $\rho_{x_j}(x)= q_i \delta(x-x_j)$. The electrostatic potential, $\overline{\phi}_{x_j}$, produced by such a Debye-shielded ion satisfies the differential equation
\begin{align}
-\nabla^2\overline{\phi}_{x_j}+\frac{1}{\lambda_D^2}\overline{\phi}_{x_j}=4\pi\rho_{x_j}.
\end{align} 
The only spherically-symmetric solution of this equation that decays as $|x|\rightarrow\infty$ is given by
\begin{align}
\overline{\phi}_{x_j}(x)=\frac{q_i}{|x-x_j|}\exp\left(-|x-x_j|/\lambda_D\right).
\end{align}
This ``raw" potential, $\overline{\phi}_{x_j}$, is not the mathematically-appropriate potential to subject to electrons in the Lorentz plasma. When an electron passes within a distance $b_o$ from the $j$'th ion, it experiences a large angle scattering event. Because such scattering events are exceedingly-rare, and because we would like to avoid infinities in our analysis, we will regularize the raw potential. We will accomplish this regularization by assuming that the potential produced by the $j$'th ion is given by $\phi_{x_j}(x)=g(|x-x_j|)\equiv\frac{q_i}{\lambda_D}\overline{g}_{\Lambda}(|x-x_j|/\lambda_D)$, where
\begin{align}
\overline{g}_\Lambda(\overline{r})=
\begin{cases}
\overline{g}_{-}(\overline{r})& \text{if }\overline{r}<\frac{1}{\Lambda}\\
\frac{1}{\overline{r}}& \text{if }\frac{1}{\Lambda}<\overline{r}<1\\
\overline{g}_{+}(\overline{r}) & \text{if }\overline{r}>1,
\end{cases}
\end{align}
and $\overline{g}_-,\overline{g}_+$ are chosen so that (i) $\overline{g}_\Lambda(\overline{r})=0$ for $\overline{r}>1+\delta$ for some small $\delta>0$ and (ii) the derivative of $\overline{g}_\Lambda$ vanishes in a neighborhood of $\overline{r}=0$. The total electrostatic potential produced by $N$ ions with centers $x_j$ is then given by 
\begin{align}
\phi=\sum_{j=1}^N\phi_{x_j}.
\end{align}

\subsubsection{Statistical properties of the ionic potential}
We will assume that the $x_j$ are i.i.d. $Q$-valued random variables with PDF $p$. We will also work in the ``thermodynamic limit."  The precise meaning of this statement is as follows. We allow the ionic PDF, $p$, to depend on the number of particles parametrically, i.e. $p(x)= p_N(x)$. Then we make two assumptions:
\\ \\
(i) $N\gg 1$  

\noindent(ii) The limit
\begin{align}
\lim_{N\rightarrow\infty}Np_N\equiv n_i
\end{align}
exists (pointwise) and is equal to the constant $n_i=\Lambda/\lambda_D^3$. 

The mean value of the ionic potential in the thermodynamic limit is given by
\begin{align}
\left<\phi\right>&=\lim_{N\rightarrow\infty}\mathbb{E}[\phi](x)\nonumber\\
&=\lim_{N\rightarrow\infty}N\int g(|x-x^\prime|)p_N(x^\prime)\,dx^|prime.\nonumber\\
&= n_i\int g(|x-x^\prime|)\,dx^\prime\nonumber\\
&=4\pi n_i\int\limits_0^\infty g(\calr)\calr^2\,d\calr,
\end{align}
where $dz$ denotes the standard volume form on $\mathbb{R}^3$. It follows that the mean electrostatic force on an electron is zero.

The covariance function of the ionic potential in the thermodynamic limit is given by
\begin{align}
C_{\phi\phi}(x,y)&=\lim_{N\rightarrow\infty}\mathbb{E}[(\phi-\mathbb{E}[\phi])(x)(\phi-\mathbb{E}[\phi])(y)]\nonumber\\
                        &=n_i\int g(|x-x^\prime|)g(|y-x^\prime|)\,dx^\prime.
\end{align}
where we have used the fact that the $x_i$ are independent and $\mathbb{E}[\phi_{x_1}]\propto N^{-1}$.
Notice that 
\begin{align}
C_{\phi\phi}(x,y)=C(|x-y|),
\end{align}
where
\begin{align}
C(d)&=\frac{Z}{2\pi}\frac{T_e}{\lambda_D}\overline{C}_\Lambda(d/\lambda_D),
\end{align}
and 
\begin{align}
&\overline{C}_\Lambda(\overline{d})=\nonumber\\
&2\pi\int\limits_{-\infty}^{\infty}\int\limits_0^\infty \overline{g}_\Lambda(\sqrt{\overline{R}^2+(\overline{Z}-\overline{d})^2})\overline{g}_\Lambda(\sqrt{\overline{R}^2+\overline{Z}^2})\,\overline{R}\,d\overline{R}\,d\overline{Z}
\end{align}
is a dimensionless covariance function. The most important properties of $C_{\phi\phi}$ and $C$ are the following.
\\ \\
(P1) $C_{\phi\phi}$ is manifestly positive semi-definite, i.e. for any finite collection of points $x_j\in Q$ and corresponding real numbers $a_j$,
\begin{align}
\sum_i\sum_ja_i\,C_{\phi\phi}(x_i,x_j)\,a_j\geq 0.
\end{align}
Thus, $C_{\phi\phi}$ is an \emph{isotropic covariance function}, and $C$ is a \emph{radial basis function}.
\\ \\
(P2) $C(d)=0$ when $d>2\lambda_{D+}$. This follows from the fact that $g(\calr\,)$ is compactly supported in the interval $[0,\lambda_{D+})$.
\\ \\
(P3) $C(d)=C(-d)$
\\ \\
Using (P3), we can also see that all odd powers of $d$ vanish in $C$'s Maclaurin series, i.e. 
\begin{align}
C(d)=C(0)+\frac{1}{2}C^{\prime\prime}(0)\,d^2+\frac{1}{24}C^{\prime\prime\prime\prime}(0)\,d^4+O(d^6),
\end{align}
as $d\rightarrow 0$. Thus,
\begin{align}\label{useful1}
\frac{C^\prime(d)}{d}-C^{\prime\prime}(d)=O(d^2),
\end{align}
and
\begin{align}\label{useful2}
C^\prime(d)=C^{\prime\prime}(0)\,d+O(d^3),
\end{align}
as $d\rightarrow 0$. 

The covariance tensor of the electrostatic \emph{field} produced by the ions is given by
\begin{align}
\mathfrak{C}_{\nabla\phi\nabla\phi}(x,y)&=\lim_{N\rightarrow\infty}\mathbb{E}\left[\nabla(\phi-\mathbb{E}[\phi])(x)\nabla(\phi-\mathbb{E}[\phi])(y)\right]\nonumber\\
                         &=\lim_{N\rightarrow\infty}\mathbb{E}\left[\nabla\phi(x)\nabla\phi(y)\right]\nonumber\\
                         &=\mathfrak{C}(x-y),
\end{align}
where
\begin{align}\label{cov_ten}
\mathfrak{C}(\bm{d})=\frac{Z}{2\pi}\frac{T_e}{\lambda_D^3}\overline{\mathfrak{C}}_\Lambda(\bm{d}/\lambda_D),
\end{align}
and
\begin{align}
\overline{\mathfrak{C}}_\Lambda(\overline{\bm{d}})=-\frac{\overline{C}_\Lambda^\prime(|\overline{\bm{d}}|)}{|\overline{\bm{d}}|}\left(\text{id}-\frac{\overline{\bm{d}}}{|\overline{\bm{d}}|}\frac{\overline{\bm{d}}}{|\overline{\bm{d}}|}\right)-\overline{C}_\Lambda^{\prime\prime}(|\overline{\bm{d}}|)\left(\frac{\overline{\bm{d}}}{|\overline{\bm{d}}|}\frac{\overline{\bm{d}}}{|\overline{\bm{d}}|}\right).
\end{align}
The most important properties of $\mathfrak{C}_{\nabla\phi\nabla\phi}$ and $\mathfrak{C}$ are the following. 
\\ \\
(F1) $\mathfrak{C}_{\nabla\phi\nabla\phi}$ is positive semi-definite, i.e. given any finite collection of points $x_j$ and corresponding vectors $v_j$,
\begin{align}
\sum_j\sum_k v_j\cdot\mathfrak{C}_{\nabla\phi\nabla\phi}(x_j,x_k)\cdot v_k\geq 0.
\end{align}
\\ \\
(F2) $\mathfrak{C}(\Delta)=0$ when $|\Delta|>2\lambda_{D+}$.
\\ \\
(F3) $\mathfrak{C}(\Delta)=\mathfrak{C}(-\Delta)$.
\\ \\
Using (F2) and integration by parts, we can simplify the following type of definite integrals involving $\mathfrak{C}$. Let $L>2\lambda_{D+}$ and choose a unit vector $e$. Set
\begin{align}
I_n(e)=\int\limits_{-L}^L|\lambda|^n\mathfrak{C}(\lambda e)\,d\lambda.
\end{align}
For integer $n\geq0$, we have
\begin{align}\label{I0}
I_n(e)=-2\left(\int\limits_0^\infty\lambda^{n-1}C^\prime(\lambda)\,d\lambda\right)(\text{id}-(n+1)ee).
\end{align}

\subsection{The argument for employing a Fokker-Planck equation to model the Lorentz plasma}
The purpose of this section is to give a qualitative motivational picture of the ideas that go into deriving a Hamiltonian Fokker-Planck equation for the Lorentz plasma, as well as to describe precisely what a Hamiltonian Fokker-Planck equation is. As a point of departure, we remind the reader of the justification and derivation of a Fokker-Planck model in terms of so-called jump moments. We then reformulate the same discussion in terms of the symplectomorphism group, $\text{Diff}_{\omega_o}(TQ)$. This reformulation offers a particularly suggestive description of how the Hamiltonian nature of the electronic equations of motion influences the structure of the Fokker-Planck equation.

\subsubsection{In terms of jump moments}
The problem of finding a formula for the trajectory of an electron in the Lorentz plasma is extremely complicated. There is not just a single ion; if there were, we would only have to solve the two-body problem. There are not just two ions either; if this were true, we would be faced with (an analogue of) the soluble Euler three-body problem. It is better to assume there are $\sim10^{23}$ ions, and therefore electron dynamics are surely chaotic. As such, when studying the dynamics of the Lorentz plasma, we must be satisfied with less detailed information than exact electron orbits.

One way to give a less detailed (and therefore simpler) description of the Lorentz plasma that still retains a great deal of dynamical information is to find the evolution equation for the single-electron PDF. This is a much more manageable task than finding the precise electron trajectories for the following reason. Whereas the large number of ions mangles the electron trajectories, it actually simplifies certain statistical properties of electron dynamics. For instance, the (Lagrangian) autocorrelation time of the force on an electron is set by the plasma period $\tau_{\text{ac}}\sim \omega_p^{-1}$. Thus, if we chop the time-axis into intervals of length $\tau\gg \tau_{\text{ac}}$ and restrict attention to an electron's phase space location at the ends of these intervals, we obtain a discrete-time Markov process in phase space, $z_n$, where $n$ is the discrete time index. If we could estimate the transition probabilities of this Markov process, we would be able to derive an equation describing the evolution of the electron PDF on time scales much longer than $\tau$. This derivation could proceed by analogy with the derivation of the late-time evolution equation for the PDF of a random walker given in \cite{Bazant_2006}. Indeed, our electron Markov process is nothing more than a random walk in phase space where the PDF of the walker's step depends on where the walker is standing. 

Provided that we make the further restriction $\tau\ll\tau_{\text{bounce}}$, where $\tau_{\text{bounce}}$ is the characteristic time for an electron trajectory to deviate significantly from free streaming, the transition probabilities can be calculated using ordinary perturbation theory \footnote{Our ability to require that $\tau$ satisfies both $\tau\ll \tau_{\text{bounce}}$ and $\tau\gg \tau_{\text{ac}}$ follows from the fact that small-angle scattering events dominate over large-angle scattering events.}. It is straightforward to show that the resulting late-time evolution equation for the single-electron PDF takes the form
\begin{align}\label{gen_FP}
\partial_tf+\text{div}(fu)=\text{div}(D\cdot df),
\end{align}
where $\text{div}$ denotes the divergence relative to the Liouville volume form $dx\,dv$, $u$ is a vector field on phase space, and $D$ is a rank-$2$ tensor on phase space with components $D^{ij}$. The drift vector $u$ is given by
\begin{align}\label{general_drift}
u&=u_o+\langle\Delta_2z\rangle/\tau-\frac{1}{2}\langle\text{div}(\Delta_1z)\Delta_1z\rangle/\tau,
\end{align}
where $\Delta_1z,\Delta_2z$ are vector fields on phase space and $u_o=v\cdot\partial_x$ is the free-streaming vector field. The diffusion tensor is given by
\begin{align}\label{general_diffusion}
D=\frac{1}{2}\langle\Delta_1z\otimes \Delta_1z\rangle/\tau.
\end{align} 
We will refer to $\Delta_1z$ and $\Delta_2z$ as the jump vectors. They are defined so that a particle starting at $z_o=(x,v)$ when $t=0$ ends at
\begin{align}
z_\tau\approx\exp(\Delta_1 z+\Delta_2 z)(x+v\tau,v)
\end{align}
when $t=\tau$, with second order accuracy\footnote{The amplitude of the fluctuating electric field can be regarded as the expansion parameter, which is essentially $\sqrt{1/\Lambda}$}. The components of $D$ are known as the jump moments. Equation\,(\ref{gen_FP}) is known as the Fokker-Planck equation.

\subsubsection{In terms of the symplectomorphism group}
The previous argument justifying the use of a Fokker-Planck equation for the late-time single-electron PDF is appealing because it draws upon only elementary facts about Markov processes. However, it has the disadvantage of obscuring a striking geometric picture that underlies the whole discussion. In order to illuminate the geometric picture, we will now give a second justification for using a Fokker-Planck equation.

Let $\mathcal{F}$ be the $t=\tau$ time-advance map associated with the full single-electron equations of motion. Given an initial condition $z\in TQ$, the approximately-Markov process that gives an electron's phase space location at the ends of $\tau$-second time intervals is given by
\begin{align}\label{basic_process}
z_n=\mathcal{F}^n(z),
\end{align}
where $n\in\mathbb{Z}$ is the discrete time variable and $\mathcal{F}^n$ denotes the $n$-fold composition of $\mathcal{F}$ with itself (e.g. when $n=2$, $\mathcal{F}^n(z)=\mathcal{F}(\mathcal{F}(z))$.) Note that the mapping $\mathcal{F}$ is random because it depends on the configuration of the ions. Because $z_n$ is a Markov process, the operator $U_n$ given by
\begin{align}
(U_nQ)(z)=\langle Q(\mathcal{F}^n(z))\rangle,
\end{align}
where $Q:TQ\rightarrow\mathbb{R}$ is an arbitrary observable, must satisfy the semi-group property
\begin{align}
U_{n_1+n_2}=U_{n_1}U_{n_2}.
\end{align}
However, by noting $U_1=\langle \mathcal{F}^*\rangle$\footnote{This formula shows that $U_1$ is a mean propagator.}, where $\mathcal{F}^*$ denotes the pullback operator along $\mathcal{F}$, we also have
\begin{align}
U_2=\langle \mathcal{F}^* \mathcal{F}^*\rangle\neq \langle \mathcal{F}^*\rangle\langle \mathcal{F}^*\rangle=U_1U_1.
\end{align}
Note that $\mathcal{F}$ and $\mathcal{F}$ are not independent random mappings -- they are identical! This contradiction tells us that the process Eq.\,(\ref{basic_process}) is not precisely Markov. 

In spite of this contradiction, we know that $z_n$ is approximately Markov because $\tau\gg \tau_{\text{ac}}$. Therefore, for the sake of modeling it is sensible to replace Eq.\,(\ref{basic_process}) with
\begin{align}\label{markov_process}
z_n=\mathcal{F}_{n}\circ \mathcal{F}_{n-1}\circ\dots\circ \mathcal{F}_1(z),
\end{align} 
where $\circ$ denotes functional composition and the $\mathcal{F}_i$ are i.i.d. random mappings each with the same PDF as $\mathcal{F}$. The effect of this replacement is that the locations of all of the ions are scrambled after each time step. While this scrambling effect is, strictly speaking, unphysical, it ought to be statistically harmless; electrons forget about the orientations of the ions after $\tau_{\text{ac}}$ seconds anyway. It is easy to check that this redefined $z_n$ is rigorously Markov.

Equation\,(\ref{markov_process}) tells us that one way to determine evolution of the single-electron PDF is to first determine the statistical behavior of the mapping-valued stochastic process
\begin{align}
g_n=\mathcal{F}_n\circ \mathcal{F}_{n-1}\circ\dots\circ \mathcal{F}_1.
\end{align}
Indeed, the expected value of $Q(z_n)$, for any observable $Q$, is given by
\begin{align}
\langle Q(z_n)\rangle=\langle g_n^*Q\rangle(z)=(\langle g_n^*\rangle Q)(z),
\end{align}
which shows that the single-electron PDF is completely determined if we know $\langle g_n^*\rangle$. But why would we want to do this? Whereas the process $z_n$ evolves in a finite-dimensional space, the process $g_n$ evolves in an infinite-dimensional space of mappings, which suggests that $g_n$ is a much more complicated object than $z_n$. The answer is that $g_n$ is simpler than $z_n$ when viewed in the right way.

First notice that $g_n$ is a diffeomorphism for all $n$, i.e. $g_n$ is smooth and invertible with a smooth inverse. This follows from the fact that $\mathcal{F}_i$ is a diffeomorphism for each $i$ (being a time-advance map for an ODE on phase space) and $g_n$ is a composition of the $\mathcal{F}_i$'s. This means that the mapping-valued process $g_n$ takes place in a very special space of mappings known as the phase space diffeomorphism group $\text{Diff}(TQ)$. $\text{Diff}(TQ)$, which is the set of all diffeomorphisms of the velocity phase space $TQ$, is a group under functional composition. In a sense that we will not discuss here, $\text{Diff}(TQ)$ is also a smooth (infinite-dimensional) manifold. Thus, the process $g_n$ evolves in a space with a very rich structure.

Next notice that \emph{the increments} $\delta g_{n_2,n_1}=g_{n_2}\circ g_{n_1}^{-1}$, for $n_1\leq n_2$, have the following simple statistical properties:
\\ \\
(RW1) If $n_1\leq n_2\leq n_3\leq n_4$, $\delta g_{n_2,n_1}$ and $\delta g_{n_4,n_3}$ are statistically independent.
\\ \\
(RW2) The PDF\footnote{Here PDF stands for probability distribution \emph{functional}.} of $\delta g_{n_2,n_1}$ depends on $n_1,n_2$ only through the difference $n_2-n_1$. 
\\ \\
If we were to replace $\text{Diff}(TQ)$ with the additive group $\mathbb{R}$, these properties would imply that $g_n$ is an ordinary random walk\footnote{The PDF of a step taken by such a walker would be arbitrary.}. More generally, if $\text{Diff}(TQ)$ is replaced by any group $G$, a discrete-time process on $G$ that satisfies properties (RW1) and (RW2) is known as a random walk on $G$. Thus, the process $g_n$ is a $\text{Diff}(TQ)$-valued random walk.

Once nice feature of thinking about the dynamics of the Lorentz plasma as a random walk on the diffeomorphism group is that we have a good intuitive understanding of the long-time behavior of random walks. In particular, we know that, under an appropriate scaling limit, an ordinary random walk is well-approximated by a Brownian motion. Therefore we can reasonably expect that the long-time behavior of the random walk $g_n$ is described by a Brownian motion on $\text{Diff}(TQ)$\cite{Baxendale_1984}, i.e. a continuous-time process $g_t\in\text{Diff}(TQ)$ that satisfies the properties
\\ \\
(BM1) If $t_1\leq t_2\leq t_3\leq t_4$, $\delta g_{t_2,t_1}$ and $\delta g_{t_4,t_3}$ are statistically-independent.
\\ \\
(BM2) The PDF of $\delta g_{t_2,t_1}$ only depends on $t_1,t_2$ through the difference $t_2-t_1$.
\\ \\
(BM3) The sample paths of $g_t$ are almost surely continuous functions of $t$.
\\ \\
Of course, properties (BM1) and (BM2) are the obvious analogues of the properties (RW1) and (RW2) that we already know $g_n$ satisfies. Property (BM3) is motivated by the dominance of small-angle scattering events over large-angle scattering events, in particular the fact that the particle jumps can be calculated using perturbation theory. 

In \cite{Baxendale_1984}, Baxendale shows that Brownian motions on $\text{Diff}(TQ)$ always arise as the stochastic time-advance maps of stochastic differential equations. If $g_t$ is such a stochastic time-advance map, then a basic fact from the theory of stochastic differential equations states that the PDF of the random variable $g_t(z)$, where $z\in TQ$ is a fixed initial condition, satisfies a Fokker-Planck equation of the form given in Eq.\,(\ref{gen_FP}). We have therefore arrived at an alternative justification for the use of a Fokker-Planck equation to model the Lorentz plasma.

Another nice feature of thinking in terms of $g_n$ instead of $z_n$ is that the Hamiltonian nature of the electron dynamical equations manifests itself in a very simple way at the level of $\text{Diff}(TQ)$. Indeed, because the electron dynamical vector field $X$ given by Eq.\,(\ref{ham_basic}) is Hamiltonian, the $t=\tau$ time-advance map $\mathcal{F}$ must preserve the symplectic form, $\mathcal{F}^*\omega_o=\omega_o$, where $\omega_o=dx^i\wedge dv_i$. Likewise, because the $\mathcal{F}_i$ have the same PDF as $\mathcal{F}$, we also have $\mathcal{F}_i^*\omega_o=\omega_o$. Therefore, by the identity $(\mathcal{F}_i\circ \mathcal{F}_j)^*=\mathcal{F}_j^*\mathcal{F}_i^*$, the process $g_n$ satisfies
\begin{align}
g_n^*\omega_o=\omega_o
\end{align}
for all $n$. In other words, $g_n$ is not free to wander everywhere in $\text{Diff}(TQ)$, but only along the constraint set defined by $g^*\omega_o=\omega_o$. Actually, the latter constraint set is a subgroup of $\text{Diff}(TQ)$ known as the symplectomorphism group, $\text{Diff}_{\omega_o}(TQ)\subset \text{Diff}(TQ)$. Because $g_n$ does not leave the symplectomorphism group, the limiting Brownian motion $g_t$ also must satisfy the same constraint. It can be shown that the latter requirement constrains the Fokker-Planck equation\,(\ref{gen_FP}) to satisfy
\begin{align}\label{ham_FP}
u&=X_{h_o}\\
D&=\sum_{k=1}X_{h_k}\otimes X_{h_k},\label{ham_FP2}
\end{align}
where $X_f$ denotes the Hamiltonian vector field with Hamiltonian $f$, the $h_k$ are arbitrary functions on phase space, and the sum over $k$ is possibly infinite. Conversely, given a Fokker-Planck equation whose drift vector $u$ and diffusion tensor $D$ take the above form, it is always possible to find a $\text{Diff}_{\omega_o}(TQ)$-valued Brownian motion that generates it.

When the drift vector and diffusion tensor of a Fokker-Planck equation are in the form prescribed by Eqs.\,(\ref{ham_FP}) and\,(\ref{ham_FP2}), we will say that the Fokker-Planck equation is Hamiltonian. Thus, an important consequence of the fact that the single-electron equations of motion are Hamiltonian in nature is that the Fokker-Planck equation for the single-electron PDF ought to be Hamiltonian. By making this observation, we can see that in passing from the microscopic single-electron equations of motion to the macroscopic Fokker-Planck equation, we move from the world of Hamiltonian mechanics into the world of stochastic Hamiltonian mechanics. Where Hamiltonian mechanics is concerned with smooth one-parameter subgroups of $\text{Diff}_{\omega_o}(TQ)$, stochastic Hamiltonian mechanics is concerned with Brownian motion on $\text{Diff}_{\omega_o}(TQ)$. The stochastic Hamiltonian nature of the Fokker-Planck equation is the moral counterpart to the Hamiltonian nature of the microscopic equations of motion alluded to in the introduction.

\subsection{Hamiltonian Fokker-Planck equation for the Lorentz plasma}

We will now apply the technique described in \cite{Burby_2013}, which we will refer to hereafter as BZQ, to derive a Hamiltonian Fokker-Planck equation for the Lorentz plasma. In BZQ's notation, we have
\begin{align}
H_o&=\frac{1}{2}v^2\\
h&=\frac{q_e}{m_e}\phi,
\end{align}
where $\phi$ is the ionic potential described earlier. The first- and second-order jump vectors are given by $\Delta_1 z=X_{s_1}$ and $\Delta_2z=X_{s_2}$, where
\begin{align}
s_1&=\int_0^\tau F_{t*}h\,dt\\
s_2&=\frac{1}{2}\int_0^\tau\int_0^{t_1}\{F_{t_2*}h,F_{t_1*}h\}\,dt_2\,dt_1.
\end{align} 
The unperturbed flow map, $F_t$, is given by
\begin{align}
F_t(x,v)=(x+vt,v).
\end{align}
Using these formulae, the drift vector $u_{HL}$ and the diffusion tensor $D_{HL}$ can be computed in terms of the covariance tensor $\mathfrak{C}$ given in Eq.\,(\ref{cov_ten}).

First we compute $u_{HL}=u_o+X_{\langle s_2\rangle}/\tau$, which amounts to computing $\langle s_2\rangle$. It is straightforward to verify that the mean of the Poisson bracket appearing in the definition of $s_2$ reduces to
\begin{align}
\langle\{F_{t_2*}h,F_{t_1*}h\}\rangle=-\bigg(\frac{q_e}{m_e}\bigg)^2[t_1-t_2]\text{tr}(\mathfrak{C}([t_1-t_2]v)),
\end{align}
where $\text{tr}$ denotes the trace. Upon substituting this expression into the double integral, changing integration variables, and applying Fubini's theorem, we then arrive at the following expression for $\langle s_2\rangle$:
\begin{align}\label{mean_s2}
\langle s_2\rangle
&=-\frac{1}{2}\bigg(\frac{q_e}{m_e}\bigg)^2\int_0^{\tau}t(\tau-t)\text{tr}(\mathfrak{C}(vt))\,dt.
\end{align}
Note that because $\langle s_2\rangle$ only depends on $(x,v)$ through $v$, the drift vector $u_{HL}=u_o+\langle\Delta_2z\rangle/\tau=u_o+X_{\langle s_2\rangle}/\tau$ only has an $x$-component.

Next we derive an expression for the diffusion tensor $D_{HL}$. For this purpose, we introduce a useful notation for contravariant second rank tensors on phase space. If $T$ is a dyad like $\mathfrak{C}$, then $T_{xv}$ is the second rank contravariant tensor on phase space given by
\begin{align}
dh_1\cdot T_{xv}\cdot dh_2=\nabla h_1\cdot T\cdot \nabla_v h_2,
\end{align}
where $h_1,h_2$ are arbitrary functions on phase space. The tensors $T_{xx},T_{vx},T_{vv}$ are similarly defined. In terms of this notation, $\langle X_{s_1}\otimes X_{s_1}\rangle$ is given by
\begin{align}
&\langle X_{s_1}\otimes X_{s_1}\rangle=\left(\frac{q_e}{m_e}\right)^2\bigg(\int\limits_0^\tau\int\limits_0^\tau\mathfrak{C}_{vv}([t_2-t_1]v)\,dt_1\,dt_2\nonumber\\
&+\int\limits_0^\tau\int\limits_0^\tau t_2\mathfrak{C}_{vx}([t_2-t_1]v)\,dt_1\,dt_2+\int\limits_0^\tau\int\limits_0^\tau t_1\mathfrak{C}_{vx}([t_2-t_1]v)\,dt_1\,dt_2\nonumber\\
&~~~~~~~~~~~~~+\int\limits_0^\tau\int\limits_0^\tau t_1t_2\mathfrak{C}_{xx}([t_2-t_1]v)\,dt_1\,dt_2\bigg).
\end{align}
After simplifying the double integrals and dividing by $2\tau$, the diffusion tensor $D$ is then given by
\begin{align}\label{hamiltonian_D}
&D_{HL}=\frac{1}{\tau}\bigg(\frac{q_e}{m_e}\bigg)^2\bigg(\int_0^\tau(\tau-t)\mathfrak{C}_{vv}(vt)\,dt\nonumber\\
&+\frac{\tau}{2}\int_0^\tau(\tau-t)(\mathfrak{C}_{xv}(vt)+\mathfrak{C}_{vx}(vt))\,dt\nonumber\\
&~~~~~~+\frac{\tau^2}{3}\int_0^\tau(\tau-\sfrac{3t}{2}+\sfrac{t^3}{2\tau^2})\mathfrak{C}_{xx}(vt)\,dt\bigg).
\end{align}
As explained in \cite{Burby_2013}, because $D_{HL}=\langle X_{s_1}\otimes X_{s_1}\rangle/(2\tau)$, there exists an expansion of $D$ in the form
\begin{align}
D_{HL}=\sum_{k=1}X_{h_k}\otimes X_{h_k},
\end{align}
where the $h_k$ form an orthonormal basis for the reproducing kernel Hilbert space associated with the phase space covariance kernel $\alpha(z_1,z_2)=\langle X_{s_1}(z_1)\otimes X_{s_2}(z_2)\rangle/(2\tau)$. We will not find the $h_k$ here. It is worth pointing out, however, that finding the $h_k$
is equivalent to the well-known problem of finding the so-called Karhunen-Lo\`eve \cite{Lord_2014} expansion of a Gaussian random field with covariance $\alpha$. 

Because $u_{HL}$ and $D_{HL}$ just calculated can be written in the form given in Eq.\,(\ref{ham_FP}), the Fokker-Planck equation
\begin{align}\label{hamiltonianLorentz_equation}
\partial_tf+\text{div}(f u_{HL})=\text{div}(D_{HL}\cdot df)
\end{align}
is Hamiltonian. In the following section, we will compare and contrast this Fokker-Planck equation with the classical result given in Eq.\,(\ref{Lorentz_equation}). In particular, we will compare the Lorentz collision operator 
\begin{align}\label{lorentz_operator}
C_L(f)=\text{div}(D_L\cdot d f),
\end{align}
where 
\begin{align}
D_L=\nu(v)\,\mathsf{U}_{vv}(v),
\end{align}
$\mathsf{U}(v)=|v|^2(\text{id}-\hat{v}\hat{v})$, and $\nu(v)=\frac{\omega_p}{8\pi}\frac{\text{ln}\Lambda}{\Lambda}\frac{v_{th}^3}{|v|^3}$,
with the Hamiltonian collision operator
\begin{align}
C_{HL}(f)=\text{div}(D_{HL}\cdot df),
\end{align}
where $D_{HL}$ is given by Eq.\,(\ref{hamiltonian_D}).

\subsection{Comparison of the Hamiltonian collision operator and the Lorentz operator}
\subsubsection{Asymptotic equivalence}
In order to demonstrate the asymptotic equivalence of Eqs.\,(\ref{Lorentz_equation}) and\,(\ref{hamiltonianLorentz_equation}), we introduce the dimensionless variables $\overline{x},\overline{v},\overline{t}$. These normalized position, velocity, and time variables are related to their unnormalized counterparts by
\begin{align}
x&=L\overline{x}\\
v&=v_{th}\overline{v}\\
t&=T\overline{t},
\end{align}
where $L$ is the length scale of the electron distribution function, $v_{th}=\lambda_D\omega_p$ is the thermal velocity, and $T$ is the temporal scale of the electron distribution function. We will set $T=\Lambda/\omega_p$, which is consistent with measuring time in units of the electron-ion collision period. The Hamiltonian Fokker-Planck equation expressed in these dimensionless variables is
\begin{align}
\partial_{\overline{t}}f+\overline{\text{div}}(f\overline{u})=\overline{\text{div}}(\overline{D}\cdot \overline{d}f),
\end{align}
where $\overline{u}=T u_{HL}$ and $\overline{D}=T D_{HL}$. We will now present expressions for $\overline{u}$ and $\overline{D}$ that exhibit their dependence on the small parameters $\epsilon_o=1/(\tau\omega_p)$, $\epsilon_1=v_{th}\tau/L$, and $1/\Lambda$.

The normalized drift vector is given by
\begin{align}\label{normalized_u}
\overline{u}=\Lambda\epsilon_o\epsilon_1\overline{v}\cdot\partial_{\bar{x}}-\frac{1}{32\pi^2}\epsilon_1\overline{\nabla}_{\overline{v}}\chi\cdot\partial_{\overline{x}},
\end{align}
where the dimensionless function $\chi$ is given by
\begin{align}
\chi(\overline{v})=\int_0^{1/\epsilon_o}\epsilon_o\overline{\lambda}\,(1-\epsilon_o\overline{\lambda})\,\text{tr}\,\overline{\mathfrak{C}}_\Lambda(\overline{v}\overline{\lambda})\,d\overline{\lambda}.
\end{align}
Note that $\chi$ depends on $\epsilon_o$ and $\Lambda$, but not $\epsilon_1$.

The normalized diffusion tensor is given by
\begin{align}
\overline{D}&=\frac{1}{16\pi^2}\bigg(\bm{\eta}_{\overline{v}\overline{v}}+\frac{\epsilon_1}{2}(\bm{\eta}_{\overline{v}\overline{x}}+\bm{\eta}_{\overline{x}\overline{v}})+\epsilon_1^2\bm{\zeta}_{\overline{x}\overline{x}}\bigg),
\end{align}
where the dimensionless dyads $\bm{\eta},\bm{\zeta}$ are given by
\begin{align}
\bm{\eta}&=\int_{0}^{1/\epsilon_o}(1-\epsilon_o\overline{\lambda})\,\overline{\mathfrak{C}}_{\Lambda}(\overline{v}\overline{\lambda})\,d\overline{\lambda}\\
\bm{\zeta}&=\int_0^{1/\epsilon_o}(\sfrac{1}{3}-\sfrac{\epsilon_o\overline{\lambda}}{2}+\sfrac{\epsilon_o^3\overline{\lambda}^3}{6})\overline{\mathfrak{C}}_{\Lambda}(\overline{v}\overline{\lambda})\,d\overline{\lambda}.
\end{align}
Note that $\bm{\eta},\bm{\zeta}$, like $\chi$, depend on $\epsilon_o$ and $\Lambda$, but not $\epsilon_1$.

These expressions for $\overline{u}$ and $\overline{D}$ can now be used to study the asymptotic behavior of the Hamiltonian Fokker-Planck equation as $\epsilon_o,\epsilon_1$ and $1/\Lambda$ tend to zero. For concreteness, we will study this limit under the assumption
\begin{align}\label{simple_scalings}
\epsilon_o&=\sfrac{1}{\sqrt{\Lambda}}\\
\epsilon_1&=\sfrac{1}{\sqrt{\Lambda}}\label{simple_scalings_a}.
\end{align}
Effectively, this assumption chooses a specific path to zero through $(\epsilon_o,\epsilon_1,1/\Lambda)$-space along which our asymptotic limit is taken. 

First we consider the limiting behavior of the quantities $\chi,\bm{\eta},\bm{\zeta}$. For this purpose, it is enough to consider the limiting behavior of the integrals
\begin{align}
\bm{I}_n&=\int_0^{\sfrac{1}{\epsilon_o}}(\epsilon_o\overline{\lambda})^n\,\overline{\mathfrak{C}}_\Lambda(\overline{v}\overline{\lambda})\,d\overline{\lambda}\nonumber\\
&=\frac{\epsilon_o^n}{|\overline{v}|^{n+1}}\int_0^{\sfrac{|\overline{v}|}{\epsilon_o}}s^n\,\overline{\mathfrak{C}}_\Lambda(\hat{\overline{v}}s)\,ds,
\end{align}
for non-negative integer $n$. It is not difficult to show that 
\begin{align}
\bm{I}_n\propto\begin{cases} 
\text{ln}\,\Lambda&\text{if }n=0\\
\epsilon^n&\text{if }n>0.
\end{cases}
\end{align}
Therefore the asymptotic limits of $\chi,\bm{\eta},\bm{\zeta}$ are given by
\begin{align}
\chi&\rightarrow0\\
\bm{\eta}&\rightarrow \bm{I}\equiv\frac{1}{|\overline{v}|}\int_0^\infty\overline{\mathfrak{C}}_\Lambda(s\hat{\overline{v}})\,ds\\
\bm{\zeta}&\rightarrow\frac{1}{3}\bm{I}.
\end{align}

The limiting drift vector and diffusion tensor are now simple to obtain. For $\overline{u}$ we have
\begin{align}
\overline{u}\rightarrow \overline{v}\cdot\partial_{\overline{x}}.
\end{align}
Similarly, $\overline{D}$ is given by
\begin{align}
\overline{D}\rightarrow \frac{1}{16\pi^2}\bm{I}_{\overline{v}\overline{v}}.
\end{align}
The dyad $\bm{I}$ can be simplified further by using the fact that when $\Lambda$ is large, $\overline{C}_{\Lambda}(\overline{d})\approx 2\pi e^{-|\overline{d}|}$ for $|d|>\Lambda^{-1}$. In fact, if the Debye screened potential was not regularized, this would not be an approximation. Thus,
\begin{align}
\bm{I}&\approx -\frac{1}{|\overline{v}|}\bigg(\int_{\Lambda^{-1}}^\infty\frac{\overline{C}_\Lambda(s)}{s}\,ds\bigg)(\text{id}-\hat{\overline{v}}\hat{\overline{v}})\nonumber\\
&=\frac{2\pi}{|\overline{v}|}\bigg(\int_{\Lambda^{-1}}^\infty\frac{e^{-s}}{s}\,ds\bigg)(\text{id}-\hat{\overline{v}}\hat{\overline{v}})\nonumber\\
&\rightarrow \frac{2\pi\,\text{ln}\,\Lambda}{|\overline{v}|}(\text{id}-\hat{\overline{v}}\hat{\overline{v}}).
\end{align}
After restoring units, we can therefore write the limiting drift vector and diffusion tensor as
\begin{align}
u&=v\cdot\partial_x\\
D&=\frac{\omega_p\,\text{ln}\,\Lambda}{8\pi\Lambda}\mathsf{U}_{vv}=D_L,
\end{align}
where the dyad $\mathsf{U}=|v|^2(\text{id}-\hat{v}\hat{v})$. Comparing these expressions with Eq.\,(\ref{lorentz_operator}) reveals that we have indeed recovered the classical result for the Lorentz plasma Fokker-Planck equation.

This result assumes the scalings given in Eqs.\,(\ref{simple_scalings}) and\,(\ref{simple_scalings_a}), but many other choices seem allowable. Therefore a natural question is whether or not the form of the limiting Fokker-Planck equation depends on how we send the small parameters to zero. It turns out that there are only two possible limiting equations, and these differ only in the limiting drift vector; either the free streaming term survives the limit or it doesn't. We chose our scaling so that the free streaming term survives because this seems to be the most interesting possible scenario.

\subsubsection{The Hamiltonian collision operator slowly produces energy}
While the Fokker-Planck equation\,(\ref{hamiltonianLorentz_equation}) is manifestly Hamiltonian, it does not conserve kinetic energy. This can be seen by direct calculation. The total electron kinetic energy is given by
\begin{align}
\mathcal{E}=\int H_o\,f\,\Omega,
\end{align}
where $\Omega=dx\,dv$ is the Liouville volume form. The rate of change of the plasma kinetic energy is therefore
\begin{align}
\frac{d}{dt}\mathcal{E}&=\int H_o\partial_tf\,\Omega\nonumber\\
&=\int H_o\,\text{div}(-fu_{HL}+D_{HL}\cdot df)\,\Omega\nonumber\\
&=-\int dH_o\cdot D_{HL}\cdot df\,\Omega\nonumber\\
&=\int\text{div}(D_{HL}\cdot dH_o)\,f\,\Omega,
\end{align}
where we have used the fact that $\langle s_2\rangle$ only depends on $v$ and $D_{HL}$ is symmetric. This expression shows that the only way kinetic energy will be conserved \emph{regardless of initial conditions} is if $\text{div}(D_{HL}\cdot dH_o)=0$. However,
\begin{align}\label{energy_growth}
\text{div}(D_{HL}\cdot dH_o)&=\bigg(\frac{q_e}{m_e}\bigg)^2\nabla_v\cdot\bigg(\int_0^\tau(1-t/\tau)v\cdot\mathfrak{C}(vt)\,dt\bigg)\nonumber\\
&=-\bigg(\frac{q_e}{m_e}\bigg)^2\nabla_v\cdot\bigg(\int_0^\tau(1-t/\tau)vC^{\prime\prime}(|v|t)\,dt\bigg)\nonumber\\
&=\bigg(\frac{q_e}{m_e}\bigg)^2\frac{1}{|v|\tau}\frac{1}{|v|}\frac{\partial}{\partial|v|}\bigg(|v|[C(0)-C(|v|\tau)]\bigg)\nonumber\\
&\approx\bigg(\frac{q_e}{m_e}\bigg)^2\frac{C(0)}{|v|^2\tau},
\end{align}
where the last line is valid when $|v|\tau>2\lambda_{D+}$. It follows that $\frac{d\mathcal{E}}{dt}\neq 0$ and that the characteristic time for energy change is 
\begin{align}
\tau_e=\frac{|\mathcal{E}|}{|d\mathcal{E}/dt|}\approx \tau\Lambda.
\end{align}
We will have more to say about why $C_{HL}$ does not conserve energy in the final section. 

\subsubsection{The Lorentz operator is not Hamiltonian}
While $C_{HL}$ does not conserve energy, the limiting collision operator $C_L$ certainly does. Therefore it is tempting to hope that the limiting procedure that transforms $C_{HL}$ into $C_L$ preserves the Hamiltonian nature of $C_{HL}$. In this section we will prove definitively that this is \emph{not} the case. Specifically we will show that there is no sequence of functions $h_k$ such that $D_L=\sum_k X_{h_k}\otimes X_{h_k}$.

The level of mathematical discourse increases substantially in this section for two reasons. First, a basic knowledge of vector bundles and operations on vector bundles is assumed. A particularly readable account of this material is given in \cite{Bott_1982}, starting on p.\,53. Second, we assume a working knowledge of the theory of reproducing kernels. Reproducing kernels are really nothing more than two-point covariance functions, but much can be said about them without referring to ideas from probability theory, and this (somewhat) justifies introducing a second name for them. The standard reference for learning about the basics of reproducing kernel theory is \cite{Aronszajn_1950}.

Suppose that $M$ is a smooth manifold and $\kappa:M\times M\rightarrow\mathbb{R}$ is a smooth reproducing kernel. Associated to $\kappa$ is a smooth section, $\alpha_\kappa$, of the symmetric tensor product $T^*M\odot T^*M$. The section $\alpha_\kappa$ is defined as follows. Let $v_1,v_2\in T_xM$ be tangent vectors at $x\in M$. Choose smooth curves $c_1,c_2:I\rightarrow M$, where $I$ is an open interval of $\mathbb{R}$ containing $0$, such that $c_1(0)=c_2(0)=x$, $c^\prime_1(0)=v_1$ and $c^\prime_2(0)=v_2$. We set 
\begin{align}
\alpha_\kappa(v_1,v_2)=\frac{d~~}{d\epsilon_1}\bigg|_0\frac{d~~}{d\epsilon_2}\bigg|_0\kappa(c_1(\epsilon_1),c_2(\epsilon_2)).
\end{align}

A section $\alpha$ of $T^*M\odot T^*M$ has a \emph{reproducing kernel primitive} if there is some reproducing kernel $\kappa$ such that $\alpha=\alpha_\kappa$. The purpose of this section is to prove a theorem that characterizes the set of $\alpha$'s with reproducing kernel primitives. We will only consider sections $\alpha$ with \emph{locally constant rank}. The \emph{rank} of a section $\alpha$ at $x\in M$ is defined as the codimension of the kernel of $\alpha_x$. More precisely, if we let $T_{\alpha x}M$ be the collection of vectors $v_x$ such that 
\begin{align}
\forall w_x\in T_xM,~\alpha(v_x,w_x)=0,
\end{align}
then the rank of $\alpha_x$ is $\text{rank}(\alpha_x)=\dim(T_{x}M)-\dim(T_{\alpha x}M)$. A section $\alpha$ has locally constant rank if for each $x\in M$ there is some open neighborhood of $x$ on which the function $u\in M\mapsto \text{rank}(\alpha_u)\in\mathbb{Z}$ is constant. 

First we will prove that if $\alpha_\kappa$ has locally-constant rank, then $L_X\alpha=0$ for any vector field $X$ that takes values in $\alpha_\kappa$'s \emph{characteristic distribution}. Given an $\alpha$ with locally-constant rank, its characteristic distribution is the subbundle of $TM$ whose fiber at $x\in M$ is given by $T_{\alpha x}M$. $\alpha$'s characteristic distribution will be denoted $T_{\alpha}M$. A general $\alpha$ with locally-constant rank need not satisfy $L_X\alpha=0$. Thus, the following theorem provides a non-trivial necessary condition for a section $\alpha$ with locally-constant rank to have a reproducing kernel primitive.
\begin{theorem}\label{t1}
Let $\kappa$ be a smooth reproducing kernel on $M$. $\alpha_\kappa$ is positive semi-definite as a bilinear form. Moreover, if $\alpha_\kappa$ has locally constant rank, then $L_X\alpha_\kappa=0$ for each vector field $X$ that takes values in $T_{\alpha_\kappa} M$. In particular $T_{\alpha}M$ is integrable in the sense of Frobenius.
\end{theorem}
\begin{proof}
Let $\{\phi_j\}$ be an orthonormal basis for the reproducing kernel Hilbert space associated with $\kappa$. The kernel $\kappa$ can be expressed in terms of these basis elements as $\kappa(x,y)=\sum_j\phi_j(x)\phi_j(y)$, which implies that $\alpha_\kappa=\sum_j(d\phi_j)^2$. This immediately implies that $\alpha_\kappa$ is positive semi-definite as a bilinear form. 

Select an $x_o\in M$ and restrict attention to an open neighborhood $U$ of $x_o$ where $u\mapsto \text{rank}(\alpha_{\kappa u})$ takes the constant value $r$. Suppose $n$ is the largest integer with the property that there exists $n$ distinct basis elements $f^1\equiv\phi_{j_1},\dots,f^n\equiv\phi_{j_n}$ that satisfy
\begin{align}\label{independent}
(d\phi_{j_1})_{x_o}\wedge\dots\wedge(d\phi_{j_n})_{x_o}\neq 0.
\end{align}
Because the space of $n$-forms with $n>m=\dim(M)$ vanishes, $n$ must satisfy $n\leq m$. Therefore we must be able to find $m-n$ additional functions $f^{n+1},\dots,f^{m}$ so that the $f^i$ comprise a coordinate chart on $U$ (it may be the case that $U$ must be shrunk). In this special coordinate system on $U$, Eq.\,(\ref{independent}) implies that each $\phi_j$ must be independent of the coordinates $f^{n+1},\dots,f^{m}$. Thus each of the $m-n$ vector fields $\partial_{m+1}\equiv\frac{\partial~~~~}{\partial f^{m+1}},\dots,\partial_{n}\equiv\frac{\partial~}{\partial f^n}$ take values in the characteristic distribution of $\alpha_\kappa$. It follows that the dimension of the characteristic distribution, $m-r$, must be greater than or equal to $m-n$, i.e. $n\geq r$. In fact, $n$ cannot be greater than $r$. To see this, note that $\alpha_\kappa=\alpha_o+\delta\alpha$, where
\begin{align}
\alpha_o&=\sum_{i=1}^{n}(df^i)^2\\
\delta\alpha&=\sum_{j\not\in \{j_1,\dots,j_n\}}(d\phi_j)^2.
\end{align}
It is straightforward to show that $\text{rank}(\alpha_o)=n$. Moreover, because $\alpha_o$ and $\delta\alpha$ are each positive semi-definite bilinear forms (being sums of squared $1$-forms), $r=\text{rank}(\alpha_\kappa)\geq\text{rank}(\alpha_o)=n$. Therefore we must have $n=r$. This shows that $\partial_{r+1},\dots,\partial_{m}$ in fact span $\alpha_\kappa$'s characteristic distribution. Moreover, being pushforwards of some of the standard basis vectors in $\mathbb{R}^m$, these $m-r$ vector fields commute. We have therefore succeeded in proving that the commutator of any pair of vector fields that take values in $T_{\alpha_{\kappa}}M$ also takes values in $T_{\alpha_\kappa}M$; this is precisely integrability in the sense of Frobenius. Actually, we have nearly proved more than this. If $X$ is any vector field that takes values in $T_{\alpha_\kappa}M$, then it must be a $C^\infty(U)$-linear combination of the vectors $\partial_{r+1},\dots,\partial_{m}$. Therefore, 
\begin{align}
L_X\alpha_\kappa&=2\sum_j(L_Xd\phi_j)(d\phi_j)\nonumber\\
&=2\sum_j(dL_X\phi_j)(d\phi_j)=0,
\end{align} 
where, on the last line, we have used the fact that each $\phi_j$ is independent of $f^{n+1},\dots,f^{m}$. In other words, for each $X$ that takes values in $T_{\alpha_\kappa}M$, $L_X\alpha_\kappa=0$. It is not hard to show that this last property actually imples $T_\alpha M$ is integrable in the sense of Frobenius.

\end{proof}
 Next we will show that the necessary conditions for a constant-rank $\alpha$ to have a reproducing kernel primitive given in Theorem\,\ref{t1} are, in a particular sense, \emph{locally} sufficient.
\begin{theorem}\label{t2}
Suppose $\alpha\in\Gamma(T^*M\odot T^*M)$ is positive-semidefinite, has locally-constant rank, and that $L_X\alpha=0$ for each vector field $X$ that takes values in $T_\alpha M$. Then for each $x_o\in M$, there is an open set $U$ containing $x_o$ such that $\alpha|U=\alpha_{\kappa_U}$, where $\kappa_U:U\times U\rightarrow\mathbb{R}$ is a smooth reproducing kernel on the open submanifold $U$.
\end{theorem}
\begin{proof}
Let $r$ be the rank of $\alpha$ near $x_o$. By the Frobenius theorem we can choose coordinates $f^i$ on a neighborhood $U$ of $x_o$ with the following property. The vector fields $\frac{\partial~}{\partial f^1},\dots,\frac{\partial~}{\partial f^r}$ do not lie in $T_\alpha M$, while the vector fields $\partial_{r+1}\equiv \frac{\partial~~~}{\partial f^{r+1}},\dots,\partial_{m}\equiv\frac{\partial~}{\partial f^m}$ do. In this coordinate system, $\alpha$ must be of the form
\begin{align}\label{t21}
\alpha=\sum_{i,j=1}^r(df^i)A_{ij}(df^j),
\end{align}
where $A_{ij}$ is an $r\times r$ symmetric, positive definite matrix of functions on $U$ (note that the upper limit of the double sum in Eq.\,(\ref{t21}) is $r\leq m$). By assumption, $L_{\partial_l}\alpha=0$ for each $l\in\{r+1,\dots,m\}$, which implies
\begin{align}
L_{\partial_l}\alpha&=\sum_{i,j=1}^{r}(df^i)(df^j)\frac{\partial A_{ij}}{\partial f^l}=0.
\end{align}
Because the $(df^i)(df^j)$ with $i\leq j$ are linearly independent and the matrix $A_{ij}$ is symmetric, the previous equation shows that 
\begin{align}
\frac{\partial A_{ij}}{\partial f^l}=0,
\end{align}
for each $i,j\in\{1,\dots,r\}$ and $l\in \{r+1,\dots,m\}$.\, i.e. the $A_{ij}$ only depend on the first $r$ coordinates in this coordinate system, $A_{ij}=A_{ij}(f^1,\dots,f^r)$. 

Let $V\subset \mathbb{R}^r$ be the image of the submersion $\pi:u\in U\mapsto (f^1(u),\dots,f^r(u))\in\mathbb{R}^r$. Without loss of generality, we can assume that $V$ is open and connected. Because the $A_{ij}$ only depend on the first $r$ coordinates on $U$, they define a positive-definite bilinear form $g$ on $V$ given by
\begin{align}
g(x^1,\dots,x^r)=\sum_{i,j=1}^{r}A_{ij}(x^1,\dots,x^r)dx^i\,dx^j.
\end{align}
In other words, $(V,g)$ is a Riemannian manifold with metric tensor $g$. By Nash's embedding theorem, there is therefore an isometric embedding $I:(V,g)\rightarrow (\mathbb{R}^M,g_o)$ where $M$ is some integer and $g_o=\sum_{i=1}^{M}(dx^i)^2$ is the standard metric tensor on $\mathbb{R}^M$. In particular, 
\begin{align}
g=I^*g_o=\sum_{i=1}^{M}(ds_i)^2,
\end{align}
where $s_i=I^*x^i$.

Now, it is simple to verify that $\alpha=\pi^*g$, and therefore 
\begin{align}
\alpha=\pi^*\sum_{i=1}^{M}(ds^i)^2=\sum_{i=1}^{M}(d\phi_i)^2,
\end{align}
where $\phi_i=\pi^*I^*x^i$. We have therefore proved that $\alpha|U=\alpha_{\kappa}$, where $\kappa(x,y)=\sum_{i=1}^{M}\phi_i(x)\phi_i(y)$ is a reproducing kernel.

\end{proof}

If we assume that the leaf space of the foliation tangent to $\alpha$'s characteristic distribution is a manifold, the previous theorem can also be globalized in the following manner.

\begin{theorem}\label{t3}
Suppose $\alpha\in\Gamma(T^*M\odot T^*M)$ is positive semi-definite, has locally-constant rank, and that $L_X\alpha=0$ for each vector field $X$ that takes values in $T_\alpha M$. Also assume that the leaf space of the foliation tangent to $T_\alpha M$ is a smooth manifold with the quotient topology. Then there is a reproducing kernel $\kappa:M\times M\rightarrow\mathbb{R}$ such that $\alpha=\alpha_\kappa$.
\end{theorem}
\begin{proof}
Let $M_\alpha$ be the leaf space of the foliation tangent to $\alpha$'s characteristic distribution. Let $\pi:M\rightarrow M_\alpha$ be the map that sends a point in $M$ to its corresponding leaf. Because $L_X\alpha=0$ for each $X$ taking values in $T_\alpha M$, there is a unique $g\in \Gamma(T(M_\alpha)\odot T(M_\alpha))$ such that $\alpha=\pi^*g$. 

To see that $g$ is uniquely determined by the formula $\alpha=\pi^*g$, consider the following. Let $w_1,w_2\in T_{x}M_\alpha$ be a pair of vectors tangent to the leaf space at $x\in M_\alpha$. Let $\tilde{w}_1,\tilde{w_2}\in T_{\tilde{x}}M$ be any pair of vectors tangent to $M$ at $\tilde{x}\in M$ that satisfy $T\pi(\tilde{w}_i)=w_i$. Suppose $\tilde{w}_1^\prime,\tilde{w}_2^\prime\in T_{\tilde{x}^\prime}M$ are also lifts of the vectors $w_1,w_2$. Then there must be a diffeomorphism $\Phi:M\rightarrow M$ that preserves the leaves of the foliation tangent to $T_\alpha M$ (i.e. $\pi\circ\Phi=\pi$) such that $\tilde{x}^\prime=\Phi(\tilde{x})$. This diffeomorphism allows us to compare the vectors $\tilde{w}_1,\tilde{w}_2$ and $\tilde{w}_1^\prime,\tilde{w}_2^\prime$. In particular, we can consider the differences $\delta_1=\tilde{w}_1^\prime-T\Phi(\tilde{w}_1)$ and $\delta_2=\tilde{w}_2^\prime-T\Phi(\tilde{w}_1)$. We have
\begin{align}
T\pi(\delta_i)&=T\pi[\tilde{w}_i^\prime-T\Phi(\tilde{w}_i)]\nonumber\\
&=w_i-T(\pi\circ\Phi)(\tilde{w}_i)\nonumber\\
&=w_i-w_i=0.
\end{align}
This identity allows us to compute the difference between $\alpha(\tilde{w}_1^\prime,\tilde{w}_2^\prime)$ and $\alpha(\tilde{w}_1,\tilde{w}_2)$. Indeed,
\begin{align}
\alpha(\tilde{w}_1^\prime,\tilde{w}_2^\prime)&=\alpha(\delta_1+T\Phi(\tilde{w}_1),\delta_2+T\Phi(\tilde{w}_2))\nonumber\\
&=\alpha(T\Phi(\tilde{w}_1),T\Phi(\tilde{w}_2))\nonumber\\
&=\Phi^*\alpha(\tilde{w}_1,\tilde{w}_2)\nonumber\\
&=\alpha(\tilde{w}_1,\tilde{w}_2),
\end{align}
where the last line follows from $L_X\alpha=0$. It follows that $\alpha(\tilde{w}_1,\tilde{w}_2)$ depends only on $\alpha$, $w_1$, and $w_2$. Thus, $g(w_1,w_2)=\alpha(\tilde{w}_1,\tilde{w}_2)$ is well-defined.

The bilinear form $g$ is positive definite for if $w\in TM_\alpha$ is tangent to the leaf space, $g(w,w)=\alpha(\tilde{w},\tilde{w})$, and the right-hand-side is zero only when $\tilde{w}$ is tangent to $T_\alpha M$, i.e. when $w=0$. The pair $(M_\alpha,g)$ is therefore a Riemannian manifold. Nash's theorem then implies that there exists an isometric embedding $I:(M_\alpha,g)\rightarrow (\mathbb{R}^{N_o},g_o)$ where $g_o$ is the standard euclidean metric $g_o=\sum_{i=1}^{N_o}(dx^i)^2$. We have therefore proved that
\begin{align}
\alpha=\pi^*g=\pi^*I^*g_o=\sum_{i=1}^{N_o}(d\phi_j)^2,
\end{align}
where $\phi_j=\pi^*I^*x^j$. Equivalently, $\alpha=\alpha_\kappa$ where the kernel $\kappa(x,y)=\sum_{i=1}^{N_o}\phi_j(x)\phi_j(y).$
\end{proof}

We will now use Theorem\,\ref{t1} to prove that the Lorentz diffusion tensor $D_L$ does not admit a decomposition of the form $D_L=\sum_k X_{h_k}\otimes X_{h_k}$. This will constitute a proof that the classical Fokker-Planck equation for the Lorentz plasma is not Hamiltonian. Suppose that $D_L$ does admit such a decomposition. Then the symmetric covariant tensor on phase space $\alpha_L$ given by
\begin{align}
\alpha_L(X,Y)&=(\text{i}_X\omega_o)\cdot D_L\cdot(\text{i}_Y\omega_o),\nonumber\\
&=\sum_k (X\cdot dh_k)(dh_k\cdot Y)
\end{align}
admits the reproducing kernel primitive $\kappa(z_1,z_2)=\sum_k h_k(z_1)\,h_k(z_2)$. Moreover, because $\alpha_L$ can also be expressed as
\begin{align}
\alpha_L=\nu(|v|)\mathsf{U}_{ij}dx^idx^j
\end{align}
we also see that $\alpha_L$ is positive semi-definite and has the constant rank $2$ whenever $|v|\neq 0$. Therefore by Theorem\,\ref{t1}, we must have $L_{X}\alpha_L=0$ for any vector field $X$ that takes values in the kernel of $\alpha_L$. On the other hand, one such $X$ is given by $X=X_{H_o}=v\cdot\partial_x$, and
\begin{align}
&(L_{X_{H_o}}\alpha_L)(Y_1,Y_2)\nonumber\\
&=L_{X_{H_o}}(\alpha_L(Y_1,Y_2))-\alpha_L(Y_1,L_{X_{H_o}}Y_2)-\alpha_L(L_{X_{H_o}}Y_1,Y_2),
\end{align}
for arbitrary vector fields $Y_1,Y_2$ on phase space. In particular, when $Y_1=Y_2=w\cdot \partial_x+w\cdot \partial_v$ where $w$ is a constant $3$-component vector,
\begin{align}
(L_{X_{H_o}}\alpha_L)(Y_1,Y_2)=2w\cdot\mathsf{U}\cdot w,
\end{align}
which is never zero everywhere in phase space. This contradiction implies that $\alpha_L$ does not admit a reproducing kernel primitive, which in turn implies that $D_L$ does not admit a decomposition of the form $D_L=\sum_k X_{h_k}\otimes X_{h_k}$. Thus, the classical Fokker-Planck equation for the Lorentz plasma is not Hamiltonian.

\subsubsection{On energy-conserving Hamiltonian collision operators}
While $C_L$ is not Hamiltonian and $C_{HL}$ does not conserve energy, perhaps there is some other collision operator $\tilde{C}(f)=\text{div}(\tilde{D}\cdot df)$ that approximates $C_L$, satisfies the Hamiltonian property, \emph{and} conserves energy. The method introduced in BZQ would not be enough to find such an operator, but it is still worthwhile to ask if such an operator exists, even in principle. In this section we will show that that the existence of such a $\tilde{C}$ is unlikely in the following sense. We will say that the collision operator, $\text{div}(D\cdot df)$ in a Fokker-Planck equation conserves energy path-wise if the corresponding diffusion tensor satisfies $D\cdot dH_o=0$, i.e. $dH_o$ is a null eigenvector of $D$. Note that $C_L$ conserves energy path-wise. We will show that any Hamiltonian collision operator $\tilde{C}$ that conserves energy path-wise is necessarily quite different from $C_L$, i.e. $\tilde{C}$ must be a poor approximation to $C_L$.

Let $\tilde{D}=\sum_k X_{h_k}\otimes X_{h_k}$ be the diffusion tensor associated with the Hamiltonian collision operator $\tilde{C}$ that conserves energy path-wise. Associated with $\tilde{D}$ is the symmetric covariant tensor $\tilde{\alpha}=\sum_k(dh_k)^2$. Because $\tilde{D}\cdot dH_o=0$, $\tilde{\alpha}\cdot X_{H_o}=0$. In particular,
\begin{align}
0=\tilde{\alpha}(X_{H_o},X_{H_o})=\sum_k\{h_k,H_o\}^2,
\end{align} 
which implies that each of the $h_k$ Poisson commute with $H_o$. Therefore,
\begin{align}
L_{X_{H_o}}\tilde{\alpha}=2\sum_kd(\{h_k,H_o\})\,dh_k=0.
\end{align}

A reading of the proof of Theorem\,\ref{t3} now shows that there must be a symmetric covariant tensor $\tilde{a}$ defined on the \emph{space of free streaming trajectories} that pulls back to give $\tilde{\alpha}$. A free streaming trajectory is a subset of $TQ$ of the form
\begin{align}
\gamma_{(x_o,v_o)}=\{(x,v)\in TQ\mid v=v_o~\text{and}~\exists t\in\mathbb{R}\text{ s.t. }x=v_ot+x_o\},
\end{align}
where $(x_o,v_o)$ is an arbitrary point in $TQ$ with $v\neq 0$. The space of free streaming trajectories, $FS$, is simply the union of all free streaming trajectories. Because the free streaming trajectories with a given velocity $v$ can be identified with points in the plane perpendicular to $v$, $FS$ has the structure of a rank-$2$ vector bundle over the $3$-dimensional velocity space with the zero velocity excluded, $\mathbb{R}^3_o=\mathbb{R}^3-\{0\}$. To be precise, $FS$ is diffeomorphic to the subbundle of $\mathbb{R}^3_o\times\mathbb{R}^3$ given by
\begin{align}
FS=\{(v,x)\in \mathbb{R}^3_o\times\mathbb{R}^3\mid x\cdot v=0\}.
\end{align}
There is also a natural projection map $\pi:TQ\rightarrow FS$ given by $\pi(x,v)=(v,x_\perp)$, where $x_\perp=x-x\cdot\hat{v}\hat{v}$. We must have $\tilde{\alpha}=\pi^*\tilde{a}$, where $\tilde{a}$ is some symmetric covariant tensor on $FS$.

We will now analyze the form of $\tilde{\alpha}$ given that it must be the pullback of $\tilde{a}$ along $\pi$. Let $e_1(v),e_2(v)$ be orthogonal unit vectors that are everywhere perpendicular to $v$, i.e. $e_1(v)\cdot v=e_2(v)\cdot v=0$. Using these unit vectors, we can introduce local coordinates on $FS$, $(v,x_1,x_2)$, given by
\begin{align}
v&=v\\
x_1&=x\cdot e_1(v)\\
x_2&=x\cdot e_2(v),
\end{align}
where $(v,x)\in FS$. If $X=X^x\cdot\partial_x+X^v\cdot\partial_v$ is a vector on the velocity phase space, its pushforward along $\pi$ is given by
\begin{align}
T\pi(X)&=\bigg(X^x\cdot e_1+\frac{X^v}{|v|}\cdot(x_\parallel e_1+|v|R \,e_2\cdot x)\bigg)\frac{\partial}{\partial x_1}\nonumber\\
&+\bigg(X^x\cdot e_2+\frac{X^v}{|v|}\cdot(x_\parallel e_2-|v|R \,e_1\cdot x)\bigg)\frac{\partial}{\partial x_2}\nonumber\\
&+X^v\cdot\frac{\partial}{\partial v},
\end{align}
where $R=(\nabla_ve_1)\cdot e_2$ and $x_\parallel=x\cdot\hat{v}$. Now because
\begin{align}
\tilde{\alpha}(X_1,X_2)=\tilde{a}(T\pi(X_1),T\pi(X_2)),
\end{align}
and $\tilde{a}$ is independent of $x_\parallel$, we see that the components of $\tilde{\alpha}$ must become large as $x_\parallel$ becomes large. Because $D_L$ does not depend on $x$, let alone $x_\parallel$, by adjusting $x_\parallel$, $\tilde{D}$ can always be made much larger than $D_L$. This rules out the possibility that $\tilde{D}$ could be a good approximation to $D_L$.

\subsection{Outlook}
We have succeeded in showing that there is a Hamiltonian Fokker-Planck equation that governs the dynamics of the Lorentz plasma. This suggests that pitch angle scattering dynamics is Hamiltonian in a stochastic sense. On the other hand, we have also shown that our Hamiltonian Fokker-Planck equation does not have an exact energy conservation law. Instead the mean kinetic energy grows on a time scale proportional to $\tau/\tau_{\text{ac}}$, where $\tau$ is the Fokker-Planck time step and $\tau_{\text{ac}}=\omega_p^{-1}$.

When inquiring as to why $C_{HL}$ produces energy, it is useful to remember the following fact about the derivation of Fokker-Planck equations.
In these derivations, one tacitly (or explicitly, as we have done) divides the time axis into $\tau$-second intervals and studies the dynamics of an electron on a typical such interval. Because the positions of the ions are assumed to be statistically independent, this problem is reduced to studying the interaction of a single electron with a single Debye screened ion on a $\tau$-second time interval. Given the location of the ion, most electrons will both begin and end their $\tau$-second journeys well outside of the support of the ion's potential. Each of these electrons will not suffer any change in its kinetic energy. However, there are some electrons that will either start or end within the support of the ion's potential. Each of these electrons will suffer a change in their kinetic energy as a result of either climbing out of or falling into the ion's potential well. Thus, an ensemble of electrons will \emph{not} precisely conserve its kinetic energy over a $\tau$-second time interval.

While this reasoning sheds some light on the energy-production problem, it is still not completely satisfactory. Yes, an ensemble of electrons will only approximately conserve its kinetic energy over $\tau$-second time intervals. However, it is not physically true that the ensemble's mean kinetic energy can grow without bound, in contrast with the long-time behavior predicted by $C_{HL}$ (see Eq.\,(\ref{energy_growth})). This would correspond to electron heating, which is impossible because there is no energy impinging on the Lorentz plasma.

The key to a complete understanding of the energetics of $C{HL}$ is a careful scrutinization of the Markov assumption. This assumption artificially eliminates any memory an electron might have of its past after each $\tau$-second time interval. As a result, within the Markov model, the following non-physical dynamical process is possible. After moving for $\tau$ seconds, an electron finds itself within the support of an ion's potential. Whereas this electron should begin the next $\tau$-second interval by climbing out of this potential, instead it forgets the locations of all ions during the previous step, and, with high probability, fails to shed any of the kinetic energy it gained. This type of unphysical behavior allowed within the context of the Markov model is ultimately the source of the artificial heating predicted by Eq.\,(\ref{energy_growth}). Moreover, it can be shown that this heating is not present in the classical Fokker-Planck equation for the Lorentz plasma because the probability that one of the unphysical processes just discussed occurs tends to zero as $\tau_{\text{ac}}/\tau$ tends to zero.

We believe this explanation of why $C_{HL}$ causes slow artificial heating suggests a way to reconcile the stochastic Hamiltonian approach with energy conservation. The idea is to slightly relax the Markov assumption. In the rare event that electron ends a $\tau$-second time interval within the support of an ion's potential, then the electrostatic field it sees in the next $\tau$-second interval should consist of the the old ion's potential plus the potential of a new ion. Each time a new ion is drawn at random, it should never been chosen so that the support of its potential intersects the electron's location. By allowing for this small memory effect, the unphysical electron trajectories that cause $C_{HL}$ to produce energy would be eliminated. Moreover, the evolution of a given electron would still be given by iterated symplectic mappings on phase space, and this ought to keep things within the realm of stochastic Hamiltonian mechanics. A challenging, yet enticing aspect of this possible route to overcoming the shortcomings of this Chapter would be developing the theory of stochastic Hamiltonian processes with memory.

\section{Acknowledgements} The results presented in the first section of this Chapter were obtained in conjunction with Andrey Zhmoginov. They are available on the arXiv at arXiv:1312.3974.



\singlespacing
\bibliographystyle{plainnatJosh}

\cleardoublepage
\ifdefined\phantomsection
  \phantomsection  
\else
\fi
\addcontentsline{toc}{chapter}{Bibliography}

\providecommand{\noopsort}[1]{}\providecommand{\singleletter}[1]{#1}%

\end{document}